\ifx \labelsONmargin\undefined
  \ifx\RequirePackage\undefined
    \expandafter\ifx\csname amsart.sty\endcsname\relax
        \documentstyle[12pt,amssymb,amscd]{amsart}
    \fi

    \def\atSign{@@}

    \def\mathbb{\Bbb}
    \def\mathfrak{\frak}
    \def\mathbf{\bold}
    \ifx\boldsymbol\undefined   
      \def\boldsymbol#1{{\bold #1}}
    \fi

    \def\mathbit{\boldsymbol}

    \makeatletter
    \newenvironment{proof}{%
         \@ifnextchar[{%
                       \expandafter\let\expandafter\end@proof
                         \csname endpf*\endcsname
                         \my@proof
                      }{\let\end@proof\endpf\pf}%
        }{\end@proof}
    \def\my@proof[#1]{\@nameuse{pf*}{#1}}
    \makeatother

    \def\xrightarrow[#1]#2{@>{#2}>{#1}>}
    \def\xleftarrow[#1]#2{@<{#2}<{#1}<}

    \def\providecommand#1{\def#1}
    \def\emph#1{{\em #1}}
    \def\textbf#1{{\bf #1}}
  \else
      \ifx\usepackage\RequirePackage
      \else
        \documentclass[12pt]{amsart}
        \usepackage{amssymb,amscd}
      \fi

      \ifx\mathbit\undefined    
        \DeclareMathAlphabet{\mathbit}{OML}{cmm}{b}{it}
      \fi

      \def\atSign{@}

      \advance\oddsidemargin -0.1\textwidth
      \evensidemargin\oddsidemargin
      \def\Sb#1\endSb{_{\substack{#1}}}
      \def\Sp#1\endSp{^{\substack{#1}}}
  \fi
\makeatletter
\@ifundefined{DeclareFontShape} 
     {\@ifundefined{selectfont}
             {
                \message{We need AmS-LaTeX, this version of LaTeX does not 
                        support it}\@@end
             }{
                \def\mathcal{\cal}
                
                \def\pcyr{%
                        \def\default@family{UWCyr}%
                        \let\oldSl@\sl
                        \def\sl{\def\default@shape{it}\oldSl@}%
                        \cyracc
                        \language\Russian\family{UWCyr}\selectfont
                }
             }}{
                \DeclareFontEncoding{OT2}{}{\noaccents@}
                \DeclareFontSubstitution{OT2}{cmr}{m}{n}
                \DeclareFontFamily{OT2}{cmr}{\hyphenchar\font45 }
                \DeclareFontShape{OT2}{cmr}{m}{n}{%
                     <5><6><7><8><9><10>gen*wncyr %
                     <10.95><12><14.4><17.28><20.74><24.88> wncyr10 %
                }{}
                \DeclareFontShape{OT2}{cmr}{m}{it}{%
                     <5><6><7><8><9><10> gen * wncyi%
                     <10.95><12><14.4><17.28><20.74><24.88> wncyi10%
                }{}
                \DeclareFontShape{OT2}{cmr}{bx}{n}{%
                     <5><6><7><8><9><10> gen * wncyb%
                     <10.95><12><14.4><17.28><20.74><24.88> wncyb10%
                }{}
                \DeclareFontShape{OT2}{cmr}{m}{sl}{%
                     <-> ssub * cmr/m/it%
                }{}
                \DeclareFontShape{OT2}{cmr}{m}{sc}{%
                     <5><6><7><8><9><10>%
                     <10.95><12><14.4><17.28><20.74><24.88> wncysc10%
                }{}
                \DeclareFontFamily{OT2}{cmss}{\hyphenchar\font45 }
                \DeclareFontShape{OT2}{cmss}{m}{n}{%
                     <8><9><10> gen * wncyss%
                     <10.95><12><14.4><17.28><20.74><24.88> wncyss10%
                }{}
                
                \def\cyrencodingdefault{OT2}
                \def\pcyr{%
                        \cyracc
                        \let\encodingdefault\cyrencodingdefault
                        \language\Russian\fontencoding{OT2}\selectfont
                }
             }   

\@ifundefined{theorembodyfont} 
     {
        \def\theorembodyfont#1{\relax}
        \makeatletter
          \let\@@th@plain\th@plain
          \def\th@plain{ \@@th@plain \slshape }
        \makeatother
        \let\normalshape\relax
     }{}   

\ifx\cprime\undefined
     \def\cprime{$'$}
\fi
\makeatletter
  \def\@sect@my#1#2#3#4#5#6[#7]#8{%
\ifnum #2>\c@secnumdepth
   \let\@svsec\@empty
 \else
   \refstepcounter{#1}%
\edef\@svsec{\ifnum#2<\@m
             \@ifundefined{#1name}{}{\csname #1name\endcsname\ }\fi
\noexpand\rom{\csname the#1\endcsname.}\enspace}\fi
 \@tempskipa #5\relax
 \ifdim \@tempskipa>\z@ 
   \begingroup #6\relax
   \@hangfrom{\hskip #3\relax\@svsec}{\interlinepenalty\@M #8\par}%
   \endgroup
   \if@article\else\csname #1mark\endcsname{%
        \ifnum \c@secnumdepth >#2\relax\csname the#1\endcsname. \fi#7}\fi
\ifnum#2>\@m \else
       \let\@tempf\\ \def\\{\protect\\}\addcontentsline{toc}{#1}%
{\ifnum #2>\c@secnumdepth \else
             \protect\numberline{%
               \ifnum#2<\@m
               \@ifundefined{#1name}{}{\csname #1name\endcsname\ }\fi
               \csname the#1\endcsname.}\fi
           #8}\let\\\@tempf
     \fi
 \else
  \def\@svsechd{#6\hskip #3\@svsec
    \@ifnotempty{#8}{\ignorespaces#8\unskip
       \ifnum\spacefactor<1001.\fi}%
        \ifnum#2>\@m \else
          \let\@tempf\\ \def\\{\protect\\}\addcontentsline{toc}{#1}%
            {\ifnum #2>\c@secnumdepth \else
              \protect\numberline{%
                \ifnum#2<\@m
                \@ifundefined{#1name}{}{\csname #1name\endcsname\ }\fi
                \csname the#1\endcsname.}\fi
             #8}\let\\\@tempf\fi}%
 \fi
\@xsect{#5}}
  \ifx\RequirePackage\undefined
  \let\@sect\@sect@my             
  \fi
  \def\th@remark@my{\theorempreskipamount6\p@\@plus6\p@
    \theorempostskipamount\theorempreskipamount
    \def\theorem@headerfont{\it}\normalshape}
  \ifx\RequirePackage\undefined
  \let\th@remark\th@remark@my
  \fi
\let\myLabel\@gobble
\def\labelsONmargin{\@mparswitchfalse\def\myLabel##1{\@bsphack\marginpar
                                  {\normalshape\tiny\rm Label ##1}\@esphack}}
\ifx \url\undefined
  \def\url#1{{\tt #1}}%
\fi
\def\cyracc{\def\u##1{
                \if \i##1\char"1A%
                \else \if I##1\char"12%
                \else \accent"24 ##1\fi\fi }%
\def\"##1{\if e##1{\char"1B}%
                \else \if E##1{\char"13}%
                \else \accent"7F ##1\fi\fi }%
\def\9##1{\if##1z\char"19 
\else\if##1Z\char"11 
\else\if##1E\char"03 
\else\if##1e\char"0B 
\else\if##1u\char"18 
\else\if##1U\char"10 
\else\if##1A\char"17 
\else\if##1a\char"1F 
\else\if##1p\char"7E 
\else\if##1P\char"5E 
\else\if##1Q\char"5F 
\else\if##1q\char"7F 
\else\if##1i\char"1A 
\else\if##1I\char"12 
\else\if##1N\char"7D 
\fi
\fi
\fi
\fi
\fi
\fi
\fi
\fi
\fi
\fi
\fi
\fi
\fi
\fi
\fi
}%
\def\cydot{{\kern0pt}}}%
\def\cydot{$\cdot$}

\ifx\Russian\undefined
        \def\Russian{0\relax
    \message{Don't know the hyphenation rules for Russian^^J
                        Please do INITeX with `input  russhyph' in the 
                        command line}%
                \gdef\Russian{0\relax}%
        }
\fi
\def\@putname#1#2#3#4{\def\@@ref{#3}\let\old@bf\bf      
        \let\old@reset@font\reset@font                  
        \def\bf##1{\old@bf\if?\noexpand##1?{#4}\else##1\fi}%
        \def\reset@font##1##2{\old@reset@font##1\if?\noexpand##2?{#4}\else##2\fi}#1{#2}%
        \let\bf\old@bf\let\reset@font\old@reset@font}
\let\my@ref=\ref
\def\ref#1{\@putname\my@ref{#1}{#1}{\tiny\rm\@@ref}}
\let\my@pageref=\pageref
\def\pageref#1{\@putname\my@pageref{#1}{#1}{\tiny\rm\@@ref}}
\let\my@cite=\cite
\def\cite#1{\@putname\my@cite{#1}{\@citeb}{\tiny\rm\@@ref}}
\makeatother
\fi
\relax 
\theoremstyle{plain} 
\theorembodyfont{\sl}

\numberwithin{equation}{section}
\theorembodyfont{\rm}
\theoremstyle{definition}

\newtheorem{definition}{Definition}[section]

\theoremstyle{remark}

\newtheorem{remark}[definition]{Remark} 

\theoremstyle{plain} 
\theorembodyfont{\sl}

\newtheorem{theorem}[definition]{Theorem}
\newtheorem{lemma}[definition]{Lemma}
\newtheorem{corollary}[definition]{Corollary}
\newtheorem{proposition}[definition]{Proposition}
\newtheorem{amplification}[definition]{Amplification}

\begin{document}
\bibliographystyle{amsplain}
\relax 

\title[Nonlinear wave equation and twistor transform of webs]{ Nonlinear wave
equation, \\
nonlinear Riemann problem, \\
and the twistor transform of Veronese webs \\
}

\author{ Ilya Zakharevich }

\address{ Department of Mathematics, Ohio State University, 231 W.~18~Ave,
Columbus, OH, 43210 }

\email {ilya\atSign{}math.ohio-state.edu}

\date{ June 2000\quad Archived as \url{math-ph/0006001} ---Preliminary version
Printed: \today }

\setcounter{section}{-1}

\maketitle
\begin{abstract}
Veronese webs are rich geometric structures with deep
relationships to various domains of mathematics. The PDEs which determine
the Veronese web are overdetermined if $ \dim >3 $, but in the case $ \dim =3 $ they
reduce to a special flavor of a non-linear wave equation. The symmetries
embedded in the definition of a Veronese web reveal themselves as
B\"acklund--Darboux transformations between these non-linear wave equations.

On the other hand, the twistor transform identifies Veronese webs
with moduli spaces of rational curves on certain complex surfaces. These
moduli spaces can be described in terms of the non-linear Riemann
problem. This reduces solutions of these non-linear wave equations to the
non-linear Riemann problem.

We examine these relationships in the particular case of
$ 3 $-dimensional Veronese webs, simultaneously investigating how these
notions relate to general notions of geometry of webs.

\end{abstract}
\tableofcontents

\section{Introduction }\label{h0}\myLabel{h0}\relax 

We denote the $ d $-dimensional coordinate vector space over the base
field by $ {\mathbb V}^{d} $, and the corresponding projective space by $ {\mathbb P}^{d-1} $. For
$ \left(v_{1},\dots ,v_{d}\right)\in{\mathbb V}^{d}\smallsetminus\left\{0\right\} $ we denote by $ \left(v_{1}:\dots :v_{d}\right) $ the corresponding element of
$ {\mathbb P}^{d-1} $. By $ {\mathbb B}_{r}^{d}\subset{\mathbb V}^{d} $ we denote the open ball of radius $ r $ centered at the
origin. Then $ \left({\mathbb B}_{r}^{1}\right)^{d} $ is a cube in the real case and a polydisk in the
complex case.

As a convention, put $ |\infty|=\infty $, so that $ \left\{|z|>1\right\} $ includes $ z=\infty $.

The word ``smooth'' can have 3 different meanings: in the case of the
base field $ {\mathbb R} $ it can mean either $ C^{\infty} $-smooth or real-analytic, in the case
of the base field $ {\mathbb C} $ it means complex-analitic. When only some of these
cases work, we use more specific terms.

In this paper we study a special family of nonlinear wave equations.
Elements of this family are parameterized by numbers $ A,B,C $ which satisfy
\begin{equation}
A\not=0\text{, }B\not=0\text{, }C\not=0\text{, }A+B+C=0.
\label{equ0.20}\end{equation}\myLabel{equ0.20,}\relax 
Given such numbers, the equation is
\begin{equation}
Aw_{x}w_{yz}+Bw_{y}w_{xz}+Cw_{z}w_{xy} =0;
\notag\end{equation}
here $ w\left(x,y,z\right) $ is a function of three variables, If we need to specify
$ A,B,C $, we may call this equation the $ \left(A,B,C\right) $-{\em equation}. Whenever we mention
an $ \left(A,B,C\right) $-equation we assume that $ A,B,C $ satisfy~\eqref{equ0.20}.

In this paper we study only those solutions of $ \left(A,B,C\right) $-equations
which are in general position, according the the following

\begin{definition} \label{def0.20}\myLabel{def0.20}\relax  Say that a function $ w\left(x,y,z\right) $ is {\em non-degenerate\/} if
$ w_{x}\not=0 $, $ w_{y}\not=0 $, $ w_{z}\not=0 $ whenever $ w\left(x,y,z\right) $ is defined. \end{definition}

\begin{definition} \label{def0.30}\myLabel{def0.30}\relax  Say that two functions $ w\left(x,y,z\right) $ and $ w'\left(x,y,z\right) $ are
{\em gauge transforms\/} of each other, if $ w=\tau\circ w' $ for an appropriate invertible
scalar function $ \tau $ of one variable. \end{definition}

The first target of this paper is the following statement:

\begin{theorem} \label{th5.50}\myLabel{th5.50}\relax  Suppose that triples $ \left(A,B,C\right) $ and $ \left(\widetilde{A},\widetilde{B},\widetilde{C}\right) $ satisfy
conditions~\eqref{equ0.20}. Consider equations
\begin{align} Aw_{x}w_{yz}+Bw_{y}w_{xz}+Cw_{z}w_{xy} & =0,
\label{equ0.10}\\
\widetilde{A}v_{x}v_{yz}+\widetilde{B}v_{y}v_{xz}+\widetilde{C}v_{z}v_{xy} & =0,
\label{equ5.50}
\end{align}\myLabel{equ0.10,equ5.50,}\relax 
the system of equations
\begin{equation}
\begin{aligned}
A\widetilde{B} w_{x}v_{y} & = \widetilde{A}B w_{y}v_{x}
\\
A\widetilde{C} w_{x}v_{z} & = \widetilde{A}C w_{z}v_{x},
\end{aligned}
\label{equ5.60}\end{equation}\myLabel{equ5.60,}\relax 
and the equation (with $ \alpha=A/\widetilde{A} $, $ \beta=B/\widetilde{B} $, $ \gamma=C/\widetilde{C} $)
\begin{equation}
\left(v_{x},v_{y},v_{z}\right) \sim \left(\alpha w_{x},\beta w_{y},\gamma w_{z}\right),
\label{equ5.70}\end{equation}\myLabel{equ5.70,}\relax 
here for two vector-functions we write $ a\sim b $ if $ a\left(x,y,z\right)=\psi\left(x,y,z\right)b\left(x,y,z\right) $
for an appropriate nowhere-0 scalar function $ \psi $. Then locally near
$ \left(x,y,z\right)=\left(0,0,0\right) $
\begin{enumerate}
\item
For non-degenerate functions $ w $, $ v $ System~\eqref{equ5.60} is equivalent to
Equation~\eqref{equ5.70};
\item
Given a solution $ \left(w,v\right) $ of System~\eqref{equ5.60} with non-degenerate $ w $ and
$ v $ and any gauge transforms $ w_{1} $ of $ w $ and $ v_{1} $ of $ v $ the pair $ \left(w_{1},v_{1}\right) $ is a
solution of System~\eqref{equ5.60};
\item
Suppose that $ A\widetilde{B}\not=\widetilde{A}B. $ Given a solution $ \left(w,v\right) $ of System~\eqref{equ5.60} with
non-degenerate $ w $ and $ v $, the function $ w $ satisfies Equation~\eqref{equ0.10}, the
function $ v $ satisfies Equation~\eqref{equ5.50};
\item
Suppose that $ A\widetilde{B}\not=\widetilde{A}B. $ Given a non-degenerate solution $ w $ of Equation
~\eqref{equ0.10}, there is a non-degenerate function $ v $ such that the pair $ \left(w,v\right) $
satisfies Equations~\eqref{equ5.60}. As a corollary, $ v $ satisfies Equation
~\eqref{equ5.50};
\item
Such a function $ v $ is defined uniquely up to a gauge transform.
\end{enumerate}
\end{theorem}

Theorem~\ref{th5.50} is proved in Section~\ref{h5}. While one could prove this
theorem purely analytically, we emphasize the geometric meaning of its
statements, thus prove it via relationship to $ 3 $-dimensional Veronese webs,
which are introduced in Sections~\ref{h1} and~\ref{h3}.

\begin{remark} \label{rem5.70}\myLabel{rem5.70}\relax  One can consider the last two statements of Theorem
~\ref{th5.50} as statements about existence of non-pointwise relationship
between Equations~\eqref{equ0.10} and~\eqref{equ5.50}. Given a solution $ w $ of Equation
~\eqref{equ0.10}, one obtains a (more or less unique) solution $ v $ of Equation
~\eqref{equ5.50} by solving Equations~\eqref{equ5.60}. Note that the latter equations
are equations of lower order than~\eqref{equ5.50} when considered as equations
in $ v $.

In other words, System~\eqref{equ5.60} provides a B\"acklund--Darboux
transform of order 1 between two equations~\eqref{equ0.10} and~\eqref{equ5.50} of order
2. Moreover, this transform is linear in $ v $. \end{remark}

The second target of this paper is to explicitly solve any
$ \left(A,B,C\right) $-equation in complex domain in terms of the {\em nonlinear Riemann
problem}. This problem is a straightforward nonlinear analogue of the
(linear) Riemann conjugation problem:

\begin{definition} \label{def11.40}\myLabel{def11.40}\relax  Consider a complex-analytic function $ g\left(\lambda,t\right) $ defined
for $ \varepsilon<|\lambda|<1/\varepsilon $ and $ |t|<\delta $, assume that for any given $ \lambda $, $ \varepsilon<|\lambda|<1/\varepsilon $, the
function $ t \mapsto g\left(\lambda,t\right) $ is invertible. Suppose that equations
\begin{equation}
\sigma_{-}\left(\lambda\right)=g\left(\lambda,\sigma_{+}\left(\lambda\right)\right)\text{ for }\varepsilon<|\lambda|<1/\varepsilon,\qquad |\sigma_{+}\left(\lambda\right)|<\delta\text{ for }|\lambda|<1/\varepsilon,
\notag\end{equation}
uniquely determine complex-analytic functions $ \sigma_{+}\left(\lambda\right) $ defined for $ |\lambda|<1/\varepsilon $,
and $ \sigma_{-}\left(\lambda\right) $ defined for $ |\lambda|>\varepsilon $. Denote the number $ \sigma_{+}\left(0\right) $ by $ {\mathfrak R}_{\varepsilon\delta}\left(g\right) $. \end{definition}

The function $ {\mathfrak R}_{\varepsilon\delta} $ sends a function $ g\left(\lambda,t\right) $ of two variables to a
complex number. Call this function the {\em non-linear Riemann transform}. Note
that changing $ \varepsilon $ and $ \delta $ cannot change the value of $ {\mathfrak R}_{\varepsilon\delta}\left(g\right) $ (though this
expression can become undefined), thus we are going to drop $ \varepsilon $, $ \delta $ and
denote this function by $ {\mathfrak R} $.

The next step is to define a special family $ g_{x,y,z}\left(\lambda,t\right) $ of functions
of two variables, given one function $ g\left(\lambda,t\right) $. Use notation $ F_{{\text M},k}\left(\lambda\right) $ for
Lagrange interpolation polynomials on points $ {\text M}=\left\{\mu_{1},\dots ,\mu_{m}\right\} $:
\begin{equation}
F_{{\text M},l}\left(\lambda\right)=F_{{\text M}\smallsetminus\left\{\mu_{l}\right\}}\left(\lambda\right)/F_{{\text M}\smallsetminus\left\{\mu_{l}\right\}}\left(\mu_{l}\right),\quad F_{{\text M}}\left(\lambda\right)=\prod_{\mu\in{\text M}}\left(\lambda-\mu_{n}\right).
\notag\end{equation}
\begin{definition} \label{def0.60}\myLabel{def0.60}\relax  Consider sets of $ k $ numbers $ \Lambda=\left\{\lambda_{1},\dots ,\lambda_{k}\right\} $ and of $ m $
numbers $ {\text M}=\left\{\mu_{1},\dots ,\mu_{m}\right\} $ satisfying $ |\lambda_{l}|>0 $, $ |\mu_{l}|>0 $. Consider a function
$ g\left(\lambda,t\right) $. Let $ \lambda_{0}=0 $, $ \Lambda_{0}=\Lambda\cup\left\{0\right\} $, $ F_{+}=F_{\Lambda_{0},0} $, $ F_{+,l}=F_{\Lambda_{0},l} $, $ F_{-}\left(\lambda\right)=F_{{\text M}}\left(\lambda\right)/\lambda^{m} $,
$ F_{-,l}=F_{{\text M},l}\left(\lambda\right)\mu_{l}^{m-1}/\lambda^{m-1} $. For collections $ \left\{a_{i}\right\} $ and $ \left\{b_{i}\right\} $ of $ k $ and $ m $ numbers
correspondingly denote
\begin{equation}
{\mathcal G}_{\Lambda{\text M},\left\{a_{i}\right\}\left\{b_{i}\right\}}\left(\lambda,t\right)=F_{-}\left(\lambda\right)^{-1}\left(g\left(\lambda,\widetilde{t}\right)-\sum_{l=1}^{m}b_{l}F_{-,l}\left(\lambda\right)\right),\qquad \widetilde{t}=tF_{+}\left(\lambda\right)+\sum_{l=1}^{k}a_{l}F_{+,l}\left(\lambda\right).
\notag\end{equation}
Given 3 numbers $ \lambda_{1},\lambda_{2},\lambda_{3} $, let $ g_{x,y,z}\left(\lambda,t\right)\buildrel{\text{def}}\over{=}{\mathcal G}_{\left\{\lambda_{1},\lambda_{2}\right\}\left\{\lambda_{3}\right\},\left\{x,y\right\}\left\{z\right\}}\left(\lambda,t\right) $. \end{definition}

Given a function $ \varphi\left(\lambda\right) $, $ |\lambda|=1 $, define $ \operatorname{ind} \varphi $ as
$ \frac{1}{2\pi i}\oint_{|\lambda|=1}\frac{d\varphi\left(\lambda\right)}{\varphi\left(\lambda\right)} $.

\begin{theorem} \label{th16.70}\myLabel{th16.70}\relax  Consider a complex-analytic function $ g\left(\lambda,t\right) $ defined for
$ \varepsilon<|\lambda|<1/\varepsilon $ and $ |t|<\delta $, such that $ g\left(\lambda,0\right)\equiv 0 $ and $ \operatorname{ind}\frac{\partial g}{\partial t}\left(\lambda,0\right)=-2 $. Fix
$ 0<r<1 $, $ \lambda_{1},\lambda_{2},\lambda_{3}\in{\mathbb P}^{1} $, $ 0<|\lambda_{1,2}|<r $, $ |\lambda_{3}|>1/r $. Then
\begin{enumerate}
\item
the function
\begin{equation}
w\left(x,y,z\right)={\mathfrak R}\left(g_{x,y,z}\right)
\label{equ11.40}\end{equation}\myLabel{equ11.40,}\relax 
is correctly defined for small $ x $, $ y $, $ z $, is complex-analytic and
nondegenerate, and satisfies the equation~\eqref{equ0.10} with
\begin{equation}
A=\lambda_{1}\left(\lambda_{2}-\lambda_{3}\right)\text{, }B=\lambda_{2}\left(\lambda_{3}-\lambda_{1}\right)\text{, }C=\lambda_{3}\left(\lambda_{1}-\lambda_{2}\right);
\label{equ11.50}\end{equation}\myLabel{equ11.50,}\relax 

\item
for any scalar functions $ \psi $, $ \varphi_{1} $, $ \varphi_{2} $, $ \varphi_{3} $ of one variable which send 0
to 0 the function $ \widehat{w}=\psi\left(w\left(\varphi_{1}\left(x\right),\varphi_{2}\left(y\right),\varphi_{3}\left(z\right)\right)\right) $ satisfies the same
$ \left(A,B,C\right) $-equation as $ w\left(x,y,z\right) $;
\item
for any triple $ \left(\widetilde{A},\widetilde{B},\widetilde{C}\right) $ which satisfies conditions~\eqref{equ0.20} one can
find $ \lambda_{1},\lambda_{2},\lambda_{3}\in{\mathbb P}^{1} $ and $ T\not=0 $ which satisfy the above inequalities and
Equation~\eqref{equ11.50} with $ A=T\widetilde{A} $, $ B=T\widetilde{B} $, $ C=T\widetilde{C} $;
\item
for any nondegenerate complex-analytic solution $ \widehat{w}\left(x,y,z\right) $ of~\eqref{equ0.10}
defined near (0,0,0) the function $ g\left(\lambda,t\right) $ constructed in Theorem~\ref{th107.40}
(for some particular value of the function $ Y\left(x\right) $) satisfies the conditions
above, and $ \widehat{w}=\psi\left(w\left(x,\varphi_{2}\left(y\right),z\right)\right) $; here $ w\left(x,y,z\right) $ is defined by~\eqref{equ11.40}, $ \varphi_{2} $
is the inverse function to $ y=Y\left(x\right) $, and $ \psi\left(t\right)=\widehat{w}\left(t,Y\left(t\right),0\right) $.
\end{enumerate}
\end{theorem}

This theorem is proved in Section~\ref{h11}.

\begin{remark} Note that Theorem~\ref{th107.40} determines the gluing function
$ g\left(\lambda,t\right) $ in terms of the values of $ w $ and the normal derivative of $ w $ on a
hypersurface. Thus Theorem~\ref{th16.70} can be considered as a procedure to
solve Equation~\eqref{equ0.10} basing on the Cauchy initial data.

Such an approach would not gain a lot if the nonlinear Riemann
problem were complicated to solve. However, in Section~\ref{h12} we are going
to show that it is as complicated as solving an ODE of high dimension. \end{remark}

{\bf Plan}. In Sections~\ref{h1} and~\ref{h3} we define Veronese webs. In Section
~\ref{h4} we show that constructing a $ 3 $-dimensional Veronese web is equivalent
to solving an $ \left(A,B,C\right) $-equation. In Section~\ref{h5} we prove Theorem~\ref{th5.50},
thus construct B\"acklund--Darboux transformations between different
$ \left(A,B,C\right) $-equations. In Sections~\ref{h5.5} and~\ref{h6} we show how Veronese webs
jump into existence given the statement of Theorem~\ref{th5.50}.

Sections~\ref{h65} and~\ref{h7} contain first encounters with the twistor
transform of the Veronese web. Although full of technical (and long but
simple) statements, these sections enable working with the twistor
transform as with a manifold (as opposed to a germ), thus remove many
linguistic complications. In Section~\ref{h10} we introduce convenient
coordinate systems on the twistor transform, in Section~\ref{h107} we describe
the gluing functions as solutions of appropriate ODEs.

Section~\ref{h75} starts dealing with the inverse problem of
reconstructing the web by its twistor transform. After recalling what are
infinitesimal deformations of submanifolds, we obtain the first solution
of the inverse problem, the solution which requires a lot of additional
data. Section~\ref{h8} contains technical results which would allow to drop
these additional data in complex-analytic cases: Kodaira--Spencer
deformation theory for sections of bundles (Theorem~\ref{th9.50}), and the
``inverse'' theory (Proposition~\ref{prop9.70}) which explicitly constructs
a small tubular neighborhood in which the deformation theory works.

Section~\ref{h85} studies in which cases the ``additional data'' of the
inverse twistor transform can be dropped. We call such webs {\em airy webs},
and show that Veronese webs are airy. This section also provides an
alternative heuristic for utility of so-called {\em Kronecker webs\/} introduced
in \cite{Zakh99Kro}: they are airy webs with the parameter space being $ {\mathbb P}^{1} $.

Section~\ref{h11} completes the full circle by proving Theorem~\ref{th16.70},
thus providing the explicit construction of the inverse twistor
transform. Given a non-degenerate solution of the $ \left(A,B,C\right) $-equation,
Section~\ref{h107} had shown how to explicitly calculate gluing functions for
the twistor transform via solutions of ODEs. Section~\ref{h11} shows how to
use these gluing data for reconstruction of the initial solution of the
$ \left(A,B,C\right) $-equation. Theorem~\ref{th16.70} provides a way to completely integrate
the $ \left(A,B,C\right) $-equation in the non-degenerate case.

The first Appendix (Section~\ref{h105}) connects the results of Section
~\ref{h107} with Turiel classification of Veronese webs of arbitrary dimension
\cite{Tur99Cla,Tur99MemB}. Additionally, we introduce terms using which one
can classify arbitrary airy webs of codimension 1. The second Appendix
(Section~\ref{h12}) shows that the nonlinear Riemann problem is not harder to
solve than Lipschitz ODEs in Hilbert spaces.

\section{Webs }\label{h1}\myLabel{h1}\relax 

Recall the definition of a foliation.

\begin{definition} A {\em prefoliation\/} $ {\mathcal F} $ of codimension $ r $ on a manifold $ M $ is a
representation of $ M $ as a disjoint union of subsets called {\em leaves}, each of
which is a connected embedded submanifold of codimension $ r $. \end{definition}

Given an open subset $ U\subset M $, one can define a {\em restriction\/} $ {\mathcal F}|_{U} $ of $ {\mathcal F} $ to
$ U $, the leaves of which are connected components of $ L\cap U $, $ L $ running through
leaves of $ {\mathcal F} $. Say that $ {\mathcal F} $ is {\em direct\/} if $ M=N\times F $ with a connected $ F $, and leaves
are $ \left\{n\right\}\times F $, $ n\in N $. In such a case $ N $ is called the {\em base\/} of $ {\mathcal F} $.

The {\em tangent space\/} $ {\mathcal T}_{m}{\mathcal F} $ to $ {\mathcal F} $ at $ m\in M $ is the tangent space $ {\mathcal T}_{m}L_{m} $ to the
leaf $ L_{m} $ of $ {\mathcal F} $ through $ m $, and the {\em normal space\/} $ {\mathcal N}_{m}{\mathcal F} $ at $ m\in M $ is $ {\mathcal T}_{m}M/{\mathcal T}_{m}{\mathcal F} $.
{\em Cotangent space\/} $ {\mathcal T}_{m}^{*}{\mathcal F} $ and {\em conormal space\/} $ {\mathcal N}_{m}^{*}{\mathcal F} $ at $ m $ are defined as dual
spaces to the tangent space and the normal space at $ m $. Clearly, $ {\mathcal N}_{m}^{*}{\mathcal F} $ can
be identified with the orthogonal complement $ \left({\mathcal T}_{m}{\mathcal F}\right)^{\perp} $ to $ {\mathcal T}_{m}{\mathcal F}\subset{\mathcal T}_{m}M $ in $ {\mathcal T}_{m}^{*}M $.

\begin{definition} Say that a prefoliation $ {\mathcal F} $ is a {\em foliation\/} if every point $ m\in M $
has a neighborhood $ U $ such that $ {\mathcal F}|_{U} $ is diffeomorphic to a direct
prefoliation. \end{definition}

Obviously, tangent, cotangent, normal and conormal spaces to a
foliation form vector bundles over $ M $, and $ {\mathcal T}{\mathcal F}\subset{\mathcal T}M $, $ {\mathcal N}^{*}{\mathcal F}\subset{\mathcal T}^{*}M $ are vector
subbundles.

\begin{definition} A {\em web\/} $ \left\{{\mathcal F}_{\lambda}\right\}_{\lambda\in\Lambda} $ of codimension $ r $ on a manifold $ M $ is a family
of foliations of codimension $ r $ on $ M $, one foliation $ {\mathcal F}_{\lambda} $ per each $ \lambda\in\Lambda $. Say
that a web is {\em smooth\/} if $ \Lambda $ is a manifold, and the vector subbundle
$ {\mathcal N}^{*}{\mathcal F}_{\lambda}\subset{\mathcal T}^{*}M $ depends smoothly on $ \lambda\in\Lambda $ (to be more precise, consider $ {\mathcal N}^{*}{\mathcal F}_{\lambda} $ as a
section of the bundle of Grassmannians $ \operatorname{Gr}_{r}\left({\mathcal T}^{*}M\right) $). \end{definition}

In what follows we use the shortcut $ {\mathcal F}_{\bullet} $ for $ \left\{{\mathcal F}_{\lambda}\right\}_{\lambda\in\Lambda} $ when we are not
interested in the set $ \Lambda $ of parameters of the web.

\begin{definition} Say that a web $ \left\{{\mathcal F}_{\lambda}\right\}_{\lambda\in\Lambda} $ on $ M $ is {\em weakly separating\/} if for any
two points $ m_{1},m_{2}\in M $ there is $ \lambda\in\Lambda $ such that $ m_{1} $ and $ m_{2} $ are on different
leaves of $ {\mathcal F}_{\lambda} $. Say that a web $ \left\{{\mathcal F}_{\lambda}\right\}_{\lambda\in\Lambda} $ on $ M $ is {\em weakly separating near\/} $ m\in M $
if $ \left\{{\mathcal F}_{\lambda}\right\}_{\lambda\in\Lambda}|_{U} $ is weakly separating for an appropriate neighborhood $ U\ni m $.

Say that a web $ \left\{{\mathcal F}_{\lambda}\right\}_{\lambda\in\Lambda} $ is {\em separating\/} at $ m $ if for any tangent vector
$ v\in{\mathcal T}_{m}M $, $ v\not=0 $, there is $ \lambda\in\Lambda $ such that $ v\notin{\mathcal T}_{m}{\mathcal F}_{\lambda} $. \end{definition}

\section{$ 3 $-dimensional Veronese webs }\label{h3}\myLabel{h3}\relax 

Recall the definition of a Veronese web (\cite{GelZakhWeb}).

\begin{definition} \label{def3.05}\myLabel{def3.05}\relax  Given a web $ \left\{{\mathcal F}_{\lambda}\right\}_{\lambda\in\Lambda} $ on $ M $ of codimension $ r $, and a point
$ m\in M $, let $ {\mathbit n}_{m}\left(\lambda\right)\subset{\mathcal T}_{m}^{*}M $, $ \lambda\in\Lambda $, be the normal subspace at $ m $ to the leaf $ L $ of $ {\mathcal F}_{\lambda} $
which passes through $ m $. In the case $ r=1 $ one can consider $ {\mathbit n}_{m} $ as a mapping
from $ \Lambda $ to the projectivization $ {\mathbb P}{\mathcal T}_{m}^{*}M $ of $ {\mathcal T}_{m}^{*}M $. \end{definition}

\begin{definition} \label{def3.07}\myLabel{def3.07}\relax  A {\em Veronese web\/} on a manifold $ M $ is a smooth separating
web $ \left\{{\mathcal F}_{\lambda}\right\}_{\lambda\in{\mathbb P}^{1}} $ on $ M $ of codimension 1, such that for any point $ m\in M $, $ {\mathbit n}_{m} $ is a
regular mapping $ {\mathbb P}^{1} \to {\mathbb P}{\mathcal T}_{m}^{*}M $ of degree $ d=\dim  M-1 $. \end{definition}

\begin{remark} The condition that $ {\mathcal F}_{\bullet} $ is separating is equivalent to $ \operatorname{Im} {\mathbit n}_{m} $
being not contained in any proper projective subspace of $ {\mathbb P}{\mathcal T}_{m}^{*}M $. Recall
that all regular mappings $ \nu\colon {\mathbb P}^{1} \to {\mathbb P}^{d} $ of degree $ d $ which satisfy this
property differ only by a projective transformation of $ {\mathbb P}^{d} $. Moreover, the
projective transformation $ T\colon {\mathbb P}^{d} \to {\mathbb P}^{d} $ such that $ T\circ\nu_{1}=\nu_{2} $ is uniquely
defined if $ \nu_{1} $ and $ \nu_{2} $ are two such mappings. A convenient model of such a
mapping is given by $ \left(x:y\right) \mapsto \left(x^{d}:x^{d-1}y:\dots :xy^{d-1}:y^{d}\right) $.

These curves are {\em Veronese curves\/} in the terminology of
\cite{GelZakhWeb}, or {\em rational normal curves\/} in the terminology of algebraic
geometry. The name {\em Veronese web\/} suggests relationship with Veronese
curves; in turn, the name {\em Veronese curve\/} was introduce in recognition of
the fact that the {\em Veronese surface\/} $ {\mathbb P}^{2} \to {\mathbb P}^{5} $ has the same property: any
deformation of it differs by a fraction-linear transformation $ {\mathbb P}^{5} \to {\mathbb P}^{5} $
only. \end{remark}

Restrict our attention to the particular case of $ 3 $-dimensional
Veronese webs. In this case the only requirement on the family $ \left\{{\mathcal F}_{\lambda}\right\}_{\lambda\in{\mathbb P}^{1}} $
is that for any $ m\in M $ the points $ {\mathbit n}_{m}\left(\lambda\right) $, $ \lambda\in{\mathbb P}^{1} $ form a smooth (parameterized)
quadric in the two-dimensional projective plane $ {\mathbb P}{\mathcal T}_{m}^{*}M $. Here the
parameterization differs from the parameterization given by any
stereographic projection by a fraction-linear transformation $ {\mathbb P}^{1} \to {\mathbb P}^{1} $
only. In what follows we consider such parameterizations of quadrics only
(any smooth parameterization is such in the complex-geometry case).

\begin{lemma} \label{lm3.10}\myLabel{lm3.10}\relax  A parameterized quadric $ \gamma\colon {\mathbb P}^{1} \to {\mathbb P}^{2} $ is uniquely determined
by $ \gamma\left(\lambda_{i}\right) $, $ i=1,2,3,4 $. Here $ \left\{\lambda_{1},\lambda_{2},\lambda_{3},\lambda_{4}\right\} $ is an arbitrary set of 4
points on $ {\mathbb P}^{1} $.

For any 4 points $ P_{i}\in{\mathbb P}^{1} $, $ i=1,2,3,4 $, on $ {\mathbb P}^{2} $ such that no 3 of these
points are on the same line one can find a parameterized quadric $ \gamma\colon {\mathbb P}^{1} \to
{\mathbb P}^{2} $ such that $ P_{i}=\gamma\left(\lambda_{i}\right) $, $ i=1,2,3,4 $. \end{lemma}

\begin{proof} Recall that given a point $ p\in{\mathbb P}^{N} $, one can consider a projection
$ \pi_{p} $ with the center at $ p $, which sends $ {\mathbb P}^{N}\smallsetminus\left\{p\right\} $ onto a projective space $ {\mathbb P}{\mathcal T}_{p}{\mathbb P}^{N} $
of tangent directions at $ p $. Here $ \pi_{p}\left(q\right) $ is the direction of the line $ \left(pq\right) $.

Consider compositions $ \pi_{P_{i}}\circ\gamma\colon {\mathbb P}^{1} \to {\mathbb P}{\mathcal T}_{p}{\mathbb P}^{2}\simeq{\mathbb P}^{1} $, $ i=1,2,3,4 $. Since $ \pi_{P_{i}}|_{\operatorname{Im}\gamma} $ is
a stereographic projection, these compositions are fraction-linear
mappings between projective lines. Thus they are determined by images of
any 3 distinct points on $ {\mathbb P}^{1} $. Thus the line $ \left(P_{i}\gamma\left(\lambda\right)\right) $ is uniquely
determined by $ P_{1,2,3,4} $. Since $ \gamma\left(\lambda\right)=\left(P_{1}\gamma\left(\lambda\right)\right)\cap\left(P_{2}\gamma\left(\lambda\right)\right) $ if $ \lambda\not=\lambda_{1,2} $, $ \gamma $ is
uniquely determined by $ P_{1,2,3,4} $.

To show the existence take any parameterized quadric $ \widetilde{\gamma}\colon {\mathbb P}^{1} \to
{\mathbb P}^{2} $, let $ \widetilde{P}_{i}=\widetilde{\gamma}\left(\lambda_{i}\right) $, $ i=1,2,3,4 $. Then no 3 points out of $ \widetilde{P}_{1},\dots ,\widetilde{P}_{4} $ are on the
same line, thus there is a projective mapping $ T\colon {\mathbb P}^{2} \to {\mathbb P}^{2} $ such that
$ \xi\left(\widetilde{P}_{i}\right)=P_{i} $, $ i=1,2,3,4 $. Then $ T\circ\widetilde{\gamma} $ is the parameterized quadric we
need. \end{proof}

\begin{corollary} \label{cor3.40}\myLabel{cor3.40}\relax  Consider a manifold $ M $, $ \dim  M=3 $. A Veronese web $ {\mathcal F}_{\lambda} $ on $ M $
can be reconstructed given 4 foliations $ {\mathcal F}_{\lambda_{i}} $, $ i=1,2,3,4 $, on $ M $. Here
$ \left\{\lambda_{1},\lambda_{2},\lambda_{3},\lambda_{4}\right\} $ is an arbitrary set of 4 points on $ {\mathbb P}^{1} $. \end{corollary}

\begin{proof} Since $ {\mathbit n}_{m}\left(\lambda_{i}\right) $, $ i=1,2,3,4 $, are known for any $ m\in M $, by Lemma
~\ref{lm3.10} one can find $ {\mathbit n}_{m}\left(\lambda\right) $ for any $ \lambda\in{\mathbb P}^{1} $ and $ m\in M $. This uniquely determines
$ {\mathcal F}_{\lambda} $ for any $ \lambda\in{\mathbb P}^{1} $. \end{proof}

Fix 4 points $ \left\{\lambda_{1},\lambda_{2},\lambda_{3},\lambda_{4}\right\}\subset{\mathbb P}^{1} $. Given a Veronese web on $ M $ and a point
$ m_{0}\in M $, consider a small neighborhood $ U $ of $ m_{0} $ in $ M $. One may assume that in
$ U $ the foliations $ {\mathcal F}_{\lambda_{i}} $, $ i=1,2,3,4 $ can be written by equations $ x=\operatorname{const} $,
$ y=\operatorname{const} $, $ z=\operatorname{const} $, $ W=\operatorname{const} $; here $ x,y,z,W $ are functions on $ U $. Moreover,
$ dx|_{m_{0}} $, $ dy|_{m_{0}} $ and $ dz|_{m_{0}} $ are linearly independent. Indeed, the directions
of these 3 vectors are 3 distinct points on a quadric in the projective
plane, thus are not on the same line.

Consider $ x,y,z $ as 3 components of a vector-function $ \varphi\colon U \to {\mathbb V}^{3} $, let
$ V=\varphi\left(U\right) $. We know that the derivative of this function at $ m_{0}\in M $ is
non-degenerate, thus decreasing $ U $ we may assume that $ \varphi $ gives a
diffeomorphism $ U \to V $. Then $ w=W\circ\varphi^{-1} $ is a function on $ V $, and
$ W\left(m\right)=w\left(x\left(m\right),y\left(m\right),z\left(m\right)\right) $ if $ m\in U $.

\begin{lemma} \label{lm3.60}\myLabel{lm3.60}\relax  The scalar function $ w $ on $ V\subset{\mathbb V}^{3} $ and $ \varphi\left(m_{0}\right)\in V $ uniquely
determine the Veronese web $ {\mathcal F}_{\lambda} $ up to a local diffeomorphism near $ m_{0}\in M $. \end{lemma}

\begin{proof} Instead of determining a web up to a local diffeomorphism near
$ m_{0}\in M $ it is enough to uniquely determine the diffeomorphic image $ \varphi_{*}\left({\mathcal F}_{\bullet}\right) $ of
this web, which is a web on a neighborhood of $ \varphi\left(m_{0}\right)\in{\mathbb V}^{3} $. By Corollary
~\ref{cor3.40} it is enough to determine $ \varphi_{*}\left({\mathcal F}_{\lambda_{i}}\right) $, $ i=1,2,3,4 $. However, leaves of
$ \varphi_{*}\left({\mathcal F}_{\lambda_{1}}\right) $, $ \varphi_{*}\left({\mathcal F}_{\lambda_{2}}\right) $, $ \varphi_{*}\left({\mathcal F}_{\lambda_{3}}\right) $ are given by equations $ x=\operatorname{const} $, $ y=\operatorname{const} $,
$ z=\operatorname{const} $; here $ \left(x,y,z\right) $ is the standard coordinate system on $ {\mathbb V}^{3} $. Similarly,
leaves of $ \varphi_{*}\left({\mathcal F}_{\lambda_{4}}\right) $ are given by the equation $ w\left(x,y,z\right)=\operatorname{const} $. \end{proof}

A change of equations $ x $, $ y $, $ z $ of foliations $ {\mathcal F}_{\lambda_{i}} $, $ i=1,2,3 $, to $ x+C_{1} $,
$ y+C_{2} $, $ z+C_{3} $ corresponds to a translation of $ V $ and $ w $ by $ \left(C_{1},C_{2},C_{3}\right) $, thus
one may assume that $ \varphi\left(m_{0}\right)=\left(0,0,0\right) $. Similarly, one may assume that
$ w\left(0,0,0\right)=0 $.

\section{Nonlinear wave equation as an integrability condition }\label{h4}\myLabel{h4}\relax 

Fix a set of 4 points $ \left\{\lambda_{1},\lambda_{2},\lambda_{3},\lambda_{4}\right\}\subset{\mathbb P}^{1} $.

\begin{definition} Say that a function $ w $ on an open subset $ M\subset{\mathbb V}^{3} $ is
$ \left(\lambda_{1},\lambda_{2},\lambda_{3},\lambda_{4}\right) $-{\em admissible\/} if there is a Veronese web $ {\mathcal F}_{\lambda} $ on $ M $ such that
foliations $ {\mathcal F}_{\lambda_{i}} $, $ i=1,2,3,4 $, are given by equations $ x=\operatorname{const} $, $ y=\operatorname{const} $,
$ z=\operatorname{const} $, $ w\left(x,y,z\right)=\operatorname{const} $; here $ \left(x,y,z\right) $ is the standard coordinate system on
$ {\mathbb V}^{3} $. \end{definition}

First of all, if $ w $ is $ \left(\lambda_{1},\lambda_{2},\lambda_{3},\lambda_{4}\right) $-admissible, Lemma~\ref{lm3.10}
implies that for any point $ m\in M $ the directions $ dx|_{m} $, $ dy|_{m} $, $ dz|_{m} $ and $ dw|_{m} $
are in general position. In other words, $ w_{x}\not=0 $, $ w_{y}\not=0 $, $ w_{z}\not=0 $ everywhere in
$ M $. Thus $ w $ is {\em non-degenerate\/} (as defined in Section~\ref{h0}).

Given non-degeneracy of $ w $, for any $ \lambda\in{\mathbb P}^{1} $ and $ m\in M $ the construction of
the proof of Corollary~\ref{cor3.40} gives a direction $ {\mathbit n}_{m}\left(\lambda\right) $ in the
projectivization of $ \left({\mathbb V}^{3}\right)^{*} $. If $ w $ is $ \left(\lambda_{1},\lambda_{2},\lambda_{3},\lambda_{4}\right) $-admissible, then $ m \mapsto
{\mathbit n}_{m}\left(\lambda\right) $ coincides with the field of normal directions of the foliation $ {\mathcal F}_{\lambda} $.

Obviously,

\begin{lemma} \label{lm4.20}\myLabel{lm4.20}\relax  Consider a non-degenerate function $ w $ defined on $ M\subset{\mathbb V}^{3} $. Suppose
that for any $ \lambda\in{\mathbb P}^{1} $ the direction field $ {\mathbit n}_{m}\left(\lambda\right) $, $ m\in M $, given by the
construction of the proof of Corollary~\ref{cor3.40} coincides with the field
of normal directions of a foliation on $ M $. Then $ w $ is
$ \left(\lambda_{1},\lambda_{2},\lambda_{3},\lambda_{4}\right) $-admissible. \end{lemma}

Thus to check whether a non-degenerate function $ w $ is
$ \left(\lambda_{1},\lambda_{2},\lambda_{3},\lambda_{4}\right) $-admissible it is enough to check whether a given direction
field coincides with a normal field to a foliation. Such direction fields
can be described by the following particular case of the {\em Frobenius
integrability condition\/} \cite{Ste68Lec}:

\begin{lemma} \label{lm4.30}\myLabel{lm4.30}\relax  Consider a $ 1 $-form $ \omega $ on a manifold $ M $ which does not vanish
at any point of $ M $. Call $ \omega $ {\em Frobenius integrable\/} if there exists a
foliation $ {\mathcal F} $ of codimension 1 on $ M $ such that $ \omega\left(m\right) $ is normal to the tangent
space at $ m $ to the leaf $ L_{m} $ of $ {\mathcal F} $ through $ m $ for any $ m\in M $.

Then $ \omega $ is Frobenius integrable iff $ \omega\wedge d\omega=0 $. \end{lemma}

\begin{proof} The ``only if'' part is simple: in an appropriate neighborhood $ U $
of any given point $ m_{0}\in M $ the foliation $ {\mathcal F}|_{U} $ can be written as $ g=\operatorname{const} $; here
$ g $ is a function on $ U $, and $ dg\not=0 $ for any $ m\in U $. Thus $ \omega=h\,dg $ for an
appropriate function $ h $ on $ U $, and $ \omega\wedge d\omega=h\,dg\wedge dh\wedge dg=0 $.

For the ``if'' part it is enough to show the existence locally on $ M $,
since the foliation is unique if it exists, thus gluing pieces together
is not a problem. We may assume that $ M $ is an open subset of $ {\mathbb V}^{n} $, and that
$ \omega|_{m_{0}}=dx_{n}|_{m_{0}} $. Say that a tangent vector $ v $ at $ m\in M $ is $ k $-{\em compatible},
$ k=1,\dots ,n-1 $, if $ \left< \omega|_{m},v \right>=0 $ and $ v $ is of the form $ \frac{\partial}{\partial x_{k}}+a\frac{\partial}{\partial x_{n}} $ with
an appropriate number $ a $. Obviously, in an appropriate neighborhood of
any point $ m_{0}\in M $ there is exactly one $ k $-compatible vector $ v_{k}\left(m\right) $ for
$ k=1,\dots ,n-1 $. Define functions $ a_{\left(k\right)}\left(m\right) $ by $ v_{k}\left(m\right)=\frac{\partial}{\partial x_{k}}+a_{\left(k\right)}\left(m\right)\frac{\partial}{\partial x_{n}} $.
Then the fundamental relationship between commutator and de
Rham differential\footnote{One can easily check this relation in local coordinates.} \cite{Ste68Lec}
\begin{equation}
\left< \omega,\left[v_{k},v_{l}\right] \right>=v_{k}\cdot\left< \omega,v_{l} \right>-v_{l}\cdot\left<\omega,v_{k}\right>+\left<d\omega,v_{k}\wedge v_{l}\right>
\notag\end{equation}
implies $ \left< \omega,\left[v_{k},v_{l}\right] \right>=\left< d\omega,v_{k}\wedge v_{l}\right> $. Since $ \omega\wedge d\omega=0 $, one can write
$ d\omega=\omega\wedge\alpha $; here $ \alpha $ is a $ 1 $-form defined near $ m_{0} $. Hence
\begin{equation}
\left< d\omega,v_{k}\wedge v_{l}\right>= \left< \omega,v_{k}\right>\left< \alpha,v_{l}\right> - \left< \alpha,v_{k}\right>\left< \omega,v_{l}\right>=0.
\notag\end{equation}

Thus $ \left< \omega,\left[v_{k},v_{l}\right] \right>=0 $. On the other hand,
$ \left[v_{k},v_{l}\right]=\left(v_{k}\cdot a_{l}-v_{l}\cdot a_{k}\right)\frac{\partial}{\partial x_{n}} $. Together with $ \left<
\omega,\left[v_{k},v_{l}\right] \right>=0 $ this implies $ \left[v_{k},v_{l}\right]=0 $. By the principal theorem of the
theory of ODE, one can find local coordinates $ \left(y_{1},\dots ,y_{n}\right) $ such that
$ v_{k}=\frac{\partial}{\partial y_{k}} $, $ k=1,\dots ,n-1 $. Since $ \omega $ is orthogonal to $ v_{k} $, $ k=1,\dots ,n-1 $, this
implies that $ \omega=h\left(y\right)dy_{n} $, thus $ y_{n}=\operatorname{const} $ gives a foliation with the required
properties. \end{proof}

The next step is to provide an explicit construction of the normal
directions $ {\mathbit n}_{m}\left(\lambda\right) $ in terms of $ w $.

\begin{lemma} Given a Veronese curve $ \gamma\left(\lambda\right) $ in $ {\mathbb P}^{n-1} $, one can find polynomials
$ p_{1}\left(\lambda\right),\dots ,p_{n}\left(\lambda\right) $ of degree $ \leq n-1 $ such that $ \gamma\left(\lambda\right)=\left(p_{1}\left(\lambda\right):\dots :p_{n}\left(\lambda\right)\right) $ for $ \lambda\not=\infty $.
Polynomials $ p_{k}\left(\lambda\right) $ are defined uniquely up to multiplication by the same
constant. \end{lemma}

\begin{proof} Any Veronese curve in $ {\mathbb P}^{n-1} $ is a projective transformation of
the closure of the image of the mapping $ \lambda \mapsto \left(1:\lambda:\dots :\lambda^{n-1}\right) $. A
consideration of the corresponding linear transformation of $ {\mathbb V}^{n} $ provides
polynomials $ p_{1},\dots ,p_{n} $.

It is enough to show uniqueness for the curve $ \left(1:\lambda:\dots :\lambda^{n-1}\right) $.
Obviously, $ p_{k}\left(\lambda\right)=\lambda^{k-1}p_{1}\left(\lambda\right) $. Moreover, since $ \deg  p_{n}\leq n-1 $, $ p_{1}\left(\lambda\right) $ is a
constant. \end{proof}

Thus any Veronese curve in $ {\mathbb P}^{2} $ is a projectivization of a polynomial
vector-function $ v\left(\lambda\right) $ of degree {\em exactly\/} 2. Note that $ v\left(\lambda\right)\not=0 $ for any $ \lambda $.

This implies that the dependence on $ \lambda $ of the directions $ {\mathbit n}_{m}\left(\lambda\right) $, $ \lambda\not=\infty $,
can be described by the direction of the $ 1 $-form $ \alpha\left(m\right)+\lambda\beta\left(m\right)+\lambda^{2}\gamma\left(m\right) $; here
$ \alpha,\beta,\gamma $ are appropriate $ 1 $-forms on $ M\subset{\mathbb V}^{3} $ which are defined up to
multiplication by the same function on $ M $. If $ \lambda\not=0 $, $ {\mathbit n}_{m}\left(\lambda\right) $ is the direction
of $ \gamma\left(m\right)+\lambda^{-1}\beta\left(m\right)+\lambda^{-2}\alpha\left(\lambda\right) $, taking the limit $ \lambda \to \infty $ implies that $ {\mathbit n}_{m}\left(\infty\right) $ is
the direction of $ \gamma\left(m\right) $.

\begin{lemma} \label{lm4.45}\myLabel{lm4.45}\relax  Consider vectors $ v_{1},v_{2},v_{3},v_{4} $ in $ {\mathbb V}^{3} $ such that $ v_{1},v_{2},v_{3} $ are
linearly independent. Fix a set of 4 points $ \left\{\lambda_{1},\lambda_{2},\lambda_{3},\lambda_{4}\right\}\subset{\mathbb P}^{1}\smallsetminus\left\{\infty\right\} $. There
is a unique polynomial vector-function $ v\left(\lambda\right) $ of degree 2 such that
$ v\left(\lambda_{4}\right)=v_{4} $, and $ v\left(\lambda_{k}\right) $ is proportional to $ v_{k} $, $ k=1,2,3 $. \end{lemma}

\begin{proof} Write $ v_{4} $ as $ av_{1}+bv_{2}+cv_{3} $. Since $ v\left(\lambda\right) $ can be written as
$ \alpha\left(\lambda\right)v_{1}+\beta\left(\lambda\right)v_{2}+\gamma\left(\lambda\right)v_{3} $, we know that $ \alpha\left(\lambda_{2}\right)=\alpha\left(\lambda_{3}\right)=0 $, $ \alpha\left(\lambda_{4}\right)=a $. This uniquely
determines the quadratic polynomial $ \alpha\left(\lambda\right) $. Proceed similarly for $ \beta\left(\lambda\right) $ and
$ \gamma\left(\lambda\right) $. \end{proof}

\begin{corollary} \label{cor4.50}\myLabel{cor4.50}\relax  Fix a set of 4 points $ \left\{\lambda_{1},\lambda_{2},\lambda_{3},\lambda_{4}\right\}\subset{\mathbb P}^{1}\smallsetminus\left\{\infty\right\} $. Given a
non-degenerate function $ w $ on $ M\subset{\mathbb V}^{3} $, the direction $ {\mathbit n}_{m}\left(\lambda\right) $ defined by the
construction of the proof of Corollary~\ref{cor3.40} coincides with
$ \left(p_{1}\left(\lambda\right)w_{x}:p_{2}\left(\lambda\right)w_{y}:p_{3}\left(\lambda\right)w_{z}\right) $; here
\begin{equation}
p_{i}\left(\lambda\right)=\left(\lambda_{4}-\lambda_{i}\right)\left(\lambda-\lambda_{j}\right)\left(\lambda-\lambda_{k}\right),
\label{equ4.50}\end{equation}\myLabel{equ4.50,}\relax 
for any permutation $ \left(i jk\right) $ of (123). \end{corollary}

\begin{corollary} \label{cor4.55}\myLabel{cor4.55}\relax  Consider distinct points $ \lambda_{i}\not=\infty $, $ i=1,2,3,4 $. Consider a
non-degenerate function $ w $ on $ M\subset{\mathbb V}^{3} $. Let
\begin{equation}
\omega_{\lambda}\buildrel{\text{def}}\over{=}p_{1}\left(\lambda\right)w_{x}dx+p_{2}\left(\lambda\right)w_{y}dy+p_{3}\left(\lambda\right)w_{z}dz;
\label{equ4.52}\end{equation}\myLabel{equ4.52,}\relax 
here $ p_{1,2,3}\left(\lambda\right) $ are from~\eqref{equ4.50}. Then the following conditions are
equivalent:
\begin{enumerate}
\item
$ w $ is $ \left(\lambda_{1},\lambda_{2},\lambda_{3},\lambda_{4}\right) $-admissible;
\item
$ \omega_{\lambda}\wedge d\omega_{\lambda}=0 $ for any $ \lambda $;
\item
$ \omega_{\lambda}\wedge d\omega_{\lambda}=0 $ for any 5 distinct values of $ \lambda $;
\item
$ \omega_{\lambda}\wedge d\omega_{\lambda}=0 $ for any $ \lambda_{0}\notin\left\{\lambda_{1},\dots ,\lambda_{4}\right\} $;
\end{enumerate}
If $ 0\notin\left\{\lambda_{1},\dots ,\lambda_{4}\right\} $, these conditions are equivalent to
\begin{equation}
\nu_{23}w_{x}w_{yz}+\nu_{31}w_{y}w_{xz}+\nu_{12}w_{z}w_{xy}=0,
\label{equ4.80}\end{equation}\myLabel{equ4.80,}\relax 
here $ \nu_{kl}=\lambda_{k}/\left(\lambda_{4}-\lambda_{k}\right)-\lambda_{l}/\left(\lambda_{4}-\lambda_{l}\right) $. \end{corollary}

\begin{proof} Obviously, $ \omega_{\lambda}|_{m}\not=0 $ for any $ \lambda\not=\infty $ and any $ m\in M $. By Lemma~\ref{lm4.30},
$ \omega_{\lambda}\wedge d\omega_{\lambda}=0 $ is equivalent to existence of a foliation to which $ \omega_{\lambda} $ is normal.
Thus by Lemma~\ref{lm4.20} the first statement implies the second one.

If $ \omega_{\lambda}\wedge d\omega_{\lambda}=0 $ for any $ \lambda $, then by Corollary~\ref{cor4.50}, the
required in Lemma~\ref{lm4.20} foliation exists for $ \lambda\not=\infty $. However, $ \widetilde{\omega}_{\lambda}=\lambda^{-2}\omega_{\lambda} $ is
defined for $ \lambda\in{\mathbb P}^{1}\smallsetminus\left\{0\right\} $, and $ \widetilde{\omega}_{\lambda}\wedge d\widetilde{\omega}_{\lambda} $ is a polynomial of degree 4 in $ \lambda^{-1} $. Thus
$ \widetilde{\omega}_{\lambda}\wedge d\widetilde{\omega}_{\lambda}=0 $, {\em including\/} $ \lambda=\infty $. Moreover, $ \widetilde{\omega}_{\infty}|_{m}\not=0 $ for any $ m $, which implies the
existence of $ {\mathcal F}_{\lambda} $ for $ \lambda=\infty $ as well. Thus the second statement implies the
first one.

Since $ \omega_{\lambda} $ is quadratic in $ \lambda $, $ \omega_{\lambda}\wedge d\omega_{\lambda} $ is a polynomial of degree 4 in $ \lambda $.
Thus the second statement is equivalent to the third one. By construction
$ \omega_{\lambda_{1,2,3,4}} $ are proportional to $ dx $, $ dy $, $ dz $, and $ dw $ correspondingly. This
implies that $ \omega_{\lambda}\wedge d\omega_{\lambda}=0 $ for $ \lambda\in\left\{\lambda_{1},\lambda_{2},\lambda_{3},\lambda_{4}\right\} $. Consequently, the fourth
statement is equivalent to the third one.

Assume that $ \lambda_{0}=0\notin\left\{\lambda_{1},\lambda_{2},\lambda_{3},\lambda_{4}\right\} $. Let $ \mu_{k}=\lambda_{4}/\lambda_{k}-1 $, $ k=1,2,3 $. Then
\begin{equation}
\omega_{0}=\lambda_{1}\lambda_{2}\lambda_{3}\widetilde{\omega},\qquad \widetilde{\omega} \buildrel{\text{def}}\over{=} \mu_{1}w_{x}dx+\mu_{2}w_{y}dy+\mu_{3}w_{z}dz,
\notag\end{equation}
and $ \widetilde{\omega}\wedge d\widetilde{\omega} $ can be written as
\begin{equation}
\mu_{1}\mu_{2}\mu_{3}\left(\left(\mu_{2}^{-1}-\mu_{3}^{-1}\right)w_{x}w_{yz}+\left(\mu_{3}^{-1}-\mu_{1}^{-1}\right)w_{y}w_{xz}+\left(\mu_{1}^{-1}-\mu_{2}^{-1}\right)w_{z}w_{xy}\right)dx\wedge dy\wedge dz.
\notag\end{equation}
(It is clear that $ \mu_{k}\not=0 $ for $ k=1,2,3 $.) Since $ \nu_{kl}=\mu_{k}^{-1}-\mu_{l}^{-1} $, the equation
$ \omega_{0}\wedge d\omega_{0}=0 $ is proportional to~\eqref{equ4.80}, which implies the last statement of
the corollary. \end{proof}

Obviously, $ \omega_{\lambda}\wedge d\omega_{\lambda}=\alpha\prod_{k=1}^{4}\left(\lambda-\lambda_{k}\right) $; here $ \alpha $ is a $ 3 $-form on $ M $ which does
not depend on $ \lambda $. Thus the equations $ \omega_{\lambda_{0}}\wedge d\omega_{\lambda_{0}}=0 $ for different values $ \lambda_{0} $
are proportional, and it does not matter much which value of $ \lambda_{0} $ one would
use. Consequently, any other choice of $ \lambda_{0} $ would lead to an equation which
is proportional to~\eqref{equ4.80}, and one can drop the conditions that
$ 0\notin\left\{\lambda_{1},\dots ,\lambda_{4}\right\} $. Moreover, it is possible to drop the condition
$ \infty\notin\left\{\lambda_{1},\dots ,\lambda_{4}\right\} $ as well:

\begin{theorem} \label{th4.65}\myLabel{th4.65}\relax  A non-degenerate function $ w $ on $ M\subset{\mathbb V}^{3} $ is
$ \left(\lambda_{1},\lambda_{2},\lambda_{3},\lambda_{4}\right) $-admissible iff it satisfies an $ \left(A,B,C\right) $-equation~\eqref{equ0.10}
with $ -A/C=\left(\lambda_{1}:\lambda_{2}:\lambda_{3}:\lambda_{4}\right) $; here $ \left(a:b:c:d\right)=\frac{d-a}{d-c}\frac{b-c}{b-a} $ is the
cross-ratio of $ a,b,c,d $. \end{theorem}

\begin{proof} Indeed, a direct calculation shows that $ \nu_{12}+\nu_{23}+\nu_{31}=0 $, $ \nu_{12}\not=0 $,
$ \nu_{23}\not=0 $, $ \nu_{31}\not=0 $, and $ -\nu_{23}/\nu_{12}= \left(\lambda_{1}:\lambda_{2}:\lambda_{3}:\lambda_{4}\right) $. Thus the statement holds for
$ \infty\notin\left\{\lambda_{1},\dots ,\lambda_{4}\right\} $. However, if $ T $ is a projective transformation, then $ w $ is
$ \left(\lambda_{1},\lambda_{2},\lambda_{3},\lambda_{4}\right) $-admissible iff it is $ \left(T\lambda_{1},T\lambda_{2},T\lambda_{3},T\lambda_{4}\right) $-admissible. Since
cross-ratio is invariant w.r.t.~projective transformations, it is enough
to prove the statement for $ \left(T\lambda_{1},T\lambda_{2},T\lambda_{3},T\lambda_{4}\right) $ with an arbitrary $ T $. By an
appropriate choice of $ T $ we can ensure that $ \infty\notin\left\{\lambda_{1},\dots ,\lambda_{4}\right\} $ (and
additionally $ 0\notin\left\{\lambda_{1},\dots ,\lambda_{4}\right\} $ if we wish). \end{proof}

\begin{remark} \label{rem4.95}\myLabel{rem4.95}\relax  Since the cross-ratio of 4 distinct points can take any
value distinct from $ 0,1,\infty $, one can momentarily see that for any triple
$ \left(A,B,C\right) $ which satisfies~\eqref{equ0.20} and for any 3 distinct points $ \lambda_{1},\lambda_{2},\lambda_{3} $
one can find $ \lambda_{4} $ such that $ -A/C=\left(\lambda_{1}:\lambda_{2}:\lambda_{3}:\lambda_{4}\right) $. Thus any $ \left(A,B,C\right) $-equation
can be interpreted as an integrability condition of a Veronese web: any
Veronese web gives rise to a non-degenerate solution of such an equation,
and any non-degenerate solution can be represented in this form. \end{remark}

\begin{remark} One can generalize Corollary~\ref{cor4.55} to the case of Veronese
webs of arbitrary dimension. In dimension $ d $ one still needs one
function $ w $ of $ d $ variables to completely determine a web up to a local
diffeomorphism. The foliation $ {\mathcal F}_{\lambda} $ can be described by a $ 1 $-form $ \omega_{\lambda} $ which is
normal to leaves of $ {\mathcal F}_{\lambda} $, and is given by a formula similar to~\eqref{equ4.52}.

The $ 1 $-form $ \omega_{\lambda} $ depends on $ \lambda $ as a polynomial of degree $ d-1 $, and the
integrability condition $ \omega_{\lambda}\wedge d\omega_{\lambda}=0 $ is a polynomial of degree $ 2d-2 $. Thus a
non-degenerate function $ w $ of $ d $ variables corresponds to a Veronese web
iff $ \omega_{\lambda}\wedge d\omega_{\lambda}=0 $ for $ 2d-1 $ different values of $ \lambda $. By its construction,
$ \omega_{\lambda}\wedge d\omega_{\lambda}=0 $ automatically holds for $ d+1 $ value of $ \lambda $. Thus a naive
generalization (as done in \cite{GelZakhWeb}) of Corollary~\ref{cor4.55} would be
that it is enough to require $ \omega_{\lambda}\wedge d\omega_{\lambda}=0 $ at $ d-2 $ ``additional'' values of $ \lambda $.

However, \cite{Nak96Cur,Nak98Cur} contain a much stronger result: if
$ \omega_{\lambda}\wedge d\omega_{\lambda}=0 $ for any ``additional'' value of $ \lambda $, then $ \omega_{\lambda}\wedge d\omega_{\lambda}=0 $ for any $ \lambda $, thus $ w $
determines a Veronese web. Unfortunately, this condition is on a $ 3 $-form
in $ d $-dimensional space, thus it is still an overdetermined system of
partial differential equations on $ w $, if $ d>3 $. It is very interesting to
investigate whether arguments of \cite{Tur99Cla,Tur99MemB} allow extraction
of one equation on $ w $ which implies $ \omega_{\lambda}\wedge d\omega_{\lambda}=0 $. \end{remark}

\section{B\"acklund--Darboux transformations }\label{h5}\myLabel{h5}\relax 

By Remark~\ref{rem4.95}, any solution of an $ \left(A,B,C\right) $-equation gives rise
to a Veronese web, which in turn leads to a solution of
$ \left(A',B',C'\right) $-equation, possibly with different $ \left(A',B',C'\right) $.

\begin{corollary} \label{cor5.30}\myLabel{cor5.30}\relax  Let $ w $ be a non-degenerate solution of $ \left(A,B,C\right) $-equation
in a neighborhood of $ \left(0,0,0\right) $, $ \left(\lambda_{1},\lambda_{2},\lambda_{3},\lambda_{4}\right) $ be numbers such that
$ -A/C=\left(\lambda_{1}:\lambda_{2}:\lambda_{3}:\lambda_{4}\right) $. Then for any number $ \lambda $
\begin{enumerate}
\item
there is a function $ v\left(x,y,z\right) $ defined in a neighborhood of $ \left(0,0,0\right) $
such that the following identity of vector-functions holds:
\begin{equation}
\left(v_{x},v_{y},v_{z}\right)=\psi\left(x,y,z\right)\left(\alpha w_{x},\beta w_{y},\gamma w_{z}\right);
\label{equ5.20}\end{equation}\myLabel{equ5.20,}\relax 
here $ \alpha,\beta,\gamma \alpha=p_{1}\left(\lambda\right) $, $ \beta=p_{2}\left(\lambda\right) $, $ \gamma=p_{3}\left(\lambda\right) $, $ p_{k} $ are polynomials given by
~\eqref{equ4.50}, and $ \psi $ is an appropriate scalar-valued function;
\item
if $ \lambda\notin\left\{\lambda_{1},\lambda_{2},\lambda_{3}\right\} $, then $ v\left(x,y,z\right) $ can be chosen to be non-degenerate;
\item
if $ v $ is non-degenerate it is $ \left(\lambda_{1},\lambda_{2},\lambda_{3},\lambda\right) $-admissible;
\item
if $ \lambda\notin\left\{\lambda_{1},\lambda_{2},\lambda_{3}\right\} $, and $ \left(\widetilde{A},\widetilde{B},\widetilde{C}\right) $ satisfy conditions~\eqref{equ0.20}, and
$ -\widetilde{A}/\widetilde{C}=\left(\lambda_{1}:\lambda_{2}:\lambda_{3}:\lambda\right) $, then the function $ v\left(x,y,z\right) $ satisfies $ \left(\widetilde{A},\widetilde{B},\widetilde{C}\right) $-equation.
\end{enumerate}
\end{corollary}

\begin{proof} By Corollary~\ref{cor4.55}, $ w $ is $ \left(\lambda_{1},\lambda_{2},\lambda_{3},\lambda_{4}\right) $-admissible, thus it
corresponds to a web $ {\mathcal F}_{\bullet} $. Write the leaves of $ {\mathcal F}_{\lambda} $ as $ v\left(x,y,z\right)=\operatorname{const} $, and
apply Corollary~\ref{cor4.55} again. \end{proof}

Obviously, the function $ v $ of the previous corollary is defined
uniquely up to a gauge transformation (see Definition~\ref{def0.30}).

Let us find relationships between 9 constants $ \left(A,B,C\right) $, $ \left(\widetilde{A},\widetilde{B},\widetilde{C}\right) $ and
$ \left(\alpha,\beta,\gamma\right) $ which appear in the statements of this section.
Construct $ \widetilde{\nu}_{kl} $ basing on the $ 4 $-tuple $ \left(\lambda_{1},\lambda_{2},\lambda_{3},\lambda\right) $ using the same formula as
used to construct $ \nu_{kl} $ basing on $ \left(\lambda_{1},\lambda_{2},\lambda_{3},\lambda_{4}\right) $. Let
\begin{equation}
\tau=\left(\lambda_{4}-\lambda_{1}\right)\left(\lambda_{4}-\lambda_{2}\right)\left(\lambda_{4}-\lambda_{3}\right)\frac{\lambda}{\lambda_{4}}.
\notag\end{equation}
Then it is easy to check that $ \alpha=\tau\nu_{23}/\widetilde{\nu}_{23} $, $ \beta=\tau\nu_{13}/\widetilde{\nu}_{13} $, $ \gamma=\tau\nu_{12}/\widetilde{\nu}_{12} $. Since
simultaneous multiplication of $ \alpha,\beta,\gamma $ by the same non-zero number does not
change the meaning of Equation~\eqref{equ5.20}, we conclude that one can take
$ \alpha=A/\widetilde{A} $, $ \beta=B/\widetilde{B} $, $ \gamma=C/\widetilde{C} $.

\begin{proof}[Proof of Theorem~\ref{th5.50} ] The first statement is obvious, and the
second one is the corollary of the first since $ dv_{1}\sim dv $ if $ v_{1} $ is a gauge
transform of $ v $. The third and the fourth statements are reformulations of
parts of Corollary~\ref{cor4.55}. The last statement is a direct corollary of
the first one and of the following obvious statement:

\begin{lemma} Given $ \left(v_{x},v_{y},v_{z}\right) \sim \left(v'_{x},v'_{y},v'_{z}\right) $ for two non-degenerate
functions $ v $ and $ v' $ defined in a neighborhood of (0,0,0) in $ {\mathbb V}^{3} $, one can
decrease the neighborhood so that the functions become gauge transforms
of each other. \end{lemma}

This finishes the proof of Theorem~\ref{th5.50}. \end{proof}

To enhance the statements about Equation~\eqref{equ5.70}, note the
following two lemmas.

\begin{lemma} Given numbers $ \alpha\not=0 $, $ \beta\not=0 $, $ \gamma\not=0 $ such that $ \alpha\not=\beta $, $ \alpha\not=\gamma $, $ \beta\not=\gamma $, there
exist two triples $ \left(A,B,C\right) $ and $ \left(\widetilde{A},\widetilde{B},\widetilde{C}\right) $ which both satisfy conditions
~\eqref{equ0.20}, and $ \alpha=A/\widetilde{A} $, $ \beta=B/\widetilde{B} $, $ \gamma=C/\widetilde{C} $. The numbers $ A,B,C,\widetilde{A},\widetilde{B},\widetilde{C} $ are defined
uniquely up to multiplication by the same constant. \end{lemma}

\begin{proof} Given $ \widetilde{A},\widetilde{B},\widetilde{C} $ put $ A=\alpha\widetilde{A} $, $ B=\beta\widetilde{B} $, $ C=\gamma\widetilde{C} $. The conditions~\eqref{equ0.20} on
$ \left(A,B,C\right) $ can be translated to an additional linear equation $ \alpha\widetilde{A}+\beta\widetilde{B}+\gamma\widetilde{C}=0 $ on
$ \widetilde{A} $, $ \widetilde{B} $, $ \widetilde{C} $. This equation is independent of $ \widetilde{A}+\widetilde{B}+\widetilde{C}=0 $, thus there is a unique
(up to proportionality) solution $ \left(\widetilde{A},\widetilde{B},\widetilde{C}\right) $ of these two equations. What
remains to check is that this solution does not contradict the conditions
$ \widetilde{A}\not=0 $, $ \widetilde{B}\not=0 $, $ \widetilde{C}\not=0 $. However, $ \widetilde{A}=0 $ contradicts $ \beta\not=\gamma $, etc. \end{proof}

\begin{lemma} Given two triples $ \left(A,B,C\right) $ and $ \left(\widetilde{A},\widetilde{B},\widetilde{C}\right) $ which both satisfy
conditions~\eqref{equ0.20}, put $ \alpha=A/\widetilde{A} $, $ \beta=B/\widetilde{B} $, $ \gamma=C/\widetilde{C} $. Then either $ \alpha=\beta=\gamma $, or $ \alpha\not=\beta $,
$ \alpha\not=\gamma $, $ \beta\not=\gamma $. \end{lemma}

This statement is elementary.

\section{Inverse construction }\label{h5.5}\myLabel{h5.5}\relax 

Of course, Theorem~\ref{th5.50} can be proven by elementary methods
without any reference to Veronese webs. However, Veronese webs are not
useful because this theorem can be proven ``naturally'' by using Veronese
webs. In fact Veronese webs appears naturally as reformulations of the
{\em statement\/} of this theorem.

Indeed, given a solution of Equation~\eqref{equ0.10}, consider Systems
~\eqref{equ5.60} for all possible triples $ \left(\widetilde{A},\widetilde{B},\widetilde{C}\right) $. Since proportional triples
$ \left(\widetilde{A},\widetilde{B},\widetilde{C}\right) $ give essentially the same systems, we can enumerate all the
triples by the ratio $ \lambda=-\widetilde{A}/\widetilde{C} $, which can be considered as an element of $ {\mathbb P}^{1} $
with the only restrictions being $ \lambda\not=\infty $, $ \lambda\not=0 $, $ \lambda\not=1 $.

For any such value of $ \lambda $ one obtains a solution $ v^{\left[\lambda\right]} $ of Equation
~\eqref{equ5.50}. This solution is defined in a neighborhood $ U_{\lambda} $ of (0,0,0), and it
is easy to show that this neighborhood may be chosen independently of $ \lambda $,
denote it by $ U $. The solution $ v^{\left[\lambda\right]} $ is not unique, but the foliation $ {\mathcal F}_{\lambda} $
of $ U $ defined by $ v^{\left[\lambda\right]}=\operatorname{const} $ is uniquely defined. Moreover, $ {\mathcal F}_{\lambda} $ depends
smoothly on $ \lambda\in{\mathbb P}^{1}\smallsetminus\left\{0,1,\infty\right\} $. What remains it to consider what happens near
$ \lambda=0 $, near $ \lambda=1 $, and near $ \lambda=\infty $.

If $ \lambda\approx0 $, then $ \widetilde{A} $ is very small, thus Equation~\eqref{equ5.70}
\begin{equation}
\left(v_{x},v_{y},v_{z}\right) \sim \left(Aw_{x},\widetilde{A}B \widetilde{B}^{-1}w_{y},\widetilde{A}C \widetilde{C}^{-1}w_{z}\right)
\notag\end{equation}
becomes close to $ \left(v_{x},v_{y},v_{z}\right) \sim \left(Aw_{x},0,0\right) $, or, in other words, to
$ \left(v_{x},v_{y},v_{z}\right) \sim $ (1,0,0). The solution to this equation is $ v=v\left(x\right) $, thus the
foliation $ {\mathcal F}_{\lambda} $ has a limit $ x=\operatorname{const} $ when $ \lambda \to $ 0. Similarly, the limit when
$ \lambda \to 1 $ is $ y=\operatorname{const} $, when $ \lambda \to \infty $ is $ z=\operatorname{const} $.

Thus an investigation of the statement of Theorem~\ref{th5.50} directly
leads to a family a foliations which depend smoothly on a parameter
$ \lambda\in{\mathbb P}^{1} $. In the following section we show that the conditions that the
normal directions to the foliations span a quadratic cone is also related
to the elementary theory of Equation~\eqref{equ0.10}.

Additionally, the following statement is easy to obtain elementary,
but it is an immediate corollary of Theorem~\ref{th5.50}:

\begin{corollary} \label{cor5.90}\myLabel{cor5.90}\relax  If $ w $ is a non-degenerate solution of Equation
~\eqref{equ0.10}, then any gauge transform of $ w $ is also a solution of Equation
~\eqref{equ0.10}. \end{corollary}

\section{Linearization }\label{h6}\myLabel{h6}\relax 

Given a solution $ \bar{\kappa} $ of a non-linear (system of) equation(s) $ F\left(\kappa\right)=0 $,
the {\em linearized\/} equation at $ \bar{\kappa} $ is the equation $ F\left(\bar{\kappa}+\varepsilon\kappa\right)=O\left(\varepsilon^{2}\right) $. It is a
(system of) linear equation(s) on $ \kappa $ with the coefficients being partial
derivatives of $ F $ at $ \bar{\kappa} $.

Obviously, given a solution $ \bar{w} $ of Equation~\eqref{equ0.10}, the
linearization is
\begin{equation}
A\bar{w}_{x}w_{yz}+B\bar{w}_{y}w_{xz}+C\bar{w}_{z}w_{xy}+A\bar{w}_{yz}w_{x}+B\bar{w}_{xz}w_{y}+C\bar{w}_{xy}w_{z}=0,
\label{equ6.10}\end{equation}\myLabel{equ6.10,}\relax 
The left-hand side is a linear differential operator of second order in
$ w $, denote this operator $ l_{\bar{w}} $ or just $ l $. The principal symbol of $ l_{\bar{w}} $ is
\begin{equation}
\Lambda\left(x,y,z,\xi,\eta,\zeta\right)=A\bar{w}_{x}\eta\zeta+B\bar{w}_{y}\xi\zeta+C\bar{w}_{z}\xi\eta.
\label{equ6.20}\end{equation}\myLabel{equ6.20,}\relax 
This is a non-degenerate quadratic form in $ \left(\xi,\eta,\zeta\right) $ iff $ \bar{w} $ is
non-degenerate. Moreover, it vanishes if $ \left(\xi,\eta,\zeta\right)=\left(1,0,0\right) $, or
$ \left(\xi,\eta,\zeta\right)=\left(0,0,1\right) $, or $ \left(\xi,\eta,\zeta\right)=\left(0,0,1\right) $. This shows that the linearization is
hyperbolic iff $ \bar{w} $ is non-degenerate. {\em This\/} is why it makes sense to call
the equation~\eqref{equ0.10} a {\em nonlinear wave equation}.

Fix a point $ \left(x,y,z\right) $. Recall that a covector $ \left(\xi,\eta,\zeta\right) $ at $ \left(x,y,z\right) $ is
{\em characteristic\/} if $ \Lambda\left(x,y,z,\xi,\eta,\zeta\right)=0 $. Characteristic covectors of a
hyperbolic linear differential equation form a cone in the cotangent
space, this cone is called a {\em wave cone}. Since our equation is of second
order, it is a quadratic cone. Recall that a surface in $ {\mathbb V}^{3} $ is called
{\em characteristic\/} if the normal direction to this surface at any point is
characteristic. One can define similar notions for square systems of
equations by taking $ \det \Lambda $ instead of $ \Lambda $.

Recall how to construct characteristic surfaces. Consider an
expression $ l\left(e^{ik\varphi\left(x,y,z\right)}\right) $ when $ k \to \infty $. It can be written as
$ \Phi_{\varphi}\left(k,x,y,z\right)e^{ik\varphi\left(x,y,z\right)} $; here $ \Phi_{\varphi} $ depends polynomially on $ k $, the degree
being 2 or less. Say that $ \varphi $ is an {\em eikonal\/} solution if $ \Phi_{\varphi} $ is a polynomial
in $ k $ of degree $ \leq1 $. If $ \Phi_{\varphi,2}\left(x,y,z\right) $ is the coefficient at $ k^{2} $ in $ \Phi_{\varphi} $, then
the equation
\begin{equation}
\Phi_{\varphi,2}\left(x,y,z\right)=0
\notag\end{equation}
is a non-linear differential equation of the first order on $ \varphi $. Call this
equation the {\em eikonal equation}.

Obviously, eikonal solutions coincide with solutions to the eikonal
equation. Moreover, it is easy to see that the eikonal equation is
equivalent to the surfaces $ \varphi=c $ being characteristic surfaces for any
constant $ c $.

A similar statements holds for square systems of differential
equations if one considers $ l\left(v\cdot e^{ik\varphi\left(x,y,z\right)}\right) $ as a linear function of a
vector $ v $. Then $ \Phi_{\varphi} $ becomes a square matrix, and we can consider the degree
of $ \det \Phi $ in $ k $ instead of the degree of $ \Phi $ in $ k $.

\begin{proposition} \label{prop6.30}\myLabel{prop6.30}\relax  Consider a non-degenerate solution $ w $ of Equation
~\eqref{equ0.10}. Then $ w $ is also a solution of the linearized equation~\eqref{equ6.10}
at $ w $. Moreover, $ w $ is also an eikonal solution for this linearized
equation. \end{proposition}

\begin{proof} To prove the first statement, apply Corollary~\ref{cor5.90}. Since $ w $
is a solution, so is $ w+\varepsilon\omega $ for any $ \varepsilon $. Similarly, since $ w+\varepsilon e^{i kw} $ is a
solution for any $ \varepsilon $ and $ k $, $ w $ is an eikonal solution as well. \end{proof}

\begin{proposition} Consider a solution $ \left(w,v\right) $ of System~\eqref{equ5.60} with
non-degenerate $ w $ and $ v $. Let $ l^{\left\{1\right\}} $, $ l^{\left\{2\right\}} $ be the linearizations of
Equations~\eqref{equ0.10},~\eqref{equ5.50} at $ w $, and $ l^{\left\{3\right\}} $ be the linearization of
Equation~\eqref{equ5.60} at $ \left(w,v\right) $. Then
\begin{enumerate}
\item
Characteristic cones of $ l^{\left\{1\right\}} $, $ l^{\left\{2\right\}} $, $ l^{\left\{3\right\}} $ coincide.
\item
The function $ v $ is a solution of the eikonal equation for $ l^{\left\{1\right\}} $.
\end{enumerate}
\end{proposition}

\begin{proof} It is easy to check the first claim by a direct calculation.
In the second claim we already know that $ v $ is a solution of the eikonal
equation for $ l^{\left\{2\right\}} $. Since characteristic cones coincide, $ v $ is also a
solution of the eikonal equation for $ l^{\left\{1\right\}} $. \end{proof}

\begin{remark} Let us provide a more conceptual heuristic proof of the first
claim of the proposition. It is enough to consider characteristic cones
for $ l^{\left\{1\right\}} $ and $ l^{\left\{3\right\}} $. If $ \varphi $ is a solution of the eikonal equation for $ l^{\left\{3\right\}} $, then
$ l^{\left\{3\right\}}\left(\widetilde{w},\widetilde{v}\right)=O\left(1\right) $ when $ k \to \infty $; here
\begin{equation}
\widetilde{w}\left(x,y,z\right)=W e^{ik\varphi\left(x,y,z\right)},\qquad \widetilde{v}\left(x,y,z\right)=V e^{ik\varphi\left(x,y,z\right)}
\notag\end{equation}
and $ W $ and $ V $ are appropriate constants. The usual arguments of calculus of
asymptotics (see, for example, \cite{GuillSte79Geo}) show that by allowing $ W $
and $ V $ depend smoothly on $ x,y,z,k^{-1} $ one can ensure that $ l^{\left\{3\right\}}\left(\widetilde{w},\widetilde{v}\right) $ is
asymptotically 0 when $ k \to \infty $.

In other words, starting with a solution of the eikonal equation for
$ l^{\left\{3\right\}} $, one can construct an asymptotic solution for $ l^{\left\{3\right\}} $. Since the
relationship between $ l^{\left\{3\right\}} $ and $ l^{\left\{1\right\}} $ is a linearization of relation between
System~\eqref{equ5.60} and Equation~\eqref{equ0.10}, we conclude that
$ W\left(x,y,z,k^{-1}\right)e^{ik\varphi\left(x,y,z\right)} $ is an asymptotic solution for $ l^{\left\{1\right\}} $ (as given this
argument is heuristic only, one needs to check that the order of taking
limits in $ k $ and in $ \varepsilon $ is correct). Thus $ \varphi $ is also a solution of the eikonal
equation for $ l^{\left\{1\right\}} $. Since the characteristic cone is spanned by
differentials of eikonal solutions, the characteristic cone for $ l^{\left\{3\right\}} $ is a
subset of a characteristic cone for $ l^{\left\{1\right\}} $.

On the other hand, characteristic cones of $ l^{\left\{1\right\}} $ and $ l^{\left\{3\right\}} $ are
quadratic cones, thus they should coincide. \end{remark}

\begin{remark} Let us repeat the arguments of Section~\ref{h5.5} in the linearized
situation. For any $ \lambda\in{\mathbb P}^{1} $ we can construct a corresponding triple $ \left(\widetilde{A},\widetilde{B},\widetilde{C}\right) $
with $ -\widetilde{A}/\widetilde{C}=\lambda $,
and a solution $ v^{\left[\lambda\right]} $ of the corresponding Equation~\eqref{equ5.50}, thus of $ l^{\left\{2\right\}} $.
Then $ v^{\left[\lambda\right]} $ is a solution of the eikonal equation for $ l^{\left\{1\right\}} $. Its level surfaces
are characteristic surfaces of $ l^{\left\{1\right\}} $. For each value of $ \lambda $ we obtain one
characteristic surface passing through a given point.

Moreover, when we vary $ \lambda $ the coefficients $ A/\widetilde{A} $, $ B/\widetilde{B} $, $ C/\widetilde{C} $ in
Equation~\eqref{equ5.70} vary as well. They cannot be proportional for
different values of $ \lambda $, thus all the above characteristic surfaces passing
through a given point have different directions.

In other words, at a given point we obtain a family of
characteristic directions parameterized by $ {\mathbb P}^{1} $. But characteristic
directions span a quadratic cone, and the base of this cone is $ {\mathbb P}^{1} $. It
easily follows that given a characteristic direction at a given point one
can find a value of $ \lambda\in{\mathbb P}^{1} $ such that $ dv^{\left[\lambda\right]} $ at the given point goes in the
prescribed direction. \end{remark}

This concludes arguments of Section~\ref{h5.5}, since using elementary
arguments we concluded that results of Theorem~\ref{th5.50} imply that normal
directions to $ {\mathcal F}_{\lambda} $ at a given point should span a quadratic cone.

\section{$ {\protect \mathcal F}_{\bullet} $-convex sets and the twistor transform }\label{h65}\myLabel{h65}\relax 

\begin{definition} Given a foliation $ {\mathcal F} $ on $ M $ and an open subset $ U\subset M $, we say that
$ U $ is $ {\mathcal F} $-{\em convex\/} if there is an open subset $ V\supset U $ such that $ {\mathcal F}|_{V} $ is direct (as
defined in Section~\ref{h1}), and for any leaf $ L $ of $ {\mathcal F}|_{V} $ the set $ L\cap U $ is
connected. Call $ U $ {\em strictly\/} $ {\mathcal F} $-{\em convex\/} if additionally the image of $ U $ under
the natural projection $ U \to {\mathfrak B}_{{\mathcal F}|_{V}} $ is homeomorphic to a ball. \end{definition}

It is obvious that any point $ m\in M $ has a strictly $ {\mathcal F} $-convex
neighborhood. For example, any direct neighborhood (see Section~\ref{h1}) goes.

\begin{definition} Given a foliation $ {\mathcal F} $ on $ M $, denote by $ {\mathfrak B}_{{\mathcal F}} $ the set of leaves of
$ {\mathcal F}|_{U} $, and by $ b\colon M \to {\mathfrak B}_{{\mathcal F}} $ the natural projection. Given an $ {\mathcal F} $-convex subset
$ U $, the set $ {\mathfrak B}_{{\mathcal F}|_{U}} $ has a natural structure of a manifold. Obviously, when
one decreases an $ {\mathcal F} $-convex subset $ U $, the base $ {\mathfrak B}_{{\mathcal F}|_{U}} $ decreases as well. In
particular, if $ m\in M $, then the germ\footnote{Given a manifold $ M $ with a closed submanifold $ N $, an open submanifold $ U\subset M $
is {\em compatible\/} with $ M $ if $ U\supset N $. Extend compatibility relation to an
equivalence relation $\sim$ between manifolds $ M_{1}\supset N $. Call equivalence classes $ \widetilde{M} $
{\em germs near\/} $ N $. A {\em mapping of germs\/} (or a {\em germ of a mapping\/}) $ \left(\widetilde{M},N\right) \to
\left(\widetilde{M}',N'\right) $ is a smooth mapping $ f\colon M \to M' $ such that $ f\left(N\right)\subset N' $; here $ M $, $ M' $ are
some representatives of classes $ \widetilde{M} $, $ \widetilde{M}' $. Such mappings are considered up to
the natural equivalence relation induced by restriction to compatible
open subsets.} of $ {\mathfrak B}_{{\mathcal F}|_{U}} $ near $ b\left(m\right) $ does not depend on
the $ {\mathcal F} $-convex neighborhood $ U $ of $ m $. Call this germ the {\em local base\/} of the
foliation $ {\mathcal F} $ near $ m $. \end{definition}

\begin{definition} Given a web $ \left\{{\mathcal F}_{\lambda}\right\}_{\lambda\in\Lambda} $ on $ M $, call an open subset $ U\subset M $ (strictly)
$ {\mathcal F}_{\bullet} $-{\em convex\/} if $ U $ is (strictly) $ {\mathcal F}_{\lambda} $-convex for all the foliations $ {\mathcal F}_{\lambda} $. \end{definition}

Recall that a {\em section\/} of a mapping $ \pi\colon M \to N $ is a right inverse to $ \pi $
mappings $ N \to M $.

\begin{definition} Consider a smooth web $ \left\{{\mathcal F}_{\lambda}\right\}_{\lambda\in\Lambda} $ on $ M $, and an $ {\mathcal F}_{\bullet} $-convex subset
$ U\subset M $. For any fixed $ \lambda\in\Lambda $ consider the manifold $ {\mathfrak B}_{{\mathcal F}_{\lambda}|_{U}} $. Taken together, they
form a manifold $ {\mathfrak T}={\mathfrak T}_{{\mathcal F}_{\bullet}}=\coprod_{\lambda\in\lambda}{\mathfrak B}_{{\mathcal F}_{\lambda}} $ equipped with a projection $ {\mathfrak T} \xrightarrow[]{\pi} \Lambda $ (which
sends $ {\mathfrak B}_{{\mathcal F}_{\lambda}} \to \left\{\lambda\right\} $). Call the pair $ \left({\mathfrak T},\pi\right) $ the {\em twistor transform\/} of $ {\mathcal F}_{\bullet}|_{U} $.

Given a point $ m\in M $, let $ \Sigma_{m}\left(\lambda\right) $ be a leaf of $ {\mathcal F}_{\lambda} $ which passes through $ m $.
Consider $ \Sigma_{m}\left(\lambda\right) $ as a point of $ {\mathfrak T} $. Then $ \Sigma_{m}\colon \Lambda \to {\mathfrak T} $ is a section of the
projection $ \pi $. If it cannot lead to a confusion, denote the image of this
map by the same symbol $ \Sigma_{m} $. \end{definition}

Describe in more details how the bases of $ {\mathcal F}_{\lambda}|_{U} $ for different $ \lambda $ fit
together inside $ {\mathfrak T} $. Call a submanifold $ S\subset M $ a {\em cross-sections\/} of a foliation
$ {\mathcal F} $ on $ M $ if $ S $ is transversal to the leaves of $ {\mathcal F} $, and each leaf of $ {\mathcal F} $
intersects $ S $ at most once. Obviously, cross-sections exist after
restriction of $ {\mathcal F} $ to an appropriate open subset $ U $, and are identified with
open subsets of the base $ {\mathfrak B}_{{\mathcal F}|_{U}} $.

Moreover, if $ {\mathcal F}_{\bullet} $ is a smooth web, and $ S $ is a cross-section to $ {\mathcal F}_{\lambda_{0}} $,
then for any point $ m\in S $ there is a neighborhood $ U\subset S $, $ U\ni m $, and a
neighborhood $ V\subset\Lambda $, $ V\ni\lambda_{0} $, such that $ U $ is a cross-section for $ {\mathcal F}_{\lambda} $, $ \lambda\in U $. This
gives a local identification of bases of $ {\mathcal F}_{\lambda} $, $ \lambda\in V $, thus a structure of a
manifold on $ {\mathfrak T} $.

\begin{remark} One can show that for a smooth web $ \left\{{\mathcal F}_{\lambda}\right\}_{\lambda\in\Lambda} $ on $ M $ with a compact
manifold $ \Lambda $, any point $ m\in M $ has an $ {\mathcal F}_{\bullet} $-convex neighborhood $ U $. Different
choices of $ U $ lead to different twistor transforms, but all of them
contain $ \Sigma_{m} $. Thus in such a case the germ of $ {\mathfrak T}_{{\mathcal F}_{\bullet}|_{U}} $ near $ \Sigma_{m} $
does not depend on $ U $.

In fact, this germ is well-defined for any smooth web $ {\mathcal F}_{\bullet} $. Indeed,
the construction with cross-sections allows gluing local bases for $ {\mathcal F}_{\lambda} $
near $ m $ into a germ of a manifold near $ \Sigma_{m} $. \end{remark}

To simplify the following exposition, we pretend that the twistor
transform is well-defined after a restriction of the web to an
appropriate small open subset of $ M $. This is always so if $ \Lambda $ is compact.
The general case can be always treated honestly by switching to the
language of germs.

\section{Explicit construction of the twistor transform }\label{h7}\myLabel{h7}\relax 

In the case of codimension 1 the construction of $ {\mathcal F} $-convex subsets
can be easily made explicit. Moreover, such an explicit
construction would make statements in the rest of the paper simpler to
formulate.

Put $ \Delta\left(a,b\right)=|a|/|b|+|b|/|a| $, $ \Delta\left(a_{1},\dots ,a_{d}\right)=\sum_{1\leq k<l\leq d}\Delta\left(a_{k},a_{l}\right) $. Consider
the following condition on a $ 1 $-form $ \alpha $ on $ U\subset{\mathbb V}^{d} $, $ 0\in U $:
\begin{equation}
\sum\Sb k=1 \\ l=1\endSb^{d}\left|\frac{\partial\alpha_{k}}{\partial x_{l}}\right|^{2} \leq \frac{1}{E\Delta\left(a_{1},\dots ,a_{d}\right)^{P}r^{2}}\sum_{k=1}^{d}\left|\alpha_{k}\right|^{2}
\label{equ7.55}\end{equation}\myLabel{equ7.55,}\relax 
here $ E $, $ r $ and $ P $ are numbers, $ \alpha_{k}\left(x_{1},\dots ,x_{d}\right) $, $ k=1,\dots ,d $, are components if
$ \alpha $, and $ a_{k}=\alpha_{k}\left(0,\dots ,0\right) $, $ 1\leq k\leq d $. This condition makes sense if $ a_{1},\dots ,a_{d}\not=0 $,
but if $ P=0 $, then it makes sense for any $ \alpha $.

The following lemma is not surprising:

\begin{lemma} \label{lm7.30}\myLabel{lm7.30}\relax  Fix an integer $ d>0 $. Consider a $ 1 $-form $ \alpha $ defined on $ {\mathbb B}_{r}^{d}\subset{\mathbb V}^{d} $
and a foliation $ {\mathcal F} $ on $ {\mathbb B}_{r}^{d} $ of codimension 1. Suppose that $ \alpha|_{0}\not=0 $, and $ \alpha\left(x\right) $
is normal to $ L_{x} $ for any $ x\in{\mathbb B}_{r}^{d} $; here $ L_{x} $ is the leaf of $ {\mathcal F} $ which passes
through $ x $. There are numbers $ D,E>0 $ (which depend on $ d $ only) such that for
any $ 0<\rho<r/D $
\begin{enumerate}
\item
if $ \alpha $ satisfies~\eqref{equ7.55} with $ P=0 $ in $ {\mathbb B}_{r}^{d} $, then $ {\mathbb B}_{\rho}^{d} $ is strictly
$ {\mathcal F} $-convex;
\item
if $ \alpha $ satisfies~\eqref{equ7.55} with $ P=2 $ in $ {\mathbb B}_{r}^{d} $, and $ a_{k}\not=0 $, $ 1\leq k\leq d $, then
$ \left({\mathbb B}_{\rho}^{1}\right)^{d} $ is strictly $ {\mathcal F} $-convex;
\end{enumerate}
\end{lemma}

\begin{proof} Transposing coordinates $ x_{k} $, one can ensure that $ |a_{d}|\geq|a_{k}| $,
$ k=1,\dots ,d-1 $. Changing $ \alpha $ to $ \alpha/a_{d} $ allows us to assume that $ a_{d}=1 $.

Obviously, one can find $ D $ and $ E $ such that the condition above
implies that in $ {\mathbb B}_{r}^{d} $ one has $ |\alpha_{d}-1|\leq1/2 $ and $ \sum_{k=1}^{d-1}|\alpha_{k}|^{2}\leq2d $. Consequently,
in $ \left({\mathbb B}_{r/D}^{d-1}\times{\mathbb V}^{1}\right)\cap{\mathbb B}_{r}^{d} $ one can write any leaf of $ {\mathcal F} $ which passes through
$ \left(0,\dots ,0,c\right) $, $ |c|<4\sqrt{d}r/D $, as $ x_{d}=\varphi_{c}\left(x_{1},\dots ,x_{d-1}\right) $, and $ \sum_{k=1}^{d-1}|\partial\varphi_{c}/\partial x_{k}|^{2} <
3\sqrt{d} $. Thus one can include $ {\mathbb B}_{\rho}^{d} $ and $ \left({\mathbb B}_{\rho}^{1}\right)^{d} $ into a chart-like subset of
$ {\mathbb B}_{r}^{d-1}\times{\mathbb V}^{1} $.

The next step is to show that the leaves intersected with $ {\mathbb B}_{\rho}^{d} $ or
$ \left({\mathbb B}_{\rho}^{1}\right)^{d} $ are connected. In the case of the ball it is enough to show that
\begin{equation}
N_{c}\left(x_{1},\dots ,x_{d-1}\right)=|\varphi_{c}\left(x_{1},\dots ,x_{d-1}\right)|^{2}+\sum_{k=1}^{d-1}|x_{k}|^{2}
\notag\end{equation}
is concave on $ {\mathbb B}_{r/D}^{d-1} $ for $ |c|<4\sqrt{d}r/D $. It is enough to show that the
Hessian $ \partial^{2}|\varphi_{c}|^{2}/\partial x_{k}\partial x_{l} $ of $ |\varphi_{c}|^{2} $ on $ {\mathbb B}_{r/D}^{d-1} $ cannot have a large negative
eigenvalue under an appropriate choice of constants $ E $ and $ D $. This Hessian
is a sum of a non-negative part $ 2\left(\partial\varphi_{c}/\partial x_{k}\right)\left(\partial\varphi_{c}/\partial x_{l}\right) $ and of
$ 2\varphi_{c}\partial^{2}\varphi_{c}/\partial x_{k}\partial x_{l} $.

In turn, it is enough to show that\footnote{In the complex-analytic case one needs to consider $ \partial\bar{\partial}/\partial x_{k}\bar{\partial}x_{l} $ as well as
$ \partial^{2}/\partial x_{k}\partial x_{l} $.} $ |\varphi_{c}|^{2}\sum\Sb k=1 \\ l=1\endSb^{d-1}|\partial^{2}\varphi_{c}/\partial x_{k}\partial x_{l}|^{2} $ can
be made bounded by 1/16. Since $ |\varphi_{c}| $ can be bounded by $ 7\sqrt{d}r/D $, it is
enough if we can bound second derivatives of $ \varphi_{c} $ as $ O\left(1/r\right) $.

However, the estimates on $ \alpha_{k} $, $ k=1,\dots ,n $, given above allow one to
estimate second derivatives of $ \varphi_{c} $ in terms of derivatives of $ \alpha_{k} $. This
finishes the proof of $ {\mathcal F} $-convexity in the case of the ball.

Investigate strict $ {\mathcal F} $-convexity in the case of the ball. It is clear
that one can invert $ \varphi_{c} $ and write $ c=\psi\left(x_{1},\dots ,x_{d}\right) $. It is enough to prove
that the $ \psi $-image of a small ball is convex, which follows from the
following simple

\begin{lemma} There is a number $ E $ (which depends on $ d $ only) such that given a
function $ \psi $ on $ {\mathbb B}_{r}^{d} $ such that $ d\psi $ satisfies~\eqref{equ7.55} with $ P=0 $ in $ {\mathbb B}_{r}^{d} $, then
the image $ \psi\left({\mathbb B}_{\rho}^{d}\right) $ is convex for $ 0<\rho<r $. \end{lemma}

Investigate the case of the polydisk. The stronger assumptions we
have in the polydisk case allow ensuring $ |\alpha_{k}-a_{k}|<|a_{k}|/F $ for any given
$ F>0 $. Now the statement follows from the following

\begin{lemma} Given $ d $, there are numbers $ F $ and $ D $ which satisfy the following
condition. Given a smooth function $ \psi\left(x_{1},\dots ,x_{d}\right) $ defined on $ \left({\mathbb B}_{r}^{1}\right)^{d} $ and any
numbers $ a_{1},\dots ,a_{d} $, and $ c $, if $ \psi $ satisfies
\begin{equation}
|\partial\psi/\partial x_{k}-a_{k}|<|a_{k}|/F\text{, }k=1,\dots ,d,
\label{equ7.57}\end{equation}\myLabel{equ7.57,}\relax 
on $ \left({\mathbb B}_{r}^{1}\right)^{d} $, then $ \psi\left(\left({\mathbb B}_{\rho}^{1}\right)^{d}\right) $ is convex, and $ \psi^{-1}\left(c\right)\cap\left({\mathbb B}_{\rho}^{1}\right)^{d} $ is connected
if non-empty for any $ 0<\rho<r/D $. \end{lemma}

\begin{proof} The statement is obvious in the real case, so assume
complex-analytic situation. Start with the case $ d=1 $. Put $ D=2 $, $ F=4 $. We may
assume $ r=1 $, $ a_{1}=1 $, then $ |\psi''|<1/2 $ on $ {\mathbb B}_{1/2}^{1} $. Thus the direction of the
tangent line $ l_{\tau} $ to the curve $ \psi\left(e^{i\tau}/2\right) $ rotates counterclockwise when $ \tau $
grows, with the angular velocity being close to 1. This implies convexity
of $ \psi\left({\mathbb B}_{r/D}^{1}\right) $. The connectivity of $ \psi^{-1}\left(c\right) $ is obvious.

In the case $ d>1 $ the convexity follows from similar arguments: the
boundary of the image of $ \left({\mathbb B}_{\rho}\right)^{d} $ is the curve
$ \Psi\left(\tau_{1}\right)=\psi\left(e^{i\tau_{1}}\rho,e^{i\tau_{2}\left(\tau_{1}\right)}\rho,\dots ,e^{i\tau_{d}\left(\tau_{1}\right)}\rho\right) $; here $ \tau_{k} $ are appropriate
functions, $ d\tau_{k}/d\tau\approx1 $, and the direction of the tangent line the curve $ \Psi\left(\tau\right) $
behaves as in the case $ d=1 $.

For connectivity proceed by induction in $ d $. We may assume that
$ |a_{d}|\geq|a_{k}| $, $ k=1,\dots ,d-1 $. Increasing $ F $ and $ D $, one can ensure that
$ \psi^{-1}\left(c\right)\cap\left(\left({\mathbb B}_{\rho}^{1}\right)^{d-1}\times{\mathbb B}_{r}^{1}\right) $ is given by $ x_{d}=\varphi_{c}\left(x_{1},\dots ,x_{d-1}\right) $ if $ c\in\psi\left(\left({\mathbb B}_{\rho}^{1}\right)^{d}\right) $, and
$ \varphi_{c} $ satisfies~\eqref{equ7.57} with $ d-1 $ taken instead of $ d $. Thus $ \varphi_{c}^{-1}\left(c_{1}\right)\cap\left({\mathbb B}_{\rho}^{1}\right)^{d-1} $
is connected if non-empty. On the other hand, $ \psi^{-1}\left(c\right)\cap\left({\mathbb B}_{\rho}^{1}\right)^{d} $ is
diffeomorphic to $ \varphi_{c}^{-1}\left({\mathbb B}_{\rho}^{1}\right)\cap\left({\mathbb B}_{\rho}^{1}\right)^{d-1} $. Since $ \varphi_{c}\left(\left({\mathbb B}_{\rho}^{1}\right)^{d-1}\right)\cap{\mathbb B}_{\rho}^{1} $ is convex, it
is connected, thus $ \psi^{-1}\left(c\right)\cap\left({\mathbb B}_{\rho}^{1}\right)^{d} $ is connected as well. \end{proof}

This finishes the proof of Lemma~\ref{lm7.30}. \end{proof}

\begin{amplification} \label{amp7.50}\myLabel{amp7.50}\relax  Consider a $ 1 $-form $ \alpha $ on $ {\mathbb B}_{r}^{d} $, and a $ 1 $-form $ \widetilde{\alpha} $ with
components $ \widetilde{\alpha}_{k}=\kappa_{k}\alpha_{k} $, $ k=1,\dots ,d $; here $ \kappa_{k} $ are arbitrary numbers, some of
which are non-0. Consider a foliation $ {\mathcal F} $ on $ {\mathbb B}_{r}^{d} $ of codimension 1. Suppose
that $ \widetilde{\alpha}\left(x\right) $ is normal to $ L_{x} $ for any $ x\in{\mathbb B}_{r}^{d} $; here $ L_{x} $ is the leaf of $ {\mathcal F} $ which
passes through $ x $. Let $ a_{k}\buildrel{\text{def}}\over{=}\alpha_{k}\left(0,\dots ,0\right)\not=0 $ for $ 1\leq k\leq d $.

There are numbers $ D,E>0 $ (which depend on $ d $ only) such that for
$ 0<\rho<r/D $
\begin{enumerate}
\item
if $ \alpha $ satisfies~\eqref{equ7.55} with $ P=2 $ in $ {\mathbb B}_{r}^{d} $, then $ {\mathbb B}_{\rho}^{d} $ is strictly
$ {\mathcal F} $-convex;
\item
if $ \alpha $ satisfies~\eqref{equ7.55} with $ P=4 $ in $ {\mathbb B}_{r}^{d} $, then $ \left({\mathbb B}_{\rho}^{1}\right)^{d} $ is strictly
$ {\mathcal F} $-convex;
\end{enumerate}
\end{amplification}

\begin{proof} Proceed similarly to the proof of Lemma~\ref{lm7.30}. One may assume
that $ \max _{k}|a_{k}|=1 $. Let $ A=\min _{k}|a_{k}| $. With the stronger conditions of the
amplification one can ensure that $ |\alpha_{k}-a_{k}|/A $ is sufficiently small in $ {\mathbb B}_{r}^{d} $.
Then the condition~\eqref{equ7.55} give absolute bounds on derivatives of $ \alpha_{k} $,
both from above and from below.

Multiplying $ \kappa_{k} $ by an appropriate constant, we may assume that
$ \max _{k}|\kappa_{k}|=1 $. Then given an estimate~\eqref{equ7.55} for $ \alpha $, we can estimate
$ \sum_{k=1}^{d}\left|\widetilde{\alpha}_{k}\right|^{2} $ from below, and $ \sum\Sb k=1 \\ l=1\endSb^{d}\left|\frac{\partial\widetilde{\alpha}_{k}}{\partial x_{l}}\right|^{2} $ from above in $ {\mathbb B}_{r}^{d} $,
loosing 2 units in $ P $. In particular, $ \widetilde{\alpha} $ satisfies~\eqref{equ7.55} with $ P=0 $ or
$ P=2 $. \end{proof}

Apply the obtained results to the nonlinear wave equation. Consider
the following condition on a function $ w $ defined on a subset $ V\subset{\mathbb V}^{3} $, $ 0\in V $:
\begin{equation}
\sum\Sb k=1 \\ l=1\endSb^{3}\left|\frac{\partial^{2}w}{\partial x_{k}\partial x_{l}}\right|^{2} \leq \frac{1}{E\Delta^{P}r^{2}}\sum_{k=1}^{3}\left|\frac{\partial w}{\partial x_{k}}\right|^{2},
\label{equ7.70}\end{equation}\myLabel{equ7.70,}\relax 
here $ E $, $ P $ and $ r $ are numbers, and $ \Delta=\Delta\left(w_{x_{1}}\left(0,0,0\right),w_{x_{2}}\left(0,0,0\right),w_{x_{3}}\left(0,0,0\right)\right) $.

\begin{theorem} \label{th8.20}\myLabel{th8.20}\relax  There are numbers $ E,D>0 $ such that given a non-degenerate
solution $ w\left(x,y,z\right) $ of Equation~\eqref{equ0.10} which satisfies~\eqref{equ7.70} in a ball
$ {\mathbb B}_{r}^{3} $, then there is a neighborhood $ U $ of (0,0,0) which is strictly
$ {\mathcal F}_{\bullet} $-convex w.r.t.~the Veronese web $ {\mathcal F}_{\bullet} $ which corresponds to $ w $; here one can
take
\begin{enumerate}
\item
$ U={\mathbb B}_{\rho}^{3} $ if $ P=2 $, $ 0<\rho<r/D $;
\item
$ U=\left({\mathbb B}_{\rho}^{1}\right)^{3} $ if $ P=4 $, $ 0<\rho<r/D $.
\end{enumerate}
\end{theorem}

\begin{proof} Obviously, any ball or polydisk is $ {\mathcal F} $-convex for 3 exceptional
foliations $ \left\{x=\operatorname{const}\right\} $, $ \left\{y=\operatorname{const}\right\} $, $ \left\{z=\operatorname{const}\right\} $ of the web. Other foliations
of the web are given by $ \left\{v_{\left(\widetilde{A},\widetilde{B},\widetilde{C}\right)}=\operatorname{const}\right\} $; here $ v_{\left(\widetilde{A},\widetilde{B},\widetilde{C}\right)} $ is a
non-degenerate solution of~\eqref{equ5.70}. Application of Amplification
~\ref{amp7.50} finishes the proof. \end{proof}

This theorem allows one to explicitly construct the twistor transform
of the Veronese web $ {\mathcal F}_{\bullet} $ associated to $ w $. Consider the set $ U $ of the
theorem, then the manifold with points enumerating leaves of all the
foliations $ {\mathcal F}_{\lambda}|_{U} $, $ \lambda\in{\mathbb P}^{1} $, is the twistor transform of $ {\mathcal F}_{\bullet} $.

Given an abstract Veronese web $ {\mathcal F}_{\bullet} $, by Lemma~\ref{lm3.60} one can describe
this web by a function $ w\left(x,y,z\right) $ which, by Theorem~\ref{th4.65}, satisfies
~\eqref{equ0.10} for appropriate $ \left(A,B,C\right) $. Thus one can apply the theorem above to
construct the twistor transform of $ {\mathcal F}_{\bullet} $.

\section{Sectional coordinates }\label{h10}\myLabel{h10}\relax 

Recall that a {\em submersion\/} is a smooth mapping of manifolds $ f\colon M \to N $
such that $ df|_{m}\colon {\mathcal T}_{m}M \to {\mathcal T}_{f\left(m\right)}N $ is surjective for any $ m\in M $.

\begin{lemma} Consider a complex manifold $ {\mathfrak T} $ with a submersion $ \pi $ onto a
manifold $ \Lambda $ and a submanifold $ S\subset{\mathfrak T} $ of codimension $ r $ such that $ \pi|_{S} $ is a
diffeomorphism. Given a covering $ \left\{V_{i}\right\} $ of $ {\mathfrak T} $ by Stein submanifolds, there
is an open subset $ U\supset S $ and identifications of $ U\cap\pi^{-1}\left(V_{i}\right) $ with $ V_{i}\times S_{i} $, $ S_{i}\subset{\mathbb C}^{r} $,
$ S_{i}\ni0 $; these identifications intertwine $ \pi $ with the projections $ V_{i}\times S_{i} \to
V_{i} $, and send $ S\cap\pi^{-1}\left(V_{i}\right) $ to $ V_{i}\times\left\{0\right\} $. \end{lemma}

\begin{proof} Suppose that $ r=1 $. Consider any function $ \widetilde{s}_{i} $ on a neighborhood of
$ S_{i}\buildrel{\text{def}}\over{=}S\cap\pi^{-1}\left(V_{i}\right) $ such that the vertical derivative of $ \widetilde{s}_{i} $ on $ S_{i} $ does not
vanish. Put $ s_{i} \buildrel{\text{def}}\over{=} \widetilde{s}_{i}-\widetilde{s}_{i}\circ\Sigma_{m}\circ\pi $. Then $ \left(\pi,s_{i}\right) $ gives the required
identification of a neighborhood of $ S_{i} $ with a subset of $ V_{i}\times{\mathbb C} $.

The existence of such a function $ \widetilde{s}_{i} $ follows from the fact that a
neighborhood of $ \pi^{-1}\left(V_{i}\right)\cap S $ is Stein if $ V_{i} $ is Stein. Indeed, any bundle
over a Stein manifold with a fiber isomorphic to a disk $ {\mathbb B}_{\varepsilon}^{1} $ is Stein
\cite{Fisch71Fib,AncSpe71Esp}.

In the case $ r>1 $ one needs to consider $ d $ functions $ \widetilde{s}_{i,k} $ instead of
one, and replaces $ {\mathbb B}_{\varepsilon}^{1} $ by $ {\mathbb B}_{\varepsilon}^{d} $ (using results of \cite{Ste75Fon}.) \end{proof}

\begin{remark} \label{rem10.10}\myLabel{rem10.10}\relax  These ``abstract nonsense'' arguments allow the following
construction: given a twistor transform $ {\mathfrak T} $ of a complex-analytic Veronese
web $ M\ni m $, cover $ {\mathbb P}^{1} $ by two disks $ V_{1,2} $, and glue a neighborhood of $ \Sigma_{m} $ from
two domains isomorphic to $ V_{i}\times{\mathbb B}_{\varepsilon}^{1} $ (with $ \pi $ compatible with projections to
$ V_{i} $). The gluing function $ g $ is going to be a mapping $ V_{1}\times{\mathbb B}_{\varepsilon}^{1}\ni\left(\lambda,t\right) \mapsto
\left(\lambda,g\left(\lambda,t\right)\right)\in V_{2}\times{\mathbb B}_{\varepsilon}^{1} $, with $ g\left(\lambda,t\right) $ defined on $ \left(V_{1}\cap V_{2}\right)\times{\mathbb B}_{\varepsilon}^{1} $. In particular, the
function $ g $ determines the germ of $ {\mathfrak T} \to {\mathbb P}^{1} $ near $ \Sigma_{m} $ up to isomorphism.
Later, in Theorem~\ref{th77.05}, we will see that this implies that the germ
of the Veronese web near $ m $ is determined by $ g $ up to isomorphism.

However, if $ {\mathfrak T} $ is a twistor transform one can achieve the same result
without applying the heavy machinery of complex analysis. One can
explicitly construct the required coordinate systems on open subsets of
$ {\mathfrak T} $. \end{remark}

\begin{definition} Consider a submanifold $ \gamma $ of a manifold $ M $ equipped with a web
$ {\mathcal F}_{\bullet} $ with a twistor transform $ {\mathfrak T} \xrightarrow[]{\pi} \Lambda $. Say that an open subset $ U\subset{\mathfrak T} $ is
{\em compatible\/} with $ \gamma $, if for any $ m\in\gamma $ and any $ \lambda\in\pi\left(U\right) $ the leaf of $ {\mathcal F}_{\lambda} $ passing
through $ m $ is in $ U $. \end{definition}

Obviously, a $ \gamma $-compatible open subset $ U\subset{\mathfrak T} $ is diffeomorphic to
$ \pi\left(U\right)\times\gamma $. In other words, such a subset defines a {\em local trivialization\/} of
the bundle $ \pi $. It is clear that $ \gamma $ and $ V\buildrel{\text{def}}\over{=}\pi\left(U\right) $ determine $ U $ uniquely.

In the rest of this section we assume that $ {\mathcal F}_{\bullet} $ is a Veronese web. As
Lemma~\ref{lm77.10} will show, for Veronese webs the normal bundles to
sections of $ \pi $ are not trivializable, thus in this case $ \pi\left(U\right) $ cannot
coincide with $ {\mathbb P}^{1} $.

Continue assuming that $ {\mathfrak T} $ is not a germ, but a bona fide manifold.

\begin{lemma} \label{lm10.30}\myLabel{lm10.30}\relax  Consider a point $ m $ on a Veronese web $ {\mathcal F}_{\bullet} $ on $ M $ and a curve
$ \gamma $ passing through $ m $. Let $ V_{m,\gamma}\subset{\mathbb P}^{1} $ consist of points $ \lambda $ such that $ \gamma $ is not
tangent to $ L_{\lambda}\left(m\right) $ at $ m $; here $ L_{\lambda}\left(m\right) $ is the leaf of $ {\mathcal F}_{\lambda} $ which passes through
$ m $. Let an open subset $ V\subset{\mathbb P}^{1} $ be compactly included into $ V_{m,\gamma} $. Then there is
a neighborhood $ \gamma_{1} $ of $ m $ in $ \gamma $ and a compatible with $ \gamma_{1} $ subset $ U\subset{\mathfrak T} $ with
$ \pi\left(U\right)=V $. \end{lemma}

\begin{proof} If $ \lambda_{0}\in V_{m,\gamma} $, there is a neighborhood $ V $ of $ \lambda_{0} $ and a neighborhood
$ W $ of $ m $ such that for $ \lambda\in V $ the leaves of $ {\mathcal F}_{\lambda} $ are not tangent to $ \gamma $ at any
point of $ W $, and each leaf intersects $ \gamma\cap W $ in at most one point. Since $ \bar{V}\subset{\mathbb P}^{1} $
is compact, one can decrease $ W $ so that this condition is satisfied for
any $ \lambda\in V $. Taking $ \gamma_{1}=\gamma\cap W $, and $ U $ to consists of leaves of $ {\mathcal F}_{\lambda} $, $ \lambda\in V $, which
intersect $ \gamma $ finishes the proof. \end{proof}

\begin{lemma} \label{lm10.40}\myLabel{lm10.40}\relax  The subset $ V_{m,\gamma}\subset{\mathbb P}^{1} $ of Lemma~\ref{lm10.30} is open, depends on
$ {\mathcal T}_{m}\gamma $ only, and $ {\mathbb P}^{1}\smallsetminus V_{m,\gamma} $ consists of at most $ \dim  M -1 $ points. Given any
subset $ Z\subset{\mathbb P}^{1} $ of at most $ \dim  M-1 $ points and $ m\in M $, one can find a curve $ \gamma $
passing through $ m $ such that $ V={\mathbb P}^{1}\smallsetminus Z $. Different possible directions $ {\mathcal T}_{m}\gamma $
correspond $ 1 $-to-1 to different ways of assigning multiplicities to points
of $ Z $ with the total being $ \dim  M-1 $. \end{lemma}

\begin{proof} The statements of this lemma concern one tangent space $ {\mathcal T}_{m}M $
only. The tangent spaces $ {\mathcal T}_{m}L_{\lambda}\left(m\right)\subset{\mathcal T}_{m}M $ are orthogonal complements to
directions $ {\mathbit n}_{m}\left(\lambda\right) $ in $ {\mathcal T}_{m}^{*}M $. Thus $ V_{m,\gamma} $ is determined by $ {\mathcal T}_{m}\gamma $ and the image of
the curve $ {\mathbit n}_{m}\colon {\mathbb P}^{1} \to {\mathbb P}{\mathcal T}_{m}^{*}M $. This is a Veronese curve, and any two such
curves are isomorphic. Thus we may replace $ {\mathcal T}_{m}^{*}M $ by an arbitrary vector
space $ S $ with a Veronese curve.

Take $ S $ to be the symmetric power $ \operatorname{Sym}^{d-1}{\mathbb V}^{2} $, $ \dim  S=d $, and let the
Veronese curve consists of $ \left(d-1\right) $st powers of elements of $ {\mathbb V}^{2} $. Then $ S^{*} $ can
be identified with homogeneous polynomials of degree $ d-1 $ of two variables
(two coordinates on $ {\mathbb V}^{2} $), thus $ {\mathcal T}_{m}\gamma\subset{\mathcal T}_{m}M=S^{*} $ provides such a polynomial $ p $ up
to a constant.

It is easy to check that $ \lambda\in V_{m,\gamma}\subset{\mathbb P}^{1}={\mathbb P}{\mathbb V}^{2} $ iff $ p $ does not vanish at the
points of $ {\mathbb V}^{2} $ in the direction of $ \lambda $. There are at most $ \deg  p=d-1 $ such
directions, and given such directions with appropriate multiplicities,
one can find a polynomial $ p\in S^{*} $ which vanishes at these points. \end{proof}

Now we can implement the program outlined in Remark~\ref{rem10.10}:

\begin{corollary} \label{cor10.50}\myLabel{cor10.50}\relax  Given $ m\in M $, one can find two curves $ \gamma_{1} $, $ \gamma_{2} $ passing
through $ m $ and two open subsets $ U_{1},U_{2}\subset{\mathfrak T} $ compatible with $ \gamma_{1} $, $ \gamma_{2} $
correspondingly such that $ U_{1}\cup U_{2} $ is a neighborhood of the section $ \Sigma_{m}\subset{\mathfrak T} $. \end{corollary}

\begin{proof} Indeed, one can find $ \gamma_{1} $, $ \gamma_{2} $ such that $ {\mathbb P}^{1}\smallsetminus V_{m,\gamma_{1}} $ is contained in
a small neighborhood of 0, and $ {\mathbb P}^{1}\smallsetminus V_{m,\gamma_{1}} $ is contained in a small
neighborhood of $ \infty $. To finish the proof, note that $ \Sigma_{m}\cap\pi^{-1}\left(\pi U\right)\subset U $ for any
subset $ U\subset{\mathfrak T} $ which is compatible with a curve $ \gamma $ passing through $ m $. \end{proof}

Consider two curves as in Corollary~\ref{cor10.50}. Let $ V_{1}=\pi U_{1} $, $ V_{2}=\pi U_{2} $.
Then $ U_{1}\simeq V_{1}\times\gamma_{1} $, $ U_{2}\simeq V_{2}\times\gamma_{2} $, thus identifications of $ \gamma_{1} $ and $ \gamma_{2} $ with $ {\mathbb B}_{\varepsilon}^{1} $ lead
the gluing function $ g\left(\lambda,t\right) $ as in the beginning of this section. The other
way to look at $ g $ is to consider it as a family of gluings $ \widehat{g}_{\lambda}\colon \gamma_{1} \to \gamma_{2} $,
$ \lambda\in V_{1}\cap V_{2} $.

Describe these gluings $ \widehat{g}_{\lambda} $ in geometric terms. This description does
not mention $ {\mathfrak T} $ as a manifold, thus one need not assume that $ {\mathfrak T} $ exists as a
manifold.

\begin{corollary} \label{cor10.60}\myLabel{cor10.60}\relax  Given a point $ m_{0} $ on a Veronese web $ M $, one can find
two curves $ \gamma_{1} $, $ \gamma_{2} $ passing through $ m_{0} $, a neighborhood $ W\subset M $ of $ m_{0} $, and two
open subsets $ V_{1},V_{2}\subset{\mathbb P}^{1} $ such that
\begin{enumerate}
\item
For any $ \lambda\in V_{j} $, $ j=1,2 $, and any $ m\in\gamma_{j} $ the leaf of $ {\mathcal F}_{\lambda}|_{W} $ which passes through
$ m $ intersects $ \gamma_{j} $ at exactly one point $ m $ and is transversal to $ \gamma_{j} $;
\item
For any $ \lambda\in V_{1}\cap V_{2} $, and any $ m\in\gamma_{1} $ the leaf of $ {\mathcal F}_{\lambda}|_{W} $ which passes through
$ m $ intersects $ \gamma_{2} $; denote the (unique) point of intersection by $ \widehat{g}_{\lambda}\left(m\right) $;
\item
$ V_{1}\cup V_{2}={\mathbb P}^{1} $.
\end{enumerate}

The germ near $ \left(V_{1}\cap V_{2}\right)\times\left\{m\right\} $ of the function $ \widehat{g}_{\bullet}\colon \left(V_{1}\cap V_{2}\right)\times\gamma_{1} \to \gamma_{2} $
uniquely determines the germ of the twistor transform $ {\mathfrak T} $ of $ M $ near the
section $ \Sigma_{m_{0}} $ and the germ of $ {\mathcal F}_{\bullet} $ near $ m $. For any $ 0<\varepsilon<1 $ one can ensure that
$ V_{1}\supset\left\{z \mid |z|>\varepsilon\right\} $, $ V_{2}\supset\left\{z \mid |z|<1/\varepsilon\right\} $. \end{corollary}

\section{Explicit construction of the gluing function }\label{h107}\myLabel{h107}\relax 

In conditions of Corollary~\ref{cor10.60} identify a neighborhood of $ m_{0} $
in $ \gamma_{1} $ with $ {\mathbb B}_{\varepsilon}^{1} $, and a neighborhood of $ m_{0} $ in $ \gamma_{2} $ with a subset of $ {\mathbb C} $. This
would make the gluing function $ g\left(\lambda,t\right) $ into a function $ \left(V_{1}\cap V_{2}\right)\times{\mathbb B}_{\varepsilon}^{1} \to {\mathbb C} $. A
different choice of identifications would lead to $ \widetilde{g}\left(\lambda,t\right)=f\left(g\left(\lambda,F\left(t\right)\right)\right) $ for
appropriate invertible functions $ f\left(z\right) $, $ F\left(z\right) $.

Describe $ g\left(\lambda,t\right) $ in terms of the function $ w\left(x,y,z\right) $ which identifies
the Veronese web. Later, in Appendix~\ref{h105}, we will see that the gluing
function should depend only on the restriction of $ w $ and first derivatives
of $ w $ to an appropriate surface. Here we prove this only in the case
of surfaces of a special form.

\begin{theorem} \label{th107.40}\myLabel{th107.40}\relax  Consider a complex-analytic non-degenerate solution
$ w\left(x,y,z\right) $ of the nonlinear wave equation~\eqref{equ0.10} defined in a
neighborhood of (0,0,0). Fix $ 0<r<1 $, $ \lambda_{1},\lambda_{2},\lambda_{3}\in{\mathbb P}^{1} $, $ |\lambda_{1,2}|<r $, $ |\lambda_{3}|>1/r $. Let
$ Y\left(x\right) $ be any function such that $ Y'\left(x\right)=dY/dx $ is nowhere 0, and $ Y\left(0\right)=0 $.
Consider the following family of ODEs with a parameter $ \mu $ on a function
$ z\left(x\right) $:
\begin{equation}
\frac{dz}{dx}=\frac{Aw_{x}\left(x,Y\left(x\right),z\right)}{\mu Cw_{z}\left(x,Y\left(x\right),z\right)}-\frac{Bw_{y}\left(x,Y\left(x\right),z\right)}{\left(\mu-1\right)Cw_{z}\left(x,Y\left(x\right),z\right)}Y'\left(x\right);
\notag\end{equation}
Let $ g_{\mu}\left(t\right) $ be $ z\left(0\right) $; here $ z\left(x\right) $ is the solution of this equation with the
initial data $ z\left(t\right)=0 $. Then for any $ \varepsilon_{1}>0 $ one can find an appropriate $ \delta>0 $ so
that the function $ g_{\mu}\left(t\right) $ is correctly defined if $ |\mu|>\varepsilon_{1} $, $ |\mu-1|>\varepsilon_{1} $, and
$ |t|<\delta $.

Consider $ \varepsilon $ such that $ r<\varepsilon<1 $. Define a surface $ \widetilde{{\mathfrak T}} $ by gluing
$ {\mathbb B}_{1/\varepsilon}^{1}\times{\mathbb B}_{\delta}^{1} $ and $ \left({\mathbb P}^{1}\smallsetminus\bar{{\mathbb B}}_{\varepsilon}^{1}\right)\times{\mathbb C} $ via $ {\mathbb B}_{1/\varepsilon}^{1}\times{\mathbb B}_{\delta}^{1}\ni\left(\lambda,t\right) \mapsto \left(\lambda,\widetilde{g}\left(\lambda,t\right)\right)\in\left({\mathbb P}^{1}\smallsetminus\bar{{\mathbb B}}_{\varepsilon}^{1}\right)\times{\mathbb C} $,
$ \varepsilon<|\lambda|<1/\varepsilon $, $ |t|<\delta $; here $ \widetilde{g}\left(\lambda,t\right)=g_{\mu}\left(t\right) $, $ \mu=\left(\lambda_{1}:\lambda_{2}:\lambda_{3}:\lambda\right) $, and $ \delta $ corresponds to
$ \varepsilon_{1} $ such that $ |\mu|>\varepsilon_{1} $ and $ |\mu-1|>\varepsilon_{1} $ if $ |\lambda|>\varepsilon $. Since $ g_{\mu}\left(0\right)\equiv 0 $, $ \widetilde{{\mathfrak T}} $ has a section
$ \widetilde{\Sigma}_{\left(0,0,0\right)}=\left\{\left(\lambda,0\right)\right\} $. Coordinates $ \lambda $ glue into a projection $ {\mathfrak T} \to {\mathbb P}^{1} $.

Suppose that
\begin{equation}
\left|\frac{Q\lambda_{1}-\lambda_{2}}{Q-1}\right| > 1/\varepsilon,\qquad Q=Y'\left(0\right)\frac{Bw_{y}\left(0,0,0\right)}{Aw_{x}\left(0,0,0\right)}.
\label{equ107.60}\end{equation}\myLabel{equ107.60,}\relax 
Then the germ of $ \widetilde{{\mathfrak T}} $ near $ \widetilde{\Sigma}_{\left(0,0,0\right)} $ is isomorphic to the germ of the
twistor transform $ {\mathfrak T} $ of the Veronese web associated\footnote{As in Remark~\ref{rem4.95}.} to $ w\left(x,y,z\right) $ near
$ \Sigma_{\left(0,0,0\right)} $. \end{theorem}

\begin{proof} Consider the $ 3 $-dimensional Veronese web $ M $ associated to
$ w\left(x,y,z\right) $ such that the foliations $ \left\{x=\operatorname{const}\right\} $, $ \left\{y=\operatorname{const}\right\} $, $ \left\{z=\operatorname{const}\right\} $ are
associated to $ \lambda=\lambda_{1} $, $ \lambda=\lambda_{2} $, $ \lambda=\lambda_{3} $. Take $ m_{0}=\left(0,0,0\right) $, $ \gamma_{2} $ to be the $ z $-axis.
Then the subset $ V_{m_{0},\gamma_{2}} $ (in notations of Lemma~\ref{lm10.30}) is
$ {\mathbb P}^{1}\smallsetminus\left\{\lambda_{1},\lambda_{2}\right\}\supset{\mathbb P}^{1}\smallsetminus\bar{{\mathbb B}}_{\varepsilon}^{1} $, since $ \gamma_{2} $ is an intersection of a leaf of $ {\mathcal F}_{\lambda_{1}} $ and of a
leaf of $ {\mathcal F}_{\lambda_{2}} $. Similarly, for a curve $ \gamma $ in $ xy $-plane the subset $ V_{m,\gamma} $ is
$ {\mathbb P}^{1}\smallsetminus\left\{\lambda_{3},\lambda\left(m\right)\right\} $; here $ \lambda\left(m\right)=\lambda_{1} $ for the curves $ x=\operatorname{const} $ in $ xy $-plane, $ \lambda\left(m\right)=\lambda_{2} $
for the curves $ y=\operatorname{const} $ in $ xy $-plane. It is clear that for a curve with any
other direction $ \lambda\left(m\right)\not=\lambda_{1} $ and $ \lambda\left(m\right)\not=\lambda_{2} $. In particular, it is so for the
curve $ \gamma_{1} $ given by $ y=Y\left(x\right) $. Thus $ V_{m_{0},\gamma_{1}}\cup V_{m_{0},\gamma_{2}}={\mathbb P}^{1} $. Thus $ \gamma_{1} $, $ \gamma_{2} $ satisfy
conditions of Corollary~\ref{cor10.60}, thus one can glue the twistor
transform $ {\mathfrak T} $ from two open subsets, one being a bundle over $ V_{m_{0},\gamma_{1}} $,
another over $ V_{m_{0},\gamma_{2}} $.

Moreover, $ V_{m_{0},\gamma_{1}}\supset{\mathbb P}^{1}\smallsetminus\bar{{\mathbb B}}_{\varepsilon}^{1} $, and if $ |\lambda\left(m_{0}\right)|>1/\varepsilon $, then $ V_{m_{0},\gamma_{2}}\supset{\mathbb B}_{1/\varepsilon}^{1} $. In
such a case $ {\mathfrak T} $ can be glued from two open subsets, one being a bundle over
$ {\mathbb P}^{1}\smallsetminus\bar{{\mathbb B}}_{\varepsilon}^{1} $, another over $ {\mathbb B}_{1/\varepsilon}^{1} $. To describe $ {\mathfrak T} $, it is enough to describe the
gluing function $ g\left(\lambda,t\right) $, $ \varepsilon<|\lambda|<1/\varepsilon $, for small $ t $. Taking $ z $ as the
coordinate on $ \gamma_{2} $ and $ x $ as the coordinate on $ \gamma_{1} $, one can describe this
gluing function in the following way: take a point $ m=\left(t,Y\left(t\right),0\right) $ on $ \gamma_{1} $,
find the leaf of $ {\mathcal F}_{\lambda} $ which passes through $ m $, and intersect this leaf with
$ \gamma_{2} $. Then $ g\left(\lambda,t\right) $ is the $ z $-coordinate of the point of intersection.

Consider the surface $ N $ given by the equation $ y=Y\left(x\right) $. The foliation
$ {\mathcal F}_{\lambda} $ can be described by the equations $ v\left(x,y,z\right)=\operatorname{const} $; here the derivative
of $ v $ is given by Corollary~\ref{cor4.50}. The curves cut out by this foliation
on $ N $ have both $ \left(-dY/dx,1,0\right) $ and $ \left(v_{x},v_{y},v_{z}\right) $ as normal vectors. Thus these
curves are tangent to directions
\begin{equation}
\left(p_{3}\left(\lambda\right)w_{z},p_{3}\left(\lambda\right)w_{z}dY/dx,-p_{1}\left(\lambda\right)w_{x}-p_{2}\left(\lambda\right)w_{y}dY/dx\right)
\notag\end{equation}
(notations as in~\eqref{equ4.50}). One can easily check that the ODE of the
theorem describes $ xz $-projections of these curves for $ \mu=\left(\lambda_{1}:\lambda_{2}:\lambda_{3}:\lambda\right) $. Thus
$ g\left(\lambda,t\right)=\widetilde{g}\left(\lambda,t\right) $.

The only thing to prove is $ |\lambda\left(m_{0}\right)|>1/\varepsilon $. In fact $ \lambda\left(m_{0}\right)=\frac{Q\lambda_{1}-\lambda_{2}}{Q-1} $.
To check this, it is enough to find the intersection of the leaf of $ {\mathcal F}_{\lambda} $
through (0,0,0) with $ z=0 $. As above, the direction of this curve is given
by $ \left(-v_{y},v_{x},0\right)=\left(-p_{2}\left(\lambda\right)w_{y},p_{1}\left(\lambda\right)w_{x},0\right) $. Again, it is easy to check that this
agrees with~\eqref{equ107.60}. \end{proof}

\begin{remark} Obviously, the condition~\eqref{equ107.60} is satisfied in $ Y'\left(0\right) $ is
inside a non-empty disk in $ {\mathbb C}{\mathbb P}^{1} $. In fact, there is a canonical choice of
$ Y\left(x\right) $ which automatically satisfies~\eqref{equ107.60}. Indeed, the condition
$ \lambda\left(m\right)=\lambda_{3} $ gives a direction field on $ xy $-plane, take an integral curve of
this direction field. Explicitly,
\begin{equation}
\frac{dY}{dx}=\frac{Aw_{x}\left(x,Y,0\right)}{Bw_{y}\left(x,Y,0\right)},\qquad Y\left(0\right)=0.
\label{equ107.30}\end{equation}\myLabel{equ107.30,}\relax 
\end{remark}

\begin{remark} \label{rem107.90}\myLabel{rem107.90}\relax  If $ \varepsilon $ with the properties required in the theorem does
not exist, by decreasing $ \delta $ one can ensure that the set of values of $ \lambda $ for
which~\eqref{equ107.60} does not hold is in a small disk $ D $ which does
not contain $ \lambda_{1} $ and $ \lambda_{2} $. If there is a circle on $ {\mathbb P}^{1} $ which separates $ \left\{\lambda_{1},\lambda_{2}\right) $
from $ \lambda_{3} $ and $ D $, then one can use this circle instead of $ \left\{|z|=1\right\} $ in Theorem
~\ref{th107.40}.

If there is no such circle, then $ \frac{B
Y'\left(0\right)w_{y}\left(0,0,0\right)}{Bw_{y}\left(0,0,0\right)Y'\left(0\right)-Aw_{x}\left(0,0,0\right)} $ is real and is between 0 and
1. In particular, by a projective transform of $ {\mathbb P}^{1} $ one can make $ \lambda_{1},\lambda_{2},\lambda_{3} $
real, and the disk $ D $ centered on the real axis between $ \lambda_{1} $ and $ \lambda_{2} $. If
additionally $ w\left(x,y,z\right) $ is real for real $ x,y,z $, and $ A,B,C $ are real, then
the real $ \left(A,B,C\right) $-equation is hyperbolic near (0,0,0) w.r.t.~the surface
$ y=Y\left(x\right) $. Thus this case is of special interest.

In such a case it is hard to describe $ {\mathfrak T} $ by representing $ {\mathbb P}^{1} $ as a
union of two disks, but one can glue $ {\mathfrak T} $ using the same function $ g_{\mu}\left(t\right) $ if
one covers $ {\mathbb P}^{1} $ by two regions of more complicated form. For example,
consider small disks $ D_{1,2,3} $ centered at $ \lambda_{1,2,3} $, consider a contour $ L $
which goes along the line $ \operatorname{Im}\lambda=0 $ with the exceptions of going around $ D_{1} $
and $ D_{2} $ from above, and around $ D $ and $ D_{3} $ from below. The function $ g\left(\lambda,t\right) $ is
still correctly defined for $ \lambda $ near $ L $, thus one can describe $ {\mathfrak T} $ by gluing
neighborhoods of the regions above $ L $ and below $ L $.

Note that for the values of $ \lambda\in L $ which are on the real axis the
function $ g\left(\lambda,t\right) $ can be defined in terms of solving a real ODE. \end{remark}

\section{Equipped twistor transforms and infinitesimal families }\label{h75}\myLabel{h75}\relax 

For a mapping $ \pi\colon M \to N $ denote by $ \Gamma\left(N,\pi\right) $ the set of {\em sections\/} of $ \pi $,
i.e., of right inverse mappings to $ \pi $.

\begin{definition} Given a web $ \left\{{\mathcal F}_{\lambda}\right\}_{\lambda\in\Lambda} $ on $ M $ with the twistor transform $ {\mathfrak T} \xrightarrow[]{\pi} \Lambda $,
consider the family $ \left\{\Sigma_{m}\right\}_{m\in M} $ of sections of $ \pi $. The {\em equipped twistor
transform\/} of $ {\mathcal F}_{\bullet} $ is the mapping $ {\mathfrak T} \xrightarrow[]{\pi} \Lambda $ together with a family of
sections $ \left\{\Sigma_{m}\right\}_{m\in M} $. \end{definition}

Given such a structure $ \left({\mathfrak T},\Lambda,\pi,M,\Sigma_{\bullet}\right) $, and $ \lambda\in\Lambda $, consider a mapping $ {\mathbit s}_{\lambda}:
M \to \pi^{-1}\left(\lambda\right)\colon m \mapsto \Sigma_{m}\left(\lambda\right) $. If this structure comes from an equipped
twistor transform, this mapping is a submersion.

\begin{lemma} \label{lm75.30}\myLabel{lm75.30}\relax  Consider a submersion $ {\mathfrak T} \xrightarrow[]{\pi} \Lambda $ together with a family of
sections $ \left\{\Sigma_{m}\right\}_{m\in M} $ parameterized by a manifold $ M $. If the rank of
differential $ d{\mathbit s}_{\lambda}|_{m} $ of the mapping $ {\mathbit s}_{\lambda} $ does not depend on $ m $ and $ \lambda $, then $ M $
is equipped with a canonically defined web structure $ \left\{{\mathcal F}_{\lambda}\right\}_{\lambda\in\Lambda} $, the leaf
$ L_{\lambda,m_{0}} $ of $ {\mathcal F}_{\lambda} $, $ \lambda\in\Lambda $, which passes through $ m_{0}\in M $ consists of points $ m\in M $ such
that $ \Sigma_{m}\left(\lambda\right)=\Sigma_{m_{0}}\left(\lambda\right) $.

If $ \left({\mathfrak T},\Lambda,\pi,M,\Sigma_{\bullet}\right) $ is a twistor transform of a web $ \widetilde{{\mathcal F}}_{\bullet} $ on $ M $, then
$ \widetilde{{\mathcal F}}_{\bullet}={\mathcal F}_{\bullet} $. \end{lemma}

\begin{proof} Indeed, mappings with constant rank of the differential are
submersions onto their images, thus preimages of points are foliations on
$ M $. The other statements are obvious. \end{proof}

It is clear that in the conditions of the lemma if $ {\mathbit s}_{\lambda} $ is not of
maximal possible rank (i.e., is not a submersion), then $ {\mathfrak T}'=\bigcup_{\lambda}\operatorname{Im} {\mathbit s}_{\lambda} $ is a
submanifold of $ {\mathfrak T} $, and $ \left({\mathfrak T}',\Lambda,\pi',M,\Sigma_{\bullet}\right) $ is the twistor transform of $ {\mathcal F}_{\bullet} $; here
$ \pi'=\pi|_{{\mathfrak T}'} $.

Lemma~\ref{lm75.30} shows that one can reconstruct a web on $ M $ by its
equipped twistor transform $ \left({\mathfrak T},\Lambda,\pi,M,\Sigma_{\bullet}\right) $. In fact in many cases to
reconstruct the web one needs much less data than $ \left({\mathfrak T},\Lambda,\pi,M,\Sigma_{\bullet}\right) $. Later, in
Section~\ref{h85}, we explain when the same information is contained in
$ \left({\mathfrak T},\Lambda,\pi\right) $, at least if one considers $ M $ up to isomorphism. Illustrate this
by several weaker statements.

Suppose that the mapping $ \Sigma_{\bullet}\colon M \to \Gamma\left(\Lambda,\pi\right)\colon m \mapsto \Sigma_{m} $ is injective, in
other words, $ {\mathcal F}_{\bullet} $ is separating. In such cases $ M $ as a set is identified
with $ \operatorname{Im}\Sigma_{\bullet} $. In fact $ \Gamma\left(\Lambda,\pi\right) $ has a natural topology, and if $ \Sigma $ is a
homeomorphism on its image, then the topology on $ M $ can be also
reconstructed basing on $ \operatorname{Im}\Sigma_{\bullet}\subset\Gamma\left(\Lambda,\pi\right) $. In such a case if we are interested
in $ \left(M,{\mathcal F}_{\bullet}\right) $ up to homeomorphism, it may be reconstructed given $ \left({\mathfrak T},\Lambda,\pi,\operatorname{Im}\Sigma_{\bullet}\right) $.

One should expect that the same argument will work for
diffeomorphisms as far as the differential of $ \Sigma_{\bullet} $ is injective. However,
in general $ \Gamma\left(\Lambda,\pi\right) $ is not finite-dimensional, thus this question is a
little bit more subtle. However, it is relatively easy to describe what
is an individual tangent space to $ \Gamma\left(\Lambda,\pi\right) $. This tangent space is going to
be the target of the differential of $ \Sigma_{\bullet} $.

\begin{definition} \label{def75.50}\myLabel{def75.50}\relax  Given a section $ \Sigma $ of submersion $ \pi\colon {\mathfrak T} \to \Lambda $, the
{\em tangent space\/} to $ \Gamma\left(\Lambda,\pi\right) $ at $ \Sigma $ is the vector space $ \Gamma\left(S,{\mathcal N}S\right) $, $ S=\operatorname{Im}\Sigma $. Call
elements of $ \Gamma\left(S,{\mathcal N}S\right) $ {\em infinitesimal deformations}. Given a family $ \left\{\Sigma_{m}\right\}_{m\in M} $ of
sections of $ \pi $, the {\em infinitesimal family\/} of $ \left\{\Sigma_{m}\right\} $ at $ m_{0}\in M $ is the naturally
defined mapping $ d\Sigma|_{m_{0}}\colon {\mathcal T}_{m_{0}}M \to \Gamma\left(\Sigma_{m_{0}},{\mathcal N}\Sigma_{m_{0}}\right) $. Say that a family $ \left\{\Sigma_{m}\right\} $ is
{\em immersive\/} if $ d\Sigma|_{m} $ is a monomorphism for any $ m\in M $. \end{definition}

Describe what is $ d\Sigma|_{m} $ and what is the geometric meaning of this
definition. To define $ d\Sigma|_{m} $, it is enough to consider the case $ \dim  M=1 $. A
{\em smooth\/} $ 1 $-{\em parametric family\/} $ \sigma_{t} $, $ t\in T\subset{\mathbb V}^{1} $, {\em of sections\/} of $ \pi $ is a mapping $ \sigma:
\Lambda\times T \to {\mathfrak T} $ such that $ \pi\circ\sigma $ coincides with the projection $ p_{1}\colon \Lambda\times T \to \Lambda $. Given
$ \sigma $ and $ t\in T $, consider the derivatives $ d\sigma|_{\left(\lambda,t\right)} $ at points of $ \Lambda\times\left\{t\right\} $. Clearly,
$ d\sigma|_{\left(\lambda,t\right)} $ maps $ {\mathcal T}_{\lambda}\Lambda\oplus{\mathcal T}_{t}{\mathbb V}^{1} $ to $ {\mathcal T}_{\sigma\left(\lambda,t\right)}{\mathfrak T} $. It can be split into a direct sum of
a mapping $ d\sigma|_{\left(\lambda,t\right)}^{\left(1\right)}\colon {\mathcal T}_{\lambda}\Lambda \to {\mathcal T}_{\sigma\left(\lambda,t\right)}{\mathfrak T} $ and $ d\sigma|_{\left(\lambda,t\right)}^{\left(2\right)}\colon {\mathcal T}_{t}{\mathbb V}^{1} \to {\mathcal T}_{\sigma\left(\lambda,t\right)}{\mathfrak T} $.

Note that the condition $ \pi\circ\sigma=p_{1} $ determines some components of
$ d\sigma|_{\left(\lambda,t\right)} $. Indeed, consider $ S=\operatorname{Im}\sigma\left(\bullet,t\right) $. It is a submanifold of $ {\mathfrak T} $. Given
$ \lambda\in\Lambda $, the vector space $ {\mathcal T}_{\sigma\left(\lambda,t\right)}{\mathfrak T} $ can be decomposed into a direct sum of
tangent spaces to $ \pi^{-1}\left(\lambda\right) $ and to $ S $. Denote components of $ v\in{\mathcal T}_{\sigma\left(\lambda,t\right)}{\mathfrak T} $ in
this decomposition by $ v^{\text{vert}} $ and $ v^{\text{hor}} $. In particular, the mappings $ d\sigma^{\left(1\right)} $,
$ d\sigma^{\left(2\right)} $ can be further subdivided into $ d\sigma^{\left(1\right)\text{vert}} $, $ d\sigma^{\left(2\right)\text{vert}} $, $ d\sigma^{\left(1\right)\text{hor}} $,
$ d\sigma^{\left(2\right)\text{hor}} $. It is clear that given two families $ \sigma $ and $ \widetilde{\sigma} $, if vertical
components of $ d\sigma $ and $ d\widetilde{\sigma} $ coincide, then $ d\sigma $ and $ d\widetilde{\sigma} $ coincide. Moreover,
$ d\sigma^{\left(1\right)\text{vert}} $ obviously vanishes. In particular, the only ``interesting'' part
of differential of $ \sigma $ is $ d\sigma^{\left(2\right)\text{vert}} $.

On the other hand, the vertical component of $ v\in{\mathcal T}_{\sigma\left(\lambda,t\right)}{\mathfrak T} $ can be also
naturally identified with an element of the quotient by the vector
subspace of horizontal sections $ {\mathcal T}_{\sigma\left(\lambda,t\right)}{\mathfrak T}/{\mathcal T}_{\sigma\left(\lambda,t\right)}S={\mathcal N}_{\sigma\left(\lambda,t\right)}S $, i.e., with a
normal vector to $ S $ at $ \sigma\left(\lambda,t\right) $. Since $ d\sigma^{\left(2\right)\text{vert}} $ sends $ \delta t\in{\mathcal T}_{t_{0}}{\mathbb V}^{1} $ to a normal
vector to $ S $ at $ \sigma\left(\lambda,t\right) $ for each $ \lambda\in\Lambda $, it associates to $ \delta t $ a section of the
normal bundle $ {\mathcal N}S $.

The following statement is obvious:

\begin{lemma} The equipped twistor transform of a web is immersive iff the
web is separating. \end{lemma}

It is clear that for an immersive family $ \Sigma_{m} $, $ m\in M $, the mappings $ {\mathbit s}_{\lambda} $,
$ \lambda\in\Lambda $, separate points on small open subsets of $ M $ (even infinitesimally).
Thus the structure of the manifold on $ M $ is reconstructed from the mapping
of the set $ M $ to $ \Gamma\left(\Lambda,\pi\right) $.

\begin{corollary} \label{cor75.70}\myLabel{cor75.70}\relax  Consider a weakly separating and separating web $ {\mathcal F}_{\bullet} $ on
$ M $. Then $ {\mathcal F}_{\bullet} $ can be reconstructed up to a diffeomorphism by the twistor
transform $ {\mathfrak T} \xrightarrow[]{\pi} \Lambda $ of $ {\mathcal F}_{\bullet} $ together with the subset $ {\mathcal M}\subset\Gamma\left(\Lambda,\pi\right) $ consisting of
sections which correspond to points of $ M $. \end{corollary}

\section{Kodaira--Spencer deformation of a section }\label{h8}\myLabel{h8}\relax 

In the classification of complex-analytic Veronese webs the central
role is played by the following corollary\footnote{Since one-dimensional Cauchy--Riemann equations are not overdetermined,
in the case $ \dim \Lambda=1 $ we are most interested in Kodaira--Spencer deformation
theory can be replaced by an argument involving an implicit function
theorem (in normed spaces).} of Kodaira--Spencer
deformation theory (for example, see \cite{Kod}).

\begin{definition} Say that a vector bundle $ E $ over a topological space $ \Lambda $ is
{\em cohomologically trivial\/} if $ H^{k}\left(\Lambda,E\right)=0 $ for $ k>0 $. \end{definition}

\begin{theorem} \label{th9.50}\myLabel{th9.50}\relax  Consider an $ n $-dimensional complex manifold $ {\mathfrak T} $ equipped
with a surjective submersion $ \pi\colon {\mathfrak T} \to \Lambda $, and with a section $ \Sigma\colon \Lambda \to {\mathfrak T} $ of
the projection $ \pi $. Let $ S=\operatorname{Im} \Sigma $, suppose that $ {\mathcal N}S $ is cohomologically
trivial, and $ \Lambda $ is compact. Then there is a connected complex manifold $ M $,
a mapping $ \sigma\colon \Lambda\times M \to {\mathfrak T} $, and a neighborhood $ U $ of $ S $ in $ {\mathfrak T} $ such that
\begin{enumerate}
\item
$ \pi\circ\sigma $ coincides with the projection $ \Lambda\times M \to \Lambda $;
\item
for any section $ s $ of $ \pi|_{U} $ there is unique $ m\in M $ such that $ s=\sigma|_{\Lambda\times\left\{m\right\}} $;
denote by $ m_{0}\in M $ the point which corresponds to $ s=\Sigma $;
\item
the infinitesimal family\footnote{See Definition~\ref{def75.50}.} $ d\sigma|_{m_{0}}\colon {\mathcal T}_{m_{0}}M \to\Gamma\left(S,{\mathcal N}S\right) $ is a bijection.
\end{enumerate}
\end{theorem}

\begin{remark} To translate to the usual formulation of deformation theory,
instead of deforming the mapping $ \Sigma $, one should deform the submanifold $ S $.
Then the first condition on $ \sigma $ disappears (is just gives a normalization
by identifying the deformed submanifold with $ \Lambda $), the second one
identifies $ M $ with the moduli set of those submanifolds in $ U\subset{\mathfrak T} $ which
project $ 1 $-to-1 to $ \Lambda $. The fact that the set $ M $ can be equipped with a
structure of a manifold is the most nontrivial part of the statement. If
$ {\mathfrak T} $ is in fact a total space of a vector bundle $ {\mathcal E} $ over $ \Lambda $, then this
statement is trivial, with $ M=\Gamma\left({\mathbb P}^{1},{\mathcal E}\right) $.

Additionally, the existence of the projection on $ \Lambda $ (thus of
retraction on $ S_{} $) removes all the bulkiness from the statement on a
deformation of an arbitrary submanifold, since one does not need to
consider the deformation of the the complex structure on $ S $. \end{remark}

\begin{remark} One should interpret the last statement of the theorem as the
fact that any infinitesimal deformation is a infinitesimal family of {\em an
actual\/} $ 1 $-{\em parameter deformation\/} of $ S $. Compare this with Definition
~\ref{def75.50}. \end{remark}

In our discussion we are most interested in the case $ \dim  {\mathfrak T}=2 $, $ \Lambda={\mathbb P}^{1} $.
Then $ {\mathcal N}S $ is a line bundle, thus is isomorphic to $ {\mathcal O}\left(d-1\right) $ with $ d\geq0 $, and $ \dim 
M=d $. In fact we need a particular case $ d=3 $, but for some time we are
going to discuss the general case of arbitrary $ d $, $ {\mathfrak T} $ and $ \Lambda $.

\begin{definition} Say that a mapping $ \pi\colon {\mathfrak T} \to \Lambda $ of complex manifolds is a {\em disk
bundle\/} if $ {\mathfrak T} $ is a manifold with $ C^{0} $-boundary, $ \dim  {\mathfrak T}=\dim  S+1 $, for any $ \lambda\in\Lambda $
there is a neighborhood $ U\ni\lambda $ such that $ \pi|_{\pi^{-1}U} $ is homeomorphic to the
projection $ p_{1}\colon U\times D \to U $; here $ D $ is $ \left\{z\in{\mathbb C} \mid |z|\leq1\right\}. $\end{definition}

\begin{proposition} \label{prop9.60}\myLabel{prop9.60}\relax  In the conditions of Theorem~\ref{th9.50} assume that $ \dim 
{\mathfrak T}=2 $, $ \Lambda={\mathbb P}^{1} $, and that $ \pi $ is a disk bundle. Consider two curves $ \gamma_{1,2}\subset{\mathfrak T} $ such
that restrictions $ \pi|_{\gamma_{1}} $ and $ \pi|_{\gamma_{2}} $ are bijections. Suppose that
$ d=\deg \left({\mathcal N}\gamma_{1}\right)+1 $, and $ \gamma_{1} $ intersects $ \gamma_{2} $ in $ \geq d $ points. Then $ \gamma_{1}=\gamma_{2} $. \end{proposition}

\begin{proof} Suppose $ \gamma_{1}\not=\gamma_{2} $. Let $ X_{1},\dots ,X_{k} $ be the points of intersection of
$ \gamma_{1} $ and $ \gamma_{2} $. Let $ \bar{{\mathfrak T}} $ be blow-up of $ {\mathfrak T} $ at these points (make repeated blow-ups
if needed to remove all the points of intersection). Removing proper
preimages of $ \pi^{-1}\pi\left(X_{i}\right) $, $ i=1,\dots ,k $, from $ \bar{{\mathfrak T}} $, we obtain a manifold $ \widetilde{{\mathfrak T}} $ with a
mapping $ \widetilde{\pi} $ to $ {\mathbb P}^{1} $ such that preimages of points of $ {\mathbb P}^{1} $ are disks, with the
exception of the points $ \pi\left(X_{i}\right) $, preimages of which are isomorphic to
$ {\mathbb P}^{1}\smallsetminus\left\{\bullet\right\}\simeq{\mathbb C} $. Cutting out far-away points of $ {\mathbb C} $ together with an appropriate
neighborhood on $ \widetilde{{\mathfrak T}} $, one may ensure that the resulting manifold is a disk
bundle over $ {\mathbb P}^{1} $.

Each blow-up decreases the degree of the normal bundle by 1, thus we
reduced the statement to the case $ d<0 $, and $ \gamma_{1}\cap\gamma_{2}=\varnothing $. Show that this leads
to contradiction.

Indeed, topological bundles with the fibers being oriented disks are
isomorphic iff their boundaries are isomorphic as bundles with a fiber
being oriented circles. In turn, any such bundle is isomorphic to a
spherical bundle of a line bundle over $ {\mathbb P}^{1} $, which is determined by its
degree up to an isomorphism. We conclude that the topological bundle $ {\mathfrak T} \to
{\mathbb P}^{1} $ is isomorphic to a neighborhood of $ 0 $-section in the total space of
$ {\mathcal O}\left(-n\right) $, $ n>0 $. However, $ {\mathcal O}\left(-n\right) $ has no continuous nowhere-0 sections: indeed,
such a section would give a trivialization of the spherical bundle of
$ {\mathcal O}\left(-n\right) $, thus, due to arguments given above, to an isomorphism of $ {\mathcal O}\left(-n\right) $
with $ {\mathcal O}\left(0\right) $. \end{proof}

\begin{remark} The condition of being a disk bundle is very essential. For
example, suppose that $ {\mathfrak T} $ is an open subset of $ {\mathbb P}^{1}\times{\mathbb P}^{1} $ with $ \pi $ being the
projection on the first $ {\mathbb P}^{1} $. It is easy to find such an $ {\mathfrak T} $ which contains
both the ``constant'' section $ x \mapsto 0 $ of $ \pi $, and the $ \operatorname{id} $-section $ x \mapsto x $.
Moreover, for most points of $ {\mathbb P}^{1} $ the preimage in $ {\mathfrak T} $ can be made a disk.
Thus a topological argument is required indeed. \end{remark}

The next step is to provide a way to find the subset $ U $ of Theorem
~\ref{th9.50} if all we new is the family $ \sigma $.

\begin{proposition} \label{prop9.70}\myLabel{prop9.70}\relax  In the conditions of Theorem~\ref{th9.50} assume that $ \dim 
{\mathfrak T}=2 $, $ \Lambda={\mathbb P}^{1} $, and that $ \pi $ is a disk bundle. Let $ \deg \left({\mathcal N}S\right)=d-1 $, $ \left\{\lambda_{1},\dots ,\lambda_{d}\right\}\subset{\mathbb P}^{1} $
be a set of $ d $ distinct points. Let $ B_{k}=\pi^{-1}\lambda_{k} $, $ U_{k} $ be an open subset of $ B_{k} $,
$ k=1,\dots ,d $. Let $ \widetilde{\sigma} $ be a mapping $ {\mathbb P}^{1}\times M \to {\mathfrak T} $ such that $ \pi\circ\widetilde{\sigma} $ is the projection
$ {\mathbb P}^{1}\times M \to {\mathbb P}^{1} $. Suppose that for any collection $ X=\left\{X_{k}\right\}_{k=1}^{d}\subset{\mathfrak T} $ such that $ X_{k}\in U_{k} $
there is $ m_{X}\in M $ such that $ \operatorname{Im}\left(\widetilde{\sigma}_{m_{X}}\right)\cap B_{k}=X_{k} $, $ k=1,\dots ,d $.

Let $ U={\mathfrak T}\smallsetminus\left(\bigcup_{k}\left(B_{k}\smallsetminus U_{k}\right)\right) $ (in other words, narrow fibers $ B_{k} $ over $ X_{k} $ to
become $ U_{k} $). Then for any curve $ \gamma\subset U $ which projects isomorphically to $ {\mathbb P}^{1} $
there is $ m\in M $ such that $ \gamma=\operatorname{Im}\widetilde{\sigma}_{m} $. \end{proposition}

\begin{proof} Take $ X_{k}=\gamma\cap B_{k} $, and apply Proposition~\ref{prop9.60} to $ \gamma $ and $ \operatorname{Im}\left(\widetilde{\sigma}_{m_{X}}\right) $. \end{proof}

\begin{remark} This proposition provides a way to check that a given family $ \widetilde{\sigma} $
and $ U\subset{\mathfrak T} $ may work as the family $ \sigma $ from Theorem~\ref{th9.50}. Note that given $ \widetilde{\sigma} $
which satisfies the last condition of Theorem~\ref{th9.50}, it is always
possible to find the subsets $ U_{k} $ with the required properties. Indeed, if
$ {\mathfrak T} $ is an open subset of the total space of $ {\mathcal O}\left(d-1\right) $, then this follows from
the fact that a section of $ {\mathcal O}\left(d-1\right) $ is uniquely determined by values in $ d $
different points (compare with Legendre interpolation formula, or
Vandermond determinant). In general one needs to apply the implicit
function theorem to the mapping $ m \mapsto \Sigma_{m}=\operatorname{Im}\widetilde{\sigma}|_{\Lambda\times\left\{m\right\}} $.

Moreover, for the manifolds we are going to consider here (twistor
transforms of Veronese webs) we can provide an explicit description of
the family $ \sigma $ and of subsets $ U_{k} $. \end{remark}

\begin{corollary} In the conditions of Theorem~\ref{th8.20}, consider the twistor
transform $ {\mathfrak T} \xrightarrow[]{\pi} {\mathbb P}^{1} $ of the $ {\mathcal F}_{\bullet} $-convex subset $ \left({\mathbb B}_{\rho}^{1}\right)^{3} $. Let $ \Sigma_{m} $ be the section
of $ {\mathfrak T} $ corresponding to $ m\in\left({\mathbb B}_{\rho}^{1}\right)^{3} $. Then the mapping $ \widetilde{\sigma}\left(m,\lambda\right)\buildrel{\text{def}}\over{=}\Sigma_{m}\left(\lambda\right) $,
$ m\in\left({\mathbb B}_{\rho}^{1}\right)^{3} $, $ \lambda\in{\mathbb P}^{1} $, satisfies the conditions of Proposition~\ref{prop9.70} with
$ U_{k}=B_{k} $. \end{corollary}

\begin{proof} Let $ \lambda_{1,2,3} $ be the values of $ \lambda $ which correspond to exceptional
foliations $ \left\{x=\operatorname{const}\right\} $, $ \left\{y=\operatorname{const}\right\} $, $ \left\{z=\operatorname{const}\right\} $ of the web. Then $ B_{k} $, $ k=1,2,3 $,
are naturally identified with $ {\mathbb B}_{\rho}^{1} $. A choice of $ X_{k}\in U_{k}=B_{k} $, $ k=1,2,3 $,
corresponds to a choice of 3 leaves of these 3 foliations, or, in other
words, to a choice of coordinates $ x $, $ y $, $ z $ such that $ |x|,|y|,|z|<\rho $. Put
$ m=\left(x,y,z\right)\in\left({\mathbb B}_{\rho}^{1}\right)^{3} $, then $ \Sigma_{m} $ passes through $ X_{k} $, $ k=1,2,3 $. \end{proof}

\section{Airy webs }\label{h85}\myLabel{h85}\relax 

Consider a smooth web $ \left\{{\mathcal F}_{\lambda}\right\}_{\lambda\in\Lambda} $ on $ M $. Suppose that the twistor
transform $ {\mathfrak T} \xrightarrow[]{\pi} \Lambda $ of $ {\mathcal F}_{\bullet} $ is well-defined as a manifold.

\begin{definition} Say that a smooth web $ \left\{{\mathcal F}_{\lambda}\right\}_{\lambda\in\Lambda} $ is {\em strictly airy\/} if for any
smooth section $ \Sigma $ of $ {\mathfrak T} \xrightarrow[]{\pi} \Lambda $ there is a point $ m\in M $ such that $ \Sigma=\Sigma_{m} $. A web is
{\em airy\/} if any point has a neighborhood $ U $ such that $ {\mathcal F}_{\bullet}|_{U} $ is strictly airy. \end{definition}

\begin{remark} This definition requires some modifications if only the germ of
$ {\mathfrak T}={\mathfrak T}_{{\mathcal F}_{\bullet}} $ near $ \Sigma_{m_{0}}\subset{\mathfrak T} $ is well-defined; here $ m_{0}\in M $. In such a case consider a
family $ \sigma_{\bullet}\colon \Lambda\times T \to {\mathfrak T} $ of sections of $ \pi $ parameterized by (a germ of) a
manifold $ T $, and such that $ \sigma_{t_{0}}=\Sigma_{m_{0}} $ for the base point $ t_{0}\in T $. We would
require that there is a family $ p_{\bullet}\colon T \to M $ of points of $ M $ such that $ \sigma_{t}=\Sigma_{p_{t}} $
for $ t\in T $ near $ t_{0} $. \end{remark}

\begin{remark} It should be clear that airy webs exist only in
complex-analytic situation, otherwise the set of sections is not a
finite-dimensional manifold. Moreover, it is reasonable to conjecture
that $ \Lambda $ cannot be a Stein manifold if $ \dim \Lambda>0 $. \end{remark}

The principal property of airy webs is the following immediate
corollary of Corollary~\ref{cor75.70}:

\begin{theorem} Consider a weakly separating and separating strictly airy
web. Locally such a web is uniquely determined (up to a local
diffeomorphism) by its twistor transform $ {\mathfrak T} \xrightarrow[]{\pi} \Lambda $. \end{theorem}

\begin{proposition} \label{prop85.50}\myLabel{prop85.50}\relax  In the conditions of Theorem~\ref{th9.50} suppose that
global sections of the vector bundle $ {\mathcal N}S $ span any fiber of $ {\mathcal N}S $. Then a
neighborhood $ M_{1} $ of $ m_{0} $ in $ M $ is equipped with a web $ \left\{{\mathcal F}_{\lambda}\right\}_{\lambda\in\Lambda} $ of codimension
equal to $ \operatorname{codim}\Sigma $. This web is separating and airy. \end{proposition}

\begin{proof} Deduce the first statement from Lemma~\ref{lm75.30}. It is enough to
calculate $ \operatorname{rk} d{\mathbit s}_{\lambda}|_{m} $. The condition on global sections is equivalent to $ \operatorname{rk}
d{\mathbit s}_{\lambda}|_{m_{0}}=\operatorname{codim}\Sigma $, thus all we need to show is that this rank does not change
if we move to a nearby section $ \Sigma_{m} $ of $ \pi $.

Consider $ {\mathcal N}S $ as a sheaf of $ {\mathcal O}_{S} $-modules. For $ P\in S $ denote by $ {\mathcal N}S\left(-P\right) $ the
sheaf of $ {\mathcal O}_{S} $-modules with local sections being sections of $ {\mathcal N}S $ which vanish
at $ P $. By the Grauert semicontinuity theorem \cite{Gra60Theo}, the Euler
characteristic $ \sum\left(-1\right)^{k}\dim  H^{k}\left(S,{\mathcal N}S\left(-P\right)\right) $ of $ {\mathcal N}S\left(-P\right) $ does not change when $ P $
changes, and the individual terms $ \dim  H^{k}\left(S,{\mathcal N}S\left(-P\right)\right) $ are semicontinuous
from above. Similar results hold for $ \dim  H^{k}\left(\Sigma_{m},{\mathcal N}\Sigma_{m}\left(-P\right)\right) $ considered as
functions of $ m\in M $ and $ P\in\Sigma_{m} $.

Consider the exact sequence of sheaves $ 0 \to {\mathcal N}S\left(-P\right) \to {\mathcal N}S \xrightarrow[]{v_{P}} \overline{{\mathcal N}_{P}S}
\to $ 0; here $ \overline{{\mathcal N}_{P}S} $ is the skyscraper sheaf with the fiber over $ P $ being
$ {\mathcal N}_{P}S $. Since the mapping $ v $ of taking the value at $ P $ is surjective on global
sections, the cohomological long exact sequence shows that $ H^{k}\left(S,{\mathcal N}S\left(-P\right)\right)=0 $
for $ k\geq1 $ and $ P\in S $. This implies $ H^{k}\left(\Sigma_{m},{\mathcal N}\Sigma_{m}\left(-P\right)\right)=0 $ for $ k>1 $ if $ m\approx m_{0} $, thus $ \dim 
H^{0}\left(\Sigma_{m},{\mathcal N}\Sigma_{m}\left(-P\right)\right) $ does not depend on $ m\approx m_{0} $ and $ P\in\Sigma_{m} $. Now a consideration of
the long exact sequence for $ 0 \to {\mathcal N}\Sigma_{m}\left(-P\right) \to {\mathcal N}\Sigma_{m} \xrightarrow[]{v_{P}} \overline{{\mathcal N}_{P}\Sigma_{m}} \to 0 $ shows
that $ v_{P} $ is surjective for $ m\approx m_{0} $ and $ P\in\Sigma_{m} $. This implies that $ d{\mathbit s}_{\lambda}|_{m} $ is a
surjection.

By Lemma~\ref{lm75.30}, a neighborhood of $ \Sigma_{m_{0}}\subset{\mathfrak T} $ is the twistor transform
of a web on an open subset $ M_{1}\subset M $, $ M_{1}\ni m_{0} $. Since $ d\sigma|_{m_{0}} $ (and, by similar
arguments, $ d\sigma|_{m} $ for any $ m\in M_{1} $) is an injection, this web is separating.

Prove airiness. One can find a neighborhood $ U $ of $ S $ in $ {\mathfrak T} $ such that
$ \Sigma_{m}\subset U $ implies $ m\in M_{1} $. Indeed, let $ U_{1}=\bigcup_{m\in M_{1}}\Sigma_{m} $. It is a neighborhood of $ S $,
thus one can apply Theorem~\ref{th9.50} to $ U_{1} $ instead of $ {\mathfrak T} $. Obviously, the
resulting neighborhood $ U\subset U_{1} $ of $ S $ satisfies the requirement above. Let
$ M_{2}=\left\{m\in M_{1} \mid \Sigma_{m}\subset U\right\} $. Now any section $ \Sigma $ of $ U \to \Lambda $ has a form $ \Sigma=\Sigma_{m} $ for $ m\in M_{1} $.
Obviously, this implies also $ m\in M_{2} $. On the other hand, a section of a
twistor transform of $ M_{2} $ induces a section of $ U \to \Lambda $, thus the restriction
of the web on $ M_{2} $ is airy. \end{proof}

By Definition~\ref{def3.05}, given a smooth web $ \left\{{\mathcal F}_{\lambda}\right\}_{\lambda\in\Lambda} $ on $ M $, each point
$ m\in M $ induces a vector bundle $ {\mathbit n}_{m} $ over $ \Lambda $, the fiber over $ \lambda $ being $ {\mathbit n}_{m}\left(\lambda\right) $.
Obviously,

\begin{lemma} Consider the section $ \Sigma_{m} $ of the twistor transform $ {\mathfrak T} \xrightarrow[]{\pi} \Lambda $ of $ {\mathcal F}_{\bullet} $.
Then $ {\mathcal N}\Sigma_{m}\simeq\pi^{*}{\mathbit n}_{m}^{*} $. \end{lemma}

\begin{proposition} \label{prop85.70}\myLabel{prop85.70}\relax  Consider a complex-analytic separating web $ \left\{{\mathcal F}_{\lambda}\right\}_{\lambda\in\Lambda} $
on $ M $ with compact $ \Lambda $. Suppose that $ {\mathbit n}_{m} $ is cohomologically trivial for any
$ m\in M $. Then there is a manifold $ M'\supset M $ with an airy separating web $ \left\{{\mathcal F}'_{\lambda}\right\}_{\lambda\in\Lambda} $
on it such that for any $ \lambda\in\Lambda $ and any leaf $ L $ of $ {\mathcal F}_{\lambda} $ there is a leaf $ L' $ of
$ {\mathcal F}'_{\lambda} $ such that $ L=M\cap L' $. The germ of $ M' $ near $ M $ is canonically defined. \end{proposition}

\begin{proof} Due to canonicity of $ M' $ it is enough to prove this statement
locally on $ M $. Thus we may assume that $ {\mathcal F}_{\bullet} $ is weakly separating and
separating. Consider the twistor transform of $ {\mathcal F}_{\bullet} $. Since $ {\mathbit n}_{m} $ may be
identified with $ {\mathcal N}\Sigma_{m} $, Proposition~\ref{prop85.50} is applicable. (The condition
on global sections is automatically satisfied if $ {\mathfrak T} $ is a twistor
transform.) This provides a construction of $ M' $ and $ {\mathcal F}'_{\bullet} $. \end{proof}

\begin{remark} This explains the choice of the term {\em airy\/}: it reasonable to
imagine that $ M' $ is obtained from $ M $ by ``blowing out'' $ M $. Here leaves of
foliations $ {\mathcal F}_{\lambda} $ work as ``walls of microscopic air cells'' in $ M $. If $ M' $ is a
Veronese web, and we consider just enough foliations $ {\mathcal F}_{\lambda_{k}} $, $ k=1,\dots ,K $, to
uniquely determine $ {\mathcal F}_{\bullet} $ (so $ K=\dim  M'+1 $) then after blow-out each cell
becomes a tiny simplex in $ M' $. Before ``expansion'' each cell is folded into
a polytop of smaller dimension. \end{remark}

Proposition~\ref{prop85.70} immediately implies

\begin{theorem} Consider a complex-analytic separating web $ \left\{{\mathcal F}_{\lambda}\right\}_{\lambda\in\Lambda} $ on a
connected manifold $ M $ such that $ {\mathbit n}_{m}\left(\lambda\right) $ is a cohomologically trivial vector
bundle over $ \Lambda $. Then $ {\mathcal F}_{\bullet} $ is airy iff $ \dim  M=\dim \Gamma\left(\Lambda,{\mathbit n}_{m}\right) $ for one (then any)
$ m\in M $. \end{theorem}

\begin{remark} Note that if $ \Lambda={\mathbb P}^{1} $, then $ {\mathbit n}_{m}\left(\lambda\right) $ is automatically cohomologically
trivial (since by definition this vector bundle is induced from
a Grassmannian). \end{remark}

\begin{remark} In Section~\ref{h105} we provide a somewhat inverse construction to
Proposition~\ref{prop85.70}: given $ M' $, we introduce a class of submanifolds
$ M\subset M' $ which are equipped with a web having the same twistor transform. \end{remark}

The arguments above give a more detailed proof of one of the
principal results of \cite{GelZakhWeb}:

\begin{theorem} \label{th77.05}\myLabel{th77.05}\relax  Complex-analytic Veronese webs are airy and are
(uniquely up to a local diffeomorphism) locally determined by their
twistor transform. \end{theorem}

This a direct corollary of

\begin{lemma} \label{lm77.10}\myLabel{lm77.10}\relax  Consider a Veronese web $ \left\{{\mathcal F}_{\lambda}\right\}_{\lambda\in{\mathbb P}^{1}} $ on a $ d $-dimensional
manifold $ M $. Then $ {\mathbit n}_{m} \simeq{\mathcal O}\left(d-1\right) $ for any $ m\in M $. \end{lemma}

\begin{proof} It is enough to show that $ {\mathcal L}\simeq{\mathcal O}\left(-d+1\right) $; here the line bundle $ {\mathcal L} $ is
induced by the Veronese inclusion $ j\colon {\mathbb P}^{1} \to {\mathbb P}^{d-1} $ from the tautological
line bundle on $ {\mathbb P}^{d-1} $ (which is isomorphic to $ {\mathcal O}\left(-1\right) $). The fiber of $ {\mathcal L} $ over
$ \lambda\in{\mathbb P}^{1} $ is the $ 1 $-dimensional subspace of $ {\mathbb V}^{d} $ corresponding to $ j\left(\lambda\right) $.

A linear function $ l $ on $ {\mathbb V}^{d} $ induces a section of $ {\mathcal L}^{*} $, zeros of this
section correspond to points on $ \operatorname{Ker} l \cap \operatorname{Im} j $. Thus it is enough to
construct a hyperplane in $ {\mathbb P}^{d-1} $ which transversally intersects $ \operatorname{Im} j $ in $ d-1 $
points. Since all the Veronese inclusions are projectively isomorphic, it
is enough to consider one given by $ \left(x:y\right) \mapsto \left(x^{d-1}:x^{d-2}y:\dots :xy^{d-2}:y^{d-1}\right) $.
Let $ \Pi_{k=1}^{d-1}\left(t-k\right)=t^{d-1}+\sum_{k=0}^{d-2}a_{k}t^{k} $. Then the functional $ l $ with coordinates
$ \left(1,a_{d-2},\dots ,a_{0}\right) $ satisfies the condition above. \end{proof}

In fact Veronese webs coincide (in complex-analytic situation) with
separating airy smooth webs of codimension 1 with $ \Lambda={\mathbb P}^{1} $. One can also
classify arbitrary separating airy smooth webs with $ \Lambda={\mathbb P}^{1} $, the result
coincides with {\em Kronecker webs\/} as defined in \cite{Zakh99Kro}.

\section{Non-linear Riemann problem }\label{h11}\myLabel{h11}\relax 

Consider $ 0<\varepsilon<1 $, $ \delta>0 $, and a complex-analytic function $ g\left(\lambda,t\right) $ defined
for $ \varepsilon<|\lambda|<1/\varepsilon $ and $ |t|<\delta $. Assume that for any given $ \lambda $, $ \varepsilon<|\lambda|<1/\varepsilon $, the
function $ g\left(\lambda,t\right) $ is invertible. Glue domains $ \left({\mathbb P}^{1}\smallsetminus\bar{{\mathbb B}}_{\varepsilon}^{1}\right)\times{\mathbb C} $ and $ {\mathbb B}_{1/\varepsilon}^{1}\times{\mathbb B}_{\delta}^{1} $
together by gluing $ \left(\lambda,t_{+}\right)\in{\mathbb B}_{1/\varepsilon}^{1}\times{\mathbb B}_{\delta}^{1} $ to $ \left(\lambda,t_{-}\right)=\left(\lambda,g\left(\lambda,t_{+}\right)\right)\in\left({\mathbb P}^{1}\smallsetminus\bar{{\mathbb B}}_{\varepsilon}^{1}\right)\times{\mathbb C} $ for
$ \varepsilon<|\lambda|<1/\varepsilon $ and $ |t|<\delta $. The result is a $ 2 $-dimensional complex manifold $ {\mathfrak T} $ with
a surjective submersive mapping $ \pi\colon {\mathfrak T} \to {\mathbb P}^{1} $ given by $ \left(\lambda,t\right) \mapsto \lambda $.

\begin{lemma} \label{lm16.10}\myLabel{lm16.10}\relax 
\begin{enumerate}
\item
If $ g\left(\lambda,0\right)\equiv 0 $, then $ \pi $ has a section given by $ \lambda \mapsto \left(\lambda,0\right) $;
\item
Sections $ \Sigma $ of $ \pi $ can be identified with pairs of functions $ \sigma_{+}:
{\mathbb B}_{1/\varepsilon}^{1} \to {\mathbb B}_{\delta}^{1} $ and $ \sigma_{-}\colon \left({\mathbb P}^{1}\smallsetminus\bar{{\mathbb B}}_{\varepsilon}^{1}\right) \to {\mathbb C} $ such that $ \sigma_{-}\left(\lambda\right)=g\left(\lambda,\sigma_{+}\left(\lambda\right)\right) $ if
$ \varepsilon<|\lambda|<1/\varepsilon $;
\item
Given a section $ \Sigma $ of $ \pi $ associated to a pair $ \sigma_{\pm} $, the degree of the
normal bundle of $ \Sigma $ in $ {\mathfrak T} $ is given by $ -\operatorname{ind} \frac{\partial g}{\partial t}\left(\lambda,\sigma_{+}\left(\lambda\right)\right) $. Here $ \operatorname{ind}
\varphi\left(\lambda\right)=\frac{1}{2\pi i}\oint_{|\lambda|=1}\frac{d\varphi\left(\lambda\right)}{\varphi\left(\lambda\right)} $.
\end{enumerate}
\end{lemma}

\begin{proof} The first two statements are obvious. On the other hand, the normal
bundle of $ \Sigma $ is canonically trivialized on $ |\lambda|<1/\varepsilon $ and on $ |\lambda|>\varepsilon $, with the
gluing function being $ \frac{\partial g}{\partial t}\left(\lambda,\sigma_{+}\left(\lambda\right)\right) $. To calculate the degree of a line
bundle $ {\mathcal L} $ it is enough to construct a section $ \tau_{+} $ in $ |\lambda|\leq1 $ and a section $ \tau_{-} $
in $ |\lambda|\geq1 $. Suppose that $ \tau_{\pm}\left(\lambda\right) $ have no zeros on $ |\lambda|=1 $. Then $ \tau_{0}=\tau_{-}\left(\lambda\right)/\tau_{+}\left(\lambda\right) $
is a well-defined function on the unit circle with values in $ {\mathbb C}^{\times} $, and
$ \deg {\mathcal L}=n_{+}+n_{-}-\operatorname{ind}\tau_{0} $; here $ n_{\pm} $ are numbers of zeros of $ \tau_{\pm} $ (inside and
outside of the unit circle correspondingly). In our case $ n_{+}=n_{-}=0 $, and
$ \tau_{0}=\frac{\partial g}{\partial t}\left(\lambda,\sigma_{+}\left(\lambda\right)\right) $. \end{proof}

\begin{theorem} \label{th11.30}\myLabel{th11.30}\relax  Consider a manifold $ K $ and a function $ g\left(\lambda,t,\kappa\right) $, $ \kappa\in K $, which
depends analytically on parameters and such that for any given $ \kappa $ the
function satisfies the condition in the beginning of this section.
Suppose that for $ \kappa_{0}\in K $ one has $ g\left(\lambda,0,\kappa_{0}\right)\equiv 0 $, and suppose that $ \operatorname{ind}
\frac{\partial g}{\partial t}\left(\lambda,0,\kappa_{0}\right)=1 $. Then there exists $ 0<\delta_{1}<\delta $ and a neighborhood $ K_{1}\ni\kappa_{0} $,
$ K_{1}\subset K $, such that for any $ \kappa\in K_{1} $ the conditions
\begin{equation}
\sigma_{-,\kappa}\left(\lambda\right)=g\left(\lambda,\sigma_{+,\kappa}\left(\lambda\right),\kappa\right)\text{ for }\varepsilon<|\lambda|<1/\varepsilon,\qquad |\sigma_{+,\kappa}\left(\lambda\right)|<\delta_{1}\text{ for }|\lambda|<1/\varepsilon,
\notag\end{equation}
uniquely determine analytic functions $ \sigma_{+,\kappa}\left(\lambda\right) $ defined for $ |\lambda|<1/\varepsilon $, and
$ \sigma_{-,\kappa}\left(\lambda\right) $ defined for $ |\lambda|>\varepsilon $. Functions $ \sigma_{\pm,\kappa}\left(\lambda\right) $ depend analytically on $ \kappa $. \end{theorem}

\begin{proof} Glue domains $ {\mathbb B}_{1/\varepsilon}^{1}\times{\mathbb B}_{\delta}\times K $ and $ \left({\mathbb P}^{1}\smallsetminus\bar{{\mathbb B}}_{\varepsilon}^{1}\right)\times{\mathbb C}\times K $ together by gluing
$ \left(\lambda,t,\kappa\right) $ to $ \left(\lambda,g\left(\lambda,t,\kappa\right),\kappa\right) $ for $ \varepsilon<|\lambda|<1/\varepsilon $, $ |t|<\delta $, and $ \kappa\in K $. Denote the
resulting manifold by $ {\mathfrak T} $, denote by $ \pi\colon {\mathfrak T} \to {\mathbb P}^{1} $ the mapping $ \left(\lambda,t,\kappa\right) \mapsto \lambda $,
by $ \Pi $ the natural projection $ {\mathfrak T} \to K $, and by $ \Sigma $ the section of $ \pi $ given by $ \lambda
\mapsto \left(\lambda,0,\kappa_{0}\right) $. Consider the normal bundle $ {\mathcal N}\Sigma $ of $ \Sigma $ inside $ {\mathfrak T} $. Let $ {\mathcal N}^{\left(0\right)}\Sigma $ be
the normal bundle of $ \Sigma $ inside $ \Pi^{-1}\left(\kappa_{0}\right) $. We know that $ \deg  {\mathcal N}^{\left(0\right)}\Sigma=-1 $. On the
other hand, $ {\mathcal N}\Sigma/{\mathcal N}^{\left(0\right)}\Sigma $ is isomorphic to $ \Pi^{*}{\mathcal T}_{\kappa_{0}}K $, thus is a trivial vector
bundle over $ \Sigma $. Thus both $ {\mathcal N}^{\left(0\right)}\Sigma $ and $ {\mathcal N}\Sigma/{\mathcal N}^{\left(0\right)}\Sigma $ are cohomologically trivial.

The exact sequence $ \dots \to H^{k}\left(\Sigma,{\mathcal N}^{\left(0\right)}\Sigma\right) \to H^{k}\left(\Sigma,{\mathcal N}\Sigma\right) \to H^{k}\left(\Sigma,{\mathcal N}\Sigma/{\mathcal N}^{\left(0\right)}\Sigma\right)
\to \dots $ shows that $ {\mathcal N}\Sigma $ is also cohomologically trivial, and $ \Gamma\left(\Sigma,{\mathcal N}\Sigma\right)\simeq{\mathcal T}_{\kappa_{0}}K $.
In other words, the Kodaira--Spencer theory (Theorem~\ref{th9.50}) is
applicable, and there is an mapping $ \Sigma_{\bullet}\colon {\mathbb P}^{1}\times M \to {\mathfrak T} $ and $ m_{0}\in M $, such that
$ \operatorname{Im}\Sigma_{m_{0}}=\Sigma $, and the associated infinitesimal family $ \delta\Sigma_{m}\colon {\mathcal T}_{m_{0}}M \to \Gamma\left(\Sigma,{\mathcal N}\Sigma\right) $ is
a bijection.

On the other hand, $ \Pi\circ\Sigma_{m}\colon {\mathbb P}^{1} \to K $ is a deformation of a constant
mapping to a point $ \kappa_{0}\in K $, thus is a constant mapping itself. Denote the
image-point of this constant mapping by $ \kappa\left(m\right) $. It is clear that the
derivative of $ m \mapsto \kappa\left(m\right) $ coincides with the composition $ {\mathcal T}_{m_{0}}M \to \Gamma\left(\Sigma,{\mathcal N}\Sigma\right) \to
H^{k}\left(\Sigma,{\mathcal N}\Sigma/{\mathcal N}^{\left(0\right)}\Sigma\right) \simeq{\mathcal T}_{\kappa_{0}}K $, thus $ \kappa\left(m\right) $ is a local diffeomorphism. Thus we can
identify $ M $ with an open subset of $ K $. We obtain a family of mappings $ \Sigma_{\kappa}:
{\mathbb P}^{1} \to {\mathfrak T} $, $ \kappa\in M\subset K $, such that $ \Pi\circ\Sigma_{\kappa} $ is the constant mapping to $ \kappa\in K $. In other
words, $ \Sigma_{\kappa} $ is a section of $ \pi|_{\Pi^{-1}\kappa} $, thus induces a pair of functions
$ \sigma_{\pm,\kappa}\left(\lambda\right) $.

This shows existence of solutions $ \sigma_{\pm,\kappa} $, as well as the analytic
dependence on parameters. Uniqueness follows from the other parts of
Kodaira--Spencer theory (Theorem~\ref{th9.50}). \end{proof}

Using Definition~\ref{def11.40}, one can restate Theorem~\ref{th11.30} in the
following way:

\begin{corollary} \label{cor11.50}\myLabel{cor11.50}\relax  Consider a function $ g\left(\lambda,t\right) $ defined for $ \varepsilon<|\lambda|<1/\varepsilon $ and
$ |t|<\delta $, such that $ g\left(\lambda,0\right)\equiv 0 $ and $ \operatorname{ind}\frac{\partial g}{\partial t}\left(\lambda,0\right)=1 $. (Obviously, $ {\mathfrak R}\left(g\right)=0 $.)
Consider an analytic family $ g_{\kappa}\left(\lambda,t\right) $, $ \varepsilon<|\lambda|<1/\varepsilon $, $ |t|<\delta $, $ \kappa\in U\subset{\mathbb C}^{n} $, such that
$ g_{0}=g $. Then there is a neighborhood $ U_{1} $ of 0 in $ U $ such that $ {\mathfrak R}\left(g_{\kappa}\right) $ is
defined for $ \kappa\in U_{1} $ and $ {\mathfrak R}\left(g_{\kappa}\right) $ depends smoothly on $ \kappa\in U_{1} $. \end{corollary}

\begin{remark} Since $ {\mathfrak R}\left(g\right) $ does not change when $ \varepsilon $ increases, it is clear that
$ \sigma_{+}\left(\mu\right) $ for $ |\mu|<1/\varepsilon $ can be written in terms of $ {\mathfrak R} $. For example, if $ |\mu|<1 $,
then $ \sigma_{+}\left(\mu\right)={\mathfrak R}\left(g\left(\frac{\lambda-\mu}{\mu\lambda-1},t\right)\right) $.

Similarly, one can calculate $ \sigma_{-}\left(\mu\right) $ by considering the inverse function
for $ g\left(\lambda,t\right) $ in $ t $ (with $ \lambda $ being a parameter) instead of $ g\left(\lambda,t\right) $. \end{remark}

Consider now what changes if one takes the gluing functions $ g\left(\lambda,t\right) $
with $ g\left(\lambda,0\right)\equiv 0 $ and non-positive values of $ \operatorname{ind} \frac{\partial g}{\partial t}\left(\lambda,0\right) $ (as opposed to
$ \operatorname{ind}=1 $). In such a case $ \deg {\mathcal N}\Sigma=d $ is non-negative, thus there is a
$ \left(d+1\right) $-parametric family of sections of $ \pi $. By Proposition~\ref{prop9.70} we
expect that a section is determined by its values at $ d+1 $ different points
of $ {\mathbb P}^{1} $.

Let us write the formula for the section in terms of $ g $ and $ {\mathfrak R} $. Use
notations of Definition~\ref{def0.60}.

\begin{proposition} \label{prop11.60}\myLabel{prop11.60}\relax  Suppose that $ k $ numbers $ \lambda_{1},\dots ,\lambda_{k} $ satisfy $ 0<|\lambda_{l}|<1 $,
$ m $ numbers $ \mu_{1},\dots ,\mu_{m} $satisfy $ |\lambda_{l}|>1 $. Consider a pair of functions
satisfying
\begin{equation}
\sigma_{-}\left(\lambda\right)=g\left(\lambda,\sigma_{+}\left(\lambda\right)\right),\qquad |\sigma_{+}\left(\lambda\right)|<\delta\text{ for }|\lambda|<1/\varepsilon,
\label{equ11.30}\end{equation}\myLabel{equ11.30,}\relax 
and conditions $ \sigma_{+}\left(\lambda_{l}\right)=a_{l} $, $ l=1,\dots ,k $, $ \sigma_{-}\left(\mu_{l}\right)=b_{l} $, $ l=1,\dots ,m $. Suppose that
$ \operatorname{ind}\frac{\partial g}{\partial t}\left(\lambda,0\right)=1-k-m $. Then $ \sigma_{+}\left(0\right)={\mathfrak R}\left({\mathcal G}_{\Lambda{\text M},\left\{a_{i}\right\}\left\{b_{i}\right\}}\right) $. \end{proposition}

\begin{proof} Indeed, one can write
\begin{equation}
\sigma_{+}\left(\lambda\right)=\widetilde{\sigma}_{+}\left(\lambda\right)F_{+}\left(\lambda\right)+\sum_{l=1}^{k}a_{l}F_{+,l}\left(\lambda\right),\qquad \sigma_{-}\left(\lambda\right)=\widetilde{\sigma}_{-}\left(\lambda\right)F_{-}\left(\lambda\right)+\sum_{l=1}^{m}b_{l}F_{-,l}\left(\lambda\right).
\notag\end{equation}
Then $ \sigma_{-}\left(\lambda\right)=g\left(\lambda,\sigma_{+}\left(\lambda\right)\right) $ can be rewritten as $ \widetilde{\sigma}_{-}\left(\lambda\right)={\mathcal G}_{\Lambda{\text M},\left\{a_{i}\right\}\left\{b_{i}\right\}}\left(\lambda,\widetilde{\sigma}_{+}\left(\lambda\right)\right) $,
and $ \sigma_{+}\left(0\right)=\widetilde{\sigma}_{+}\left(0\right) $. The only thing one needs to prove is that
$ \operatorname{ind}\frac{\partial{\mathcal G}_{\Lambda{\text M},\left\{a_{i}\right\}\left\{b_{i}\right\}}}{\partial t}=1 $, which follows from $ \operatorname{ind} F_{+}\left(\lambda\right)=k $, $ \operatorname{ind} F_{-}\left(\lambda\right)=-m $. \end{proof}

\begin{proof}[Proof of Theorem~\ref{th16.70} ] Correctness follows from Proposition
~\ref{prop11.60}. Show that $ w $ is non-degenerate. Suppose that $ \partial w/\partial x=0 $ for some
value of $ \left(x,y,z\right) $. Recall that $ w\left(m\right) $, $ m=\left(x,y,z\right) $, is the value of $ \sigma\left(0,m\right) $;
here $ \sigma\left(\lambda,m\right) $ is a Kodaira--Spencer family of sections of $ {\mathfrak T} $, and $ x $, $ y $, $ z $
are $ \sigma\left(\lambda_{1,2,3},m\right) $. If $ \partial w/\partial x=0 $, this would mean that there is a
one-parametric family of sections such that the infinitesimal family is
non-vanishing, but infinitesimal family vanishes for $ \lambda\in\left\{0,\lambda_{2},\lambda_{3}\right\} $.
However, by Lemma~\ref{lm77.10}, the infinitesimal family is a section of
$ {\mathcal O}\left(2\right) $, thus cannot vanish at 3 distinct points.

Show that $ w $ is $ \left(\lambda_{1},\lambda_{2},\lambda_{3},0\right) $-admissible. Glue domains $ {\mathbb B}_{1/\varepsilon}^{1}\times{\mathbb B}_{\delta} $ and
$ \left({\mathbb P}^{1}\smallsetminus\bar{{\mathbb B}}_{\varepsilon}^{1}\right)\times{\mathbb C} $ together by gluing $ \left(\lambda,t_{+}\right)\in{\mathbb B}_{1/\varepsilon}^{1}\times{\mathbb B}_{\delta} $ to
$ \left(\lambda,t_{-}\right)=\left(\lambda,g\left(\lambda,t_{+}\right)\right)\in\left({\mathbb P}^{1}\smallsetminus\bar{{\mathbb B}}_{\varepsilon}^{1}\right)\times{\mathbb C} $ for $ \varepsilon<|\lambda|<1/\varepsilon $, $ |t_{+}|<\delta $. Call the resulting
$ 2 $-dimensional manifold $ {\mathfrak T} $. It is equipped with a projection $ \pi $ to $ {\mathbb P}^{1} $ and a
section $ S=\left\{\left(\lambda,0\right)\right\} $ of this projection. As in the proof of Lemma~\ref{lm16.10},
one can show that degree of $ {\mathcal N}S $ is 2. Since $ \deg {\mathcal N}S\geq0 $, $ {\mathcal N}S $ is cohomologically
trivial, and fibers of $ {\mathcal N}S $ are generated by global sections. By
Proposition~\ref{prop85.50}, a neighborhood $ U $ of $ S $ in $ {\mathfrak T} $ is a twistor transform
of a web of codimension 1 on a manifold $ M $. Since $ \dim \Gamma\left(S,{\mathcal N}S\right)=3 $, $ \dim  M=3 $.
Again, $ \deg {\mathcal N}S=2 $ implies that $ {\mathbit n}_{m}\left(\lambda\right) $ of this web spans a quadratic cone in
$ {\mathcal T}_{m}^{*}M $, thus this web is a Veronese web.

Taking a point $ u\in U $, $ \pi\left(u\right)=\lambda\in{\mathbb P}^{1} $, gives a leaf of the foliation $ {\mathcal F}_{\lambda} $ on
$ M $. Since fibers of $ U $ over $ \lambda_{1} $, $ \lambda_{2} $, $ \lambda_{3} $ and $ \lambda_{4}=0 $ are identified with subsets
of $ {\mathbb C} $ by the construction of $ {\mathfrak T} $, this gives 4 functions $ x $, $ y $, $ z $, $ W $ on $ M $,
each constant on leaves of $ {\mathcal F}_{\lambda_{1,2,3,4}} $. We may assume that $ W=W\left(x,y,z\right) $ for
an appropriate function $ W $ defined in a neighborhood of $ 0\in{\mathbb C}^{3} $. By
definition, the latter function is $ \left(\lambda_{1},\lambda_{2},\lambda_{3},0\right) $-admissible.

On the other hand, a point $ m\in M $ induces a section of $ \pi $. A section of
$ \pi $ which is close to $ S $ is determined by two functions $ \sigma_{+} $ and $ \sigma_{-} $ which
satisfy~\eqref{equ11.30}. By Proposition~\ref{prop11.60} this section is determined
by $ x=\sigma_{+}\left(\lambda_{1}\right) $, $ y=\sigma_{+}\left(\lambda_{2}\right) $, and $ z=\sigma_{-}\left(\lambda_{3}\right) $, moreover, $ \sigma_{+}\left(0\right)=w\left(x,y,z\right) $. We
conclude that $ w\left(x,y,z\right)=W\left(x,y,z\right) $.

This implies the first statement of the theorem. The second
statement is a direct corollary of the first one.

Given $ \left(A,B,C\right) $, find $ \lambda_{1,2,3,4}\in{\mathbb P}^{1} $ as in Remark~\ref{rem4.95}. By a
projective transform of $ {\mathbb P}^{1} $ one can make $ \lambda_{4}=0 $, and $ \lambda_{3}=\infty $. By
transformations $ \lambda \mapsto c\lambda $ one can make $ \lambda_{1,2} $ arbitrarily small. After this a
transformation $ \lambda \to \frac{\lambda}{1+\lambda/N} $ with $ N\gg 0 $ would produce a triple $ \lambda_{1,2,3} $
with desired properties.

Given a non-degenerate solution $ \widehat{w}\left(x,y,z\right) $ of the $ \left(A,B,C\right) $-equation and
$ \lambda_{1,2,3} $ as found above, consider the $ 3 $-dimensional Veronese web defined by
Theorem~\ref{th4.65}. Let $ \widetilde{{\mathfrak T}} $ is the twistor transform of this web. Then $ \widetilde{\pi}\colon \widetilde{{\mathfrak T}} \to
{\mathbb P}^{1} $ is (locally near $ \Sigma_{\left(0,0,0\right)} $) isomorphic to $ \pi\colon {\mathfrak T} \to {\mathbb P}^{1} $; here $ {\mathfrak T} $ is glued
using the function $ g\left(\lambda,t\right) $ defined in Theorem~\ref{th107.40}. It is clear that
$ \operatorname{ind} g=-\deg {\mathcal N}\Sigma_{\left(0,0,0\right)}=-2 $.

Functions $ x $, $ y $, $ z $, $ \widehat{w}\left(x,y,z\right) $ on the manifold of this Veronese web
define local coordinates on the fibers $ \widetilde{\pi}^{-1}\left(\lambda\right) $, $ \lambda\in\left\{\lambda_{1},\lambda_{2},\lambda_{3},0\right\} $, thus on
$ \pi^{-1}\left(\lambda\right) $. Denote these coordinates by the same symbols $ x $, $ y $, $ z $, $ \widehat{w} $.
Investigate how these coordinates are related to coordinates $ t_{+} $, $ t_{-} $ on
two pieces of $ {\mathfrak T} $ it is glued of.

Recall that Theorem~\ref{th107.40} defined the coordinate $ t_{-} $ on
$ \left({\mathbb P}^{1}\smallsetminus\bar{{\mathbb B}}_{\varepsilon}^{1}\right)\times{\mathbb C} $ by taking $ z $-coordinates of the intersection point of leaves of
the foliations with $ \gamma_{2} $, which is the $ z $-axis. The leaves of the foliation
$ {\mathcal F}_{\lambda_{3}} $ are $ z=\operatorname{const} $, thus the coordinates $ z $ and $ t_{-} $ on $ \pi^{-1}\left\{\lambda_{3}\right\} $ coincide, and
there is no translation of the argument $ z $ of the function $ \widehat{w} $. The
coordinate $ t_{+} $ over $ {\mathbb B}_{1/\varepsilon}^{1} $ is induced by taking $ x $-coordinate of the
intersection point of leaves with $ \gamma_{1} $. Thus the coordinates $ x $ and $ t_{+} $ on
$ \pi^{-1}\left(\lambda_{1}\right) $ coincide, and there is no translation of the argument $ x $.
Similarly, the coordinates $ y $ and $ t_{+} $ on $ \pi^{-1}\left(\lambda_{2}\right) $ differ by the
transformation $ y=Y\left(t_{+}\right) $; here $ y=Y\left(x\right) $, $ z=0 $ are the equations of the curve
$ \gamma_{1} $.

Finally, the coordinate $ \widehat{w} $ on $ \pi^{-1}\left(0\right) $ is given by taking the value of
$ \widehat{w}\left(x,y,z\right) $ on the leaf of $ {\mathcal F}_{0} $. The leaf which corresponds to a given value
of $ t_{+} $ passes through the point $ \left(t_{+},Y\left(t_{+}\right),0\right) $, thus the corresponding value
of $ \widehat{w} $ is $ \widehat{w}\left(t_{+},Y\left(t_{+}\right),0\right) $. \end{proof}

\begin{remark} Consider the case when $ w\left(x,y,z\right) $ is real for real values of
$ x,y,z $, and $ \lambda_{1,2,3} $ are real. In such a case it is a meaningful question to
reconstruct $ w $ basing on the Cauchy data on a hypersurface w.r.t.~
which the linearization is hyperbolic. As we have seen in Remark
~\ref{rem107.90}, the nonlinear Riemann problem with gluing data provided on a
neighborhood of a circle is not enough to treat such a problem. One
should be able to treat gluing data on more general regions.

However, the results of \cite{Tur99Cla,Tur99MemB} suggest that
providing the gluing data of some kind {\em on the real axis alone\/} should
provide enough information. Note that this gluing data should be more
general than one we consider here, since a literal application of our
arguments leads to a function $ \frac{\partial g}{\partial t} $ with zeros and/or poles on the real
axis. \end{remark}

\section{Appendix on transversal sections of webs }\label{h105}\myLabel{h105}\relax 

Some statements of this section are stated in complex-analytic case
only. To restate them in real-analytic case is straightforward.
Additionally, there is a $ C^{\infty} $-treatment of some of these statements as
well, see \cite{Tur99Cla,Tur99MemB}.

\begin{definition} Say that a submanifold $ N\subset M $ is {\em transversal\/} to the web $ {\mathcal F}_{\bullet} $ on $ M $
if it is transversal to any leaf of any foliation of the web. \end{definition}

If $ \operatorname{codim}{\mathcal F}_{\bullet}=r $, then the usual count of dimensions shows that there
are transversal varieties with dimensions down to $ r+\dim \Lambda $. However, one
should not expect them to exist is smaller dimensions, for example, Lemma
~\ref{lm10.40} shows that there are no curves transversal to a complex-analytic
Veronese web.

Obviously, a web $ {\mathcal F}_{\bullet} $ on $ M $ cuts out a smooth web $ {\mathcal F}_{\bullet}^{\left[N\right]} $ on a
transversal submanifold $ N\subset M $. Call this web the {\em transversal section\/} of $ {\mathcal F}_{\bullet} $
by $ N $. By definition of transversality, the germs of twistor transforms of
$ {\mathcal F}_{\bullet} $ and of $ {\mathcal F}_{\bullet}^{\left[N\right]} $ near $ m\in N $ coincide. This implies

\begin{theorem} Consider a separating airy web $ {\mathcal F}_{\bullet} $ on $ M $ and a transversal to
$ {\mathcal F}_{\bullet} $ submanifold $ N\subset M $. Then the transversal section web $ {\mathcal F}_{\bullet}^{\left[N\right]} $ on $ N $
determines the germ of $ M $ near $ N $ and the web $ {\mathcal F}_{\bullet} $ on this germ (uniquely up
to diffeomorphisms $ M \to \widetilde{M} $ which preserve $ N $). \end{theorem}

\begin{remark} Note that Theorem~\ref{th107.40} and taken together with Theorem
~\ref{th77.05} imply a particular case of this statement: the Veronese web is
locally determined by its restriction on the surface $ N $ given by $ y=Y\left(x\right) $
(in terms of Theorem~\ref{th107.40}). \end{remark}

On the other hand, classification of $ {\mathcal F}_{\bullet}^{\left[N\right]} $ up to diffeomorphism can
be much easier than classification of $ {\mathcal F}_{\bullet} $, since leaves of $ {\mathcal F}_{\bullet}^{\left[N\right]} $ have a
smaller dimension. If $ \dim  N=r+\dim \Lambda $, and $ \dim \Lambda=1 $, then leaves of $ {\mathcal F}_{\bullet}^{\left[N\right]} $ have
dimension 1. But to specify a foliation on $ N $ of codimension $ \dim  N-1 $ is
exactly the same as to specify a direction $ {\mathbit d}_{n}\in{\mathbb P}{\mathcal T}_{n}N $ in a tangent space at
every point $ n\in N $, there is {\em no\/} integrability condition involved (as, for
example, one in Lemma~\ref{lm4.30}). A family of foliations induces a family
of directions $ {\mathbit d}_{n}\left(\lambda\right) $. (Note that $ {\mathbit d}_{n}\left(\lambda\right) $ is a direction in a tangent space,
not in a cotangent space, as is $ {\mathbit n}_{n}\left(\lambda\right) $ in the case of webs of codimension
1.)

If $ \Lambda $ is a compact complex curve, then a mapping $ {\mathbit d}\colon \Lambda \to {\mathbb P}{\mathcal T}_{n}N $ of
given degree in general position is uniquely determined by images of $ P $
points on $ \Lambda $ for an appropriate $ P>0 $. The standard arguments of algebraic
geometry of curves show that for $ {\mathbit d}_{n} $ one should expect this for
$ P\geq r+2+\frac{g+\left(r+1\right)\left(d-r-1\right)}{r} $; here $ g $ is the genus of $ \Lambda $, $ d=\dim  M $.
Additionally, if this inequality is an equality, then the images of these
$ P $ points in $ {\mathbb P}{\mathcal T}_{n}N $ should be expected to be arbitrary. Moreover, in the
case $ g=0 $, $ r=1 $ these expectations can be easily checked to be true, as far
as among these $ P $ images no more than $ d-1 $ glue into any point of $ {\mathbb P}^{1} $.
Additionally, the condition of general position can be removed, if one
allows the degree of the mapping $ {\mathbit d}\colon {\mathbb P}^{1} \to {\mathbb P}^{1} $ to drop. This leads to

\begin{theorem} \label{th105.30}\myLabel{th105.30}\relax  Consider a complex-analytic Veronese web $ {\mathcal F}_{\bullet} $ on $ M $ and a
transversal surface $ N\subset M $, $ \dim  N=2 $, $ \dim  M=d $. Consider $ 2d-1 $ distinct points
$ \lambda_{1},\dots ,\lambda_{2d-1}\in{\mathbb P}^{1} $. Then the germ of $ {\mathcal F}_{\bullet} $ near $ N $ is determined (uniquely up to
a diffeomorphism preserving $ N $) by $ 2d-1 $ foliations $ {\mathcal F}^{\left(k\right)}={\mathcal F}_{\lambda_{k}}^{\left[N\right]} $,
$ k=1,\dots ,2d-1 $, of codimension 1 on $ N $. The foliations $ {\mathcal F}^{\left(k\right)} $ on $ N $ can be
taken arbitrarily with the restriction that at any point of $ N $ no more
than $ d-1 $ foliation have any given tangent direction, and there is no
mapping $ {\mathbit f}\colon {\mathbb P}^{1} \to {\mathbb P}^{1} $ of degree less than $ d-1 $ such that $ {\mathbit f}\left(\lambda_{k}\right)={\mathcal T}_{n}{\mathcal F}^{\left(k\right)} $,
$ k=1,\dots ,2d-1 $. \end{theorem}

By a choice of coordinates on $ N $ one can take last two of these
foliations to be $ \left\{x=\operatorname{const}\right\} $, $ \left\{y=\operatorname{const}\right\} $, the rest to be $ \left\{w_{k}\left(x,y\right)=\operatorname{const}\right\} $.
Thus the collection $ \left(M,N,{\mathcal F}_{\bullet}\right) $ (up to the same transformations as in the
theorem) is determined by $ 2d-3 $ functions on $ N $.

The next step is to use the freedom in the choice of $ N $ to reduce the
number of parameters. Accidentally, a proper choice of $ N $ also allows to
ensure that no mapping like $ {\mathbit f} $ exists.

Recall Lemma~\ref{lm10.40}: given a Veronese web on $ M $, a choice of a
subset $ T\subset{\mathbb P}^{1} $ with multiplicities and the total count $ \dim  M-1 $ determines a
direction at each point of $ M $. In particular, given $ T $ and $ m_{0}\in M $, there is a
canonically defined curve $ \gamma_{m_{0},T}\ni m_{0} $ on $ M $ (one with the prescribed
directions). Taking another subset $ T' $, one can put a curve $ \gamma_{m,T'} $ through
every point $ m $ of $ \gamma $. Taken together, these curves $ \gamma_{m,T'} $, $ m\in\gamma_{m_{0},T} $ sweep a
surface $ N_{m_{0},T,T'} $ in $ M $.

One can check that if $ T\cap T'=\varnothing $, then $ N_{m,T,T'} $ is transversal to $ {\mathcal F}_{\bullet} $ at
$ m $, thus in a neighborhood of $ m $. A proof of the following statement is
straightforward:

\begin{lemma} Suppose that $ N $ is transversal to a Veronese web $ {\mathcal F}_{\bullet} $ and contains
a curve $ \gamma_{m,T} $. Then $ \gamma $ is a leaf of $ {\mathcal F}_{\lambda}^{\left[N\right]} $ for any $ \lambda\in T $. \end{lemma}

In particular, for $ N=N_{m,T,T'} $ the foliations $ {\mathcal F}_{\lambda}^{\left[N\right]} $, $ \lambda\in T' $, coincide.
Since $ T' $ contains $ d-1 $ points, this condition ensures that no mapping $ {\mathbit f} $ of
degree smaller than $ d-1 $ can exist. Additionally, the leaves of the
foliations $ {\mathcal F}_{\lambda}^{\left[N\right]} $, $ \lambda\in T $, which pass through $ m $ coincide. This leads to the
following

\begin{corollary} Consider a complex-analytic surface $ N $, a point $ n\in N $, $ 2d-1 $
distinct points $ \lambda_{k} $ on $ {\mathbb P}^{1} $, and $ 2d-1 $ foliations $ \widetilde{{\mathcal F}}_{k} $, $ k=1,\dots ,P $, on $ N $. Let $ \gamma_{k} $
be the leaf of $ \widetilde{{\mathcal F}}_{k} $ through $ n $. Suppose that $ \widetilde{{\mathcal F}}_{k}=\widetilde{{\mathcal F}}_{k'} $ if $ 1\leq k,k'\leq d-1 $, $ \gamma_{k}=\gamma_{k'} $ if
$ d\leq k,k'\leq2d-2 $, that for any fixed $ k $, $ d\leq k\leq2d-1 $, the foliations $ \widetilde{{\mathcal F}}_{1} $ and $ \widetilde{{\mathcal F}}_{k} $,
$ d\leq k\leq2d-1 $, have distinct directions at any point of $ N $, and that directions
of the foliations $ \widetilde{{\mathcal F}}_{k} $, $ d\leq k\leq2d-1 $, are not all the same at any point of $ N $.
Then there is a complex-analytic Veronese web $ \left(M,{\mathcal F}_{\bullet}\right) $ of dimension $ d $ and
an embedding $ f\colon N\hookrightarrow M $ such that $ \operatorname{Im} f=N_{f\left(n\right),T,T'} $, and that $ f $ identifies the
foliations $ \widetilde{{\mathcal F}}_{k} $ on $ N $ with foliations $ {\mathcal F}_{\lambda_{k}}^{\left[\operatorname{Im} f\right]} $ on $ \operatorname{Im} f $; here
$ T=\left\{\lambda_{d},\dots ,\lambda_{2d-2}\right\} $, $ T'=\left\{\lambda_{1},\dots ,\lambda_{d-1}\right\} $. The germ of $ \left(M,{\mathcal F}_{\bullet}\right) $ near $ \operatorname{Im} f $ is
determined uniquely up to isomorphism. \end{corollary}

This is a geometric local classification of complex-analytic
Veronese webs: given $ M\ni n $ and a Veronese web on $ M $, $ N=N_{n,T,T'} $ is
canonically defined, thus foliations $ {\mathcal F}_{\lambda_{k}}^{\left[N\right]} $ are canonically defined.
These foliations satisfy the conditions of the corollary, and allow
reconstruction of the web on $ M $. In addition to the restriction that $ M $ is
defined only as a germ near $ N $, there is another direction of locality in
this result: $ N $ can be embedded into $ M $, not included into $ M $.

One can make appropriate modifications to this statements if some of
the points $ \lambda_{k} $, $ 1\leq k\leq d-1 $ or $ d\leq k\leq2d-2 $ can collide. In any case, count the
number of parameters in this representation. It is enough to specify $ \widetilde{{\mathcal F}}_{1} $
and $ \widetilde{{\mathcal F}}_{k} $, $ d\leq k\leq2d-1 $, one can suppose that $ \widetilde{{\mathcal F}}_{1} $ is $ \left\{x=\operatorname{const}\right\} $, $ \widetilde{{\mathcal F}}_{d} $ is $ \left\{y=\operatorname{const}\right\} $.
Then one can write $ \widetilde{{\mathcal F}}_{k} $ as $ \left\{w_{k}\left(x,y\right)=\operatorname{const}\right\} $, $ d+1\leq k\leq2d-1 $. Assume that the
point $ n $ is given by $ x=y=0 $. As in \cite{GelZakhWeb,GelZakh93,%
GelZakh99Web}, one can normalize $ w_{2d-1}\left(x,y\right)=x+y+xy\,u_{2d-1}\left(x,y\right) $.
Additionally, one can normalize $ w_{k}\left(x,y\right) $ by $ w_{k}\left(0,y\right)=y $. Since
$ \frac{dw_{k}\left(x,0\right)}{dx}=0 $, one can write $ w_{k}\left(x,y\right)=y+xy\,u_{k}\left(x,y\right) $. The functions
$ u_{k}\left(x,y\right) $ are defined uniquely up to a transformation
\begin{equation}
\widetilde{u}_{k}\left(x,y\right)=C^{-1}u_{k}\left(Cx,Cy\right),\qquad d+1\leq k\leq2d-1.
\notag\end{equation}
Thus an analytic $ d $-dimensional Veronese web on $ M $ near a point $ m\in M $ is
locally uniquely determined by $ d-1 $ functions $ u_{k}\left(x,y\right) $ up the
transformation above.

Note the similarity of this description with the
Turiel classification of Veronese webs \cite{Tur99Cla,Tur99MemB}. In fact
what we did above is just a geometric reformulation of this result.
Unfortunately, our approach works in an analytic situation only, and does
not imply the $ C^{\infty} $-case of the Turiel classification.

\begin{remark} In the case of arbitrary separating airy webs and $ \dim  N=r+\dim \Lambda $,
here $ r=\operatorname{codim}{\mathcal F}_{\bullet} $, it is not feasible to describe transversal sections of
webs by specifying several foliations on $ N $, since these foliations should
satisfy too many conditions. However, if $ \operatorname{codim}{\mathcal F}_{\bullet}=1 $ (so there are no
integrability conditions on $ {\mathcal F}_{\bullet}^{\left[N\right]} $) one can make a substitution. The
mapping $ {\mathbit n}_{n}\colon \Lambda \to {\mathcal T}_{n}^{*}N $ induces a line bundle $ {\mathcal L}_{n}={\mathbit n}_{n}^{*}{\mathcal O}\left(1\right) $ over $ \Lambda $, and an
inclusion $ \iota_{n}\colon {\mathcal T}_{n}N \hookrightarrow \Gamma\left(\Lambda,{\mathcal L}_{n}^{*}\right) $.

Suppose that $ \Lambda $ is a compact curve, then the latter space is
finite-dimensional. Since $ \iota_{n} $ (up to multiplication by a constant)
determines $ {\mathbit n}_{n} $, it is enough to provide enough information to describe $ {\mathcal L}_{n} $
and $ \iota_{n} $. Since we are free to multiply $ \iota_{n} $ by a constant, it is enough to
know $ {\mathcal L}_{n} $ up to isomorphism. We can see that to describe $ {\mathcal F}_{\bullet}^{\left[N\right]} $, it is
enough to describe the degree $ \delta $ of $ {\mathcal L}_{\bullet} $, provide the mapping $ l_{\bullet}\colon N \to
\operatorname{Pic}^{\delta}\left(\Lambda\right) $ which sends $ n $ to the class of $ {\mathcal L}_{n} $ inside the Picard variety, the
mapping $ \tau\colon N \to \operatorname{Gr}_{2}\left(\Gamma\left(\Lambda,l_{n}\right)\right)\colon n \mapsto \operatorname{Im}\iota_{n}\subset\Gamma\left(\Lambda,l_{n}\right) $, and an identification of
$ {\mathcal T}_{n}N $ with the $ 2 $-dimensional vector subspace described by $ \tau\left(n\right) $ up to a
constant. In general position, given $ l $ and $ \tau $, one needs to know $ {\mathcal F}_{\lambda}^{\left[N\right]} $ for
3 values of $ \lambda $ to provide such an identification.

If $ \Lambda={\mathbb P}^{1} $, then $ \delta $ is a number, and $ \operatorname{Pic}^{\delta}\left(\Lambda\right) $ has one point only. It is
clear that $ {\mathcal L}_{n}\simeq{\mathcal O}\left(\dim  M-1\right) $, so it is enough to provide 3 foliations on $ N $,
and a mapping $ N \to \operatorname{Gr}_{2}\left({\mathbb V}^{\dim  M}\right) $. It is easy to see that this data is
equivalent to the data of Theorem~\ref{th105.30}. \end{remark}

Consider now a transversal submanifold $ N $ to a web $ {\mathcal F}_{\bullet} $, and a
submanifold $ \gamma\subset N $ of codimension $ \operatorname{codim}{\mathcal F}_{\bullet} $. Let $ m\in\gamma $. If $ {\mathcal T}_{m}\gamma\subset{\mathcal T}_{m}N $ is in general
position, then $ {\mathcal T}_{m}\gamma $ is transversal to $ {\mathcal T}_{m}{\mathcal F}_{\lambda} $ for $ \lambda\in\Lambda\smallsetminus Z $, here $ Z $ is a proper
analytic subset of $ \Lambda $. Consequently, a small neighborhood of $ m $ in $ \gamma $ is
transversal to $ {\mathcal F}_{\lambda} $ for $ \lambda $ in an open subset $ U\subset\Lambda $. Reducing $ M $ to a
neighborhood of $ m $, we obtain the corresponding sectional coordinate
system on $ {\mathfrak T} $. Given two such submanifolds $ \gamma_{1} $, $ \gamma_{2} $, $ m\in\gamma_{1}\cap\gamma_{2} $ we obtain two
subsets $ U_{1,2}\subset\Lambda $, and the corresponding local identifications $ g_{\lambda}\colon \gamma_{1} \to \gamma_{2} $,
$ \lambda\in U_{1}\cap U_{2} $. This identification are obtained in the same way as in Section
~\ref{h10}, the principal difference being that the whole construction is
performed on $ N $ instead of $ M $.

Taking enough $ \gamma_{k} $ to cover $ \Lambda $, the corresponding pairwise gluing
functions determine the germ of $ {\mathfrak T} $ near $ \Sigma_{m} $ up to isomorphism, thus the
germ of $ {\mathcal F}_{\bullet} $ near $ m $ up to isomorphism (assuming $ {\mathcal F}_{\bullet} $ is airy). Note that to
construct $ g_{\lambda} $, we need to find a leaf of $ {\mathcal F}_{\lambda}^{\left[N\right]} $ which passes through a
given point of $ \gamma_{1} $, and find the intersection of this leaf with $ \gamma_{2} $.
Obviously, to do this it is enough to solve some ordinary differential
equations.

Consequently, the construction of Theorem~\ref{th107.40} can be
generalized to arbitrary webs.

\section{Appendix on computational complexity of the nonlinear Riemann
transform }\label{h12}\myLabel{h12}\relax 

Continue using notations of Section~\ref{h11}. Consider not the mapping
$ {\mathfrak R}\colon g\left(\lambda,t\right) \mapsto \sigma_{+}\left(0\right) $, but a more general mappings $ \widetilde{{\mathfrak R}}_{\pm}\colon g\left(\lambda,t\right) \mapsto \sigma_{\pm}\left(\lambda\right) $. Let
us introduce operators solving the linear Riemann problem: given a
function $ \varphi\left(\lambda\right) $ defined for $ \varepsilon<|\lambda|<1/\varepsilon $, define functions $ {\mathbb H}_{+}\varphi $ and $ {\mathbb H}_{-}\varphi $ by the
conditions $ \varphi\left(\lambda\right)=\lambda{\mathbb H}_{+}\varphi\left(\lambda\right)+{\mathbb H}_{-}\varphi\left(\lambda\right) $ and the conditions that $ {\mathbb H}_{+}\varphi\left(\lambda\right) $ and $ {\mathbb H}_{-}\varphi\left(\lambda\right) $
can be holomorphically extended on $ |\lambda|<1/\varepsilon $ and $ |\lambda|>\varepsilon $ correspondingly.
Similarly, if $ \varphi\left(\lambda\right) $ is nowhere 0, and $ \operatorname{ind} \varphi=0 $, define $ {\mathbb M}_{+}\varphi $ and $ {\mathbb M}_{-}\varphi $ by
$ \varphi\left(\lambda\right)={\mathbb M}_{+}\varphi\left(\lambda\right)/{\mathbb M}_{-}\varphi\left(\lambda\right) $ and the conditions that $ {\mathbb M}_{+}\varphi\left(\lambda\right) $ and $ {\mathbb M}_{-}\varphi\left(\lambda\right) $ can be
holomorphically extended on $ |\lambda|<1/\varepsilon $ and $ |\lambda|>\varepsilon $ correspondingly, these
extensions are nowhere 0, and $ {\mathbb M}_{+}\varphi\left(0\right)=1 $.

Uniqueness of $ {\mathbb M}_{+}\varphi $ and $ {\mathbb M}_{-}\varphi $ is obvious, existence follows from the
theory of the linear Riemann problem---or, what is the same,
classification of line bundles over $ {\mathbb P}^{1} $. Uniqueness of $ {\mathbb H}_{+}\varphi $ and $ {\mathbb H}_{-}\varphi $ is
obvious, existence follows from existence of $ \log  {\mathbb M}_{+}e^{\varphi} $ and $ \log  {\mathbb M}_{-}e^{\varphi} $.

\begin{lemma} Denote by $ S^{1} $ the circle $ |\lambda|=1 $. Then $ {\mathbb H}_{\pm}\varphi|_{S^{1}} $ is uniquely
determined by $ \varphi|_{S^{1}} $. This induces two linear operators on real-analytic
complex-valued functions on $ S^{1} $. These operators can be extended to
continuous linear operators in Sobolev spaces $ H^{s}\left(S^{1}\right) $ for any $ s\in{\mathbb R} $. \end{lemma}

\begin{proof} The first statement is obvious, since $ \varphi|_{S^{1}} $ uniquely determines
$ \varphi $. The second statement follows from $ {\mathbb H}_{+}\lambda^{k}=c_{k}\lambda^{k-1} $ and $ {\mathbb H}_{-}\lambda^{k}=c'_{k}\lambda^{k-1} $ with $ c_{k} $
and $ c'_{k} $ being 0 or 1, and from the fact that $ \left(\lambda^{k}\right)_{k\in{\mathbb Z}} $ is an orthogonal
basis in the Sobolev spaces $ H^{s}\left(S^{1}\right) $. \end{proof}

Denote the continuations of operators $ {\mathbb H}_{\pm} $ into $ H^{s}\left(S^{1}\right) $ by the same
symbols. Similarly, if $ s>1/2 $, then the mappings $ {\mathbb M}_{\pm} $ can be considered as
continuous mappings from an open subset of $ H^{s}\left(S^{1}\right) $ into $ H^{s}\left(S^{1}\right) $. Indeed, if
$ s>1/2 $, then $ \varphi \mapsto e^{\varphi} $ is a continuously differentiable mapping $ H^{s}\left(S^{1}\right) \to
H^{s}\left(S^{1}\right) $ with an open image.

\begin{lemma} \label{lm12.20}\myLabel{lm12.20}\relax  Consider $ \varepsilon,\delta>0 $ and a family $ g_{\kappa}\left(\lambda,t\right) $, $ \kappa\in K $, of functions
such that $ {\mathfrak R}_{\varepsilon\delta}\left(g_{\kappa}\right) $ is well-defined for any $ \kappa\in K $. Let $ \sigma_{\pm,\kappa}\left(\lambda\right)=\widetilde{{\mathfrak R}}_{\pm}\left(g_{\kappa}\right) $. Then
\begin{equation}
\frac{\partial}{\partial\kappa}\widetilde{{\mathfrak R}}_{\pm}\left(g_{\kappa}\right)= a_{\pm,\kappa}^{-1}\frac{\partial b_{\pm,\kappa}}{\partial\kappa},
\notag\end{equation}
here
\begin{equation}
a_{\pm,\kappa}\left(\lambda\right)={\mathbb M}_{\pm}\left(\lambda^{-1}\frac{\partial g_{\kappa}}{\partial t}\left(\lambda,\sigma_{+,\kappa}\left(\lambda\right)\right)\right),\qquad b_{\pm,\kappa}\left(\lambda\right)={\mathbb H}_{\pm}\left(a_{-,\kappa}\left(\lambda\right)g_{\kappa}\left(\lambda,\sigma_{+,\kappa}\left(\lambda\right)\right)\right).
\notag\end{equation}
\end{lemma}

\begin{proof} Fix $ \kappa_{0}\in K $. We may assume that $ \dim  K=1 $, for example, $ K={\mathbb B}_{r}^{1} $. Let
$ \sigma_{\pm}=\widetilde{{\mathfrak R}}_{\pm}\left(g_{\kappa_{0}}\right) $, $ \delta_{\pm}=\frac{d}{d\kappa}\widetilde{{\mathfrak R}}_{\pm}\left(g_{\kappa}\right)|_{\kappa_{0}} $. Then
\begin{equation}
\frac{\partial g_{\kappa_{0}}}{\partial t}\left(\lambda,\sigma_{+}\left(\lambda\right)\right)\delta_{+}\left(\lambda\right) + \frac{\partial g_{\kappa}}{\partial\kappa}|_{\kappa_{0}}\left(\lambda,\sigma_{+}\left(\lambda\right)\right) = \delta_{-}\left(\lambda\right).
\notag\end{equation}
Since $ \operatorname{ind}\frac{\partial g}{\partial t}=1 $, one can write $ \frac{\partial g_{\kappa_{0}}}{\partial t}\left(\lambda,\sigma_{+}\left(\lambda\right)\right) $ as $ \lambda a_{+}\left(\lambda\right)/a_{-}\left(\lambda\right) $,
here $ a_{+}\left(\lambda\right) $ and $ a_{-}\left(\lambda\right) $ have invertible holomorphic continuations into
$ |\lambda|<1/\varepsilon $ and $ |\lambda|\geq\varepsilon $ correspondingly. Similarly, write
\begin{equation}
a_{-}\left(\lambda\right)\frac{\partial g_{\kappa}}{\partial\kappa}|_{\kappa_{0}}\left(\lambda,\sigma_{+}\left(\lambda\right)\right) = \lambda B_{+}\left(\lambda\right)+B_{-}\left(\lambda\right),
\notag\end{equation}
here $ B_{+}\left(\lambda\right) $ and $ B_{-}\left(\lambda\right) $ have holomorphic continuations into $ |\lambda|<1/\varepsilon $ and
$ |\lambda|\geq\varepsilon $ correspondingly. Then $ \sigma_{+}\left(\lambda\right)=a_{+}^{-1}B_{+} $, $ \sigma_{-}\left(\lambda\right)=a_{-}^{-1}B_{-} $. Since operators
$ {\mathbb H}_{\pm} $ are linear and continuous in an appropriate topology, it is easy to
see that $ B_{\pm}=\frac{db_{\pm,\kappa}}{d\kappa}|_{\kappa_{0}} $. \end{proof}

Consider a vector space $ V=H^{s}\left(S^{1}\right)\times H^{s}\left(S^{1}\right) $, $ s\geq1/2 $. Denote the element
of $ V $ by $ \left(\sigma_{+},\sigma_{-}\right) $. In the conditions of Lemma~\ref{lm12.20} suppose that $ \dim 
K=1 $. Define a mapping $ v_{\kappa}\colon U \to V\colon \left(\sigma_{+},\sigma_{-}\right) \mapsto \left(\delta_{+},\delta_{-}\right) $, here $ U $ is an
appropriate open subset of $ V $, and $ \delta_{\pm}\left(\lambda\right) = a_{\pm,\kappa}\left(\lambda\right)^{-1}B_{\pm,\kappa}\left(\lambda\right) $,
\begin{equation}
a_{\pm,\kappa}\left(\lambda\right)={\mathbb M}_{\pm}\left(\lambda^{-1}\frac{\partial g_{\kappa}}{\partial t}\left(\lambda,\sigma_{+}\left(\lambda\right)\right)\right),\qquad B_{\pm,\kappa}\left(\lambda\right)={\mathbb H}_{\pm}\left(a_{-}\left(\lambda\right)\frac{dg_{\kappa}}{d\kappa}\left(\lambda,\sigma_{+}\left(\lambda\right)\right)\right).
\notag\end{equation}
Since one can take value of elements of $ H^{s}\left(S^{1}\right) $, $ s>1/2 $, at points, it
makes sense to require that $ |\sigma_{+}\left(\lambda\right)|<\delta $ for any $ \lambda $, thus $ v_{\kappa}\left(\sigma_{+},\sigma_{-}\right) $ is indeed
well-defined on an open subset of $ H^{s}\left(S^{1}\right) $. Moreover, $ v_{\kappa} $ is Lipschitz on
$ K\times{\mathbb B} $, here $ {\mathbb B} $ is any ball in $ H^{s}\left(S^{1}\right) $.

\begin{corollary} Given a family $ g_{\kappa}\left(\lambda,t\right) $, $ \kappa\in K $, $ \dim  K=1 $, of functions as in
Definition~\ref{def11.40}, one can define a Lipschitz family $ v_{\kappa} $ of vector
fields on an open subset $ U $ of $ H^{s}\left(S^{1}\right)\times H^{s}\left(S^{1}\right) $ such that if $ \widetilde{{\mathfrak R}}_{\pm}\left(g_{\kappa}\right) $ makes
sense for any $ \kappa\in K $, then the curve $ \left(\widetilde{{\mathfrak R}}_{+}\left(g_{\kappa}\right),\widetilde{{\mathfrak R}}_{-}\left(g_{\kappa}\right)\right) $ is an integral curve of
the ODE $ \frac{d\Phi\left(\kappa\right)}{d\kappa}=v_{\kappa}\left(\Phi\right) $. \end{corollary}

Now Lipschitz ODEs in Banach spaces enjoy most of the properties of
finite-dimensional ODEs, and are not harder to solve. We conclude that in
the setting of Corollary~\ref{cor11.50} one can calculate $ {\mathfrak R}\left(g_{\kappa}\right) $ by solving a
Lipschitz ODE in a Hilbert space. In particular, Theorem~\ref{th16.70} reduces
solution of Cauchy problem for the nonlinear wave equation to solution of
such an ODE.

\bibliography{ref,outref,mathsci}
\end{document}